\newcommand{\Msun}{\ensuremath{\textrm{\,M}_{\odot}}}
\newcommand{\Mzams}{\ensuremath{\textrm{\,M}_{ZAMS}}}
\newcommand{\Mcsm}{\ensuremath{\textrm{\,M}_{CSM}}}
\newcommand{\Rsun}{\ensuremath{\textrm{\,R}_{\odot}}}
\newcommand{\Rcsm}{\ensuremath{\textrm{\,R}_{CSM}}}
\newcommand{\Rstar}{\ensuremath{\textrm{\,R}_{\star}}}

\newcommand{\kms}{\ensuremath{\textrm{\,km s}^{-1}}}

\newcommand{\ha}{\ensuremath{\mathrm{H}\alpha}}
\newcommand{\hb}{\ensuremath{\mathrm{H}\beta}}
\newcommand{\hg}{\ensuremath{\mathrm{H}\gamma}}
\newcommand{\hd}{\ensuremath{\mathrm{H}\delta}}
\newcommand{\Feii}{\ensuremath{\mbox{\ion{Fe}{2}}}}
\newcommand{\NaiD}{\ensuremath{\mbox{\ion{Na}{1}\,D}}}
\newcommand{\ebv}{E(B--V)}
\newcommand{\Rad}{\ensuremath{\textrm{\,R}}}
\newcommand{\Rext}{\ensuremath{\textrm{\,R}_{ext}}}
\newcommand{\rv}{\ensuremath{\mathrm{R_V}}}
\newcommand{\Vs}{\ensuremath{\textrm{\,V}_s}}
\newcommand{\Menv}{\ensuremath{\textrm{\,M}_{env}}}
\newcommand{\DL}{\ensuremath{\textrm{\,d}_{L}}}
\newcommand{\too}{\ensuremath{\textrm{\,t}_{0}}}
\newcommand{\frho}{\ensuremath{\mbox{f}_{\rho}}}
\newcommand{\cms}{\ensuremath{\textrm{\,cm s}^{-1}}}

\newcommand{\Ni}{\ensuremath{\mbox{Ni}}}

\newcommand{\he}{\ensuremath{\mbox{\ion{He}{1}}}}
\newcommand{\N}{\ensuremath{\mbox{\ion{N}{3}/\ion{N}{4}/\ion{N}{5}}}}

\newcommand{\gsim}{\raisebox{-0.13cm}{~\shortstack{$>$ \\[-0.07cm] $\sim$}}~}
\DeclareUnicodeCharacter{2212}{-}

\documentclass[twocolumn]{aastex631} 
\begin{document}

\newcommand{\vdag}{(v)^\dagger}
\newcommand\aastex{AAS\TeX}

\title{SN\,2020bij and a Possible Slow-Rise High-Velocity Subclass of Type IIP Supernovae}

\correspondingauthor{Sondos Mohsen-Tanev}
\email{sondosmohsen@tauex.tau.ac.il}

\author[0009-0006-7842-2125]{Sondos Mohsen-Tanev}
\affiliation{The School of Physics and Astronomy, Tel Aviv University, Tel Aviv 69978, Israel}

\author[0000-0001-7090-4898]{Iair Arcavi}
\affiliation{The School of Physics and Astronomy, Tel Aviv University, Tel Aviv 69978, Israel}

\author[0000-0001-5320-3352]{Shahar Bracha}
\affiliation{McGovern Institute for Brain Research, Massachusetts Institute of Technology, Cambridge, MA}

\author[0000-0002-4924-444X]{K. Azalee Bostroem}
\altaffiliation{LSSTC Catalyst Fellow}
\affiliation{Steward Observatory, University of Arizona, 933 North Cherry Avenue, Tucson, AZ 85721-0065, USA}

\author[0000-0002-0832-2974]{Griffin Hosseinzadeh}
\affiliation{Department of Astronomy \& Astrophysics, University of California, San Diego, 9500 Gilman Drive, MC 0424, La Jolla, CA 92093-0424, USA}

\author[0000-0003-1546-6615]{Jesper Sollerman}
\affiliation{The Oskar Klein Centre, Department of Astronomy, Stockholm University AlbaNova, SE-106 91 Stockholm, Sweden}

\author[0000-0003-2375-2064]{Claudia P. Guti\'errez}
\affiliation{Institut d'Estudis Espacials de Catalunya (IEEC), Edifici RDIT, Campus UPC, 08860 Castelldefels (Barcelona), Spain}
\affiliation{Institute of Space Sciences (ICE, CSIC), Campus UAB, Carrer de Can Magrans, s/n, E-08193 Barcelona, Spain}

\author[0000-0002-8041-8559]{Priscila J. Pessi}
\affiliation{Astrophysics Division, National Centre for Nuclear Research, Pasteura 7, 02-093 Warsaw, Poland}

\author[0000-0003-0227-3451]{Joseph Anderson}
\affiliation{European Southern Observatory, Alonso de Córdova 3107, Vitacura, Casilla 19001, Santiago, Chile}

\author[0000-0002-1650-1518]{Mariusz Gromadzki}
\affiliation{Astronomical Observatory, University of Warsaw, Al. Ujazdowskie 4, 00-478 Warszawa, Poland}

\author[0000-0002-3653-5598]{Avishay Gal-Yam}
\affiliation{Department of Particle Physics and Astrophysics, Weizmann Institute of Science, 7610001 Rehovot, Israel.}

\author[0000-0002-1125-9187]{Daichi Hiramatsu}
\affiliation{Department of Astronomy, University of Florida, 211 Bryant Space Science Center, Gainesville, FL 32611-2055 USA}

\author[0000-0003-0035-6659]{Jamison Burke}
\affiliation{Shady Side Academy, 423 Fox Chapel Road, Pittsburgh, PA 15238, USA}

\author{Koichi Itagaki}
\affiliation{Itagaki Astronomical Observatory, Yamagata 990-2492, Japan}

\author[0000-0002-1066-6098]{Ting-Wan Chen}
\affiliation{Graduate Institute of Astronomy, National Central University, 300 Jhongda Road, 32001 Jhongli, Taiwan}

\author[0000-0003-4253-656X]{D. Andrew Howell}
\affiliation{Las Cumbres Observatory, 6740 Cortona Dr, Suite 102, Goleta, CA 93117-5575, USA}
\affiliation{Department of Physics, University of California, Santa Barbara, CA 93106-9530, USA}

\author[0000-0001-5807-7893]{Curtis McCully}
\affiliation{Las Cumbres Observatory, 6740 Cortona Dr, Suite 102, Goleta, CA 93117-5575, USA}

\author[0000-0001-9570-0584]{Megan Newsome}
\affiliation{University of Texas at Austin, 1 University Station C1400, Austin, TX 78712-0259, USA}

\author[0000-0003-0209-9246]{Estefania Padilla Gonzalez}
\affiliation{Space Telescope Science Institute, Baltimore, MD 21218, USA}

\author[0000-0002-7472-1279]{Craig Pellegrino}
\affiliation{Goddard Space Flight Center, 8800 Greenbelt Road, Greenbelt, MD 20771, USA}

\begin{abstract}
Mapping how the explosion properties of Type II supernovae (SNe II) relate to the properties of their progenitors can provide strong constraints for understanding the final evolutionary stages of massive stars. Type IIP SNe, linked to the explosions of single red super-giant (RSG) stars, have recently been found to require some form of interaction with circumstellar material (CSM) to reproduce the rapid rise to the plateau often seen in their light curves. In this work, we present observations and analysis of the Type IIP SN\,2020bij, characterized by a slow rise to its plateau as well as high expansion velocities. We identify four other SNe IIP from the literature (ASASSN-14kg, SN\,2018fif, SN\,2021yja and SN\,2023axu) with similarly slowly rising light curves and find that they also show high expansion velocities. Using both analytical and numerical models, all five events can be explained with weak to no CSM interaction. We therefore propose that these events constitute a new subclass of Type IIP SNe which could be associated with relatively confined CSM. Early and dense photometric coverage of future SNe IIP together with early spectroscopic observations will further map this subclass and its physical properties. Understanding such rare events could be key to constraining the diversity of late-stage mass-loss in RSGs.
\end{abstract}


\section{Introduction}
\label{sec:intro}

Massive stars (with zero-age-main-sequence masses \Mzams\,\gsim 8\,\Msun) are relatively rare but play important roles in the Universe. For example, they provide energy and chemical feedback through their deaths as core-collapse supernovae (SNe), which are also the birth places of neutron stars and black holes \citep[e.g.][]{Woosley_2002,Heger_2003}. As such, understanding the physics of SNe, and specifically the connection between the diversity of massive-star evolution and the diversity of core-collapse SNe, is an important goal.

Core-collapse SNe are classified mainly according to their spectral properties \citep{Filippenko_1997,Gal-Yam2017}.
They are traditionally divided into two main types based on the presence of hydrogen in their spectra: Type I (H-free) and Type II (H-rich), of which the subclass IIP (named after the distinctive plateau in their light curve) is the most common type of SN realized in nature (e.g. \citealt{Li_2011,Smith_2011,Shivvers_2017,Perley_2020}; see \citealt{Arcavi_2017} for a review of H-rich SN types). Using direct detections of their pre-explosion progenitors, Type IIP SNe were found to be the explosions of single hydrogen-rich red super-giant (RSG) stars \citep[see, for example,][and references therein]{Smartt_2009,Leonard_2011,Smartt_2015,Dyk_2017}.

In recent years, advancements in high-cadence transient surveys and in follow-up facilities have enabled the detection and rapid follow-up of SNe mere hours after their first light \citep[e.g.][]{Gal-Yam_2011, Nugent_2011, Gal-Yam_2014,Arcavi2017gkg, Tartaglia_2017,Yaron_2017, Jacobson_2023, Jacobson_2024}.
Early emission from core-collapse SNe can provide constraints on the SN progenitor properties and on the explosion mechanism. 
One such property is mass-loss, which is an important, yet poorly understood process in massive-star evolution \citep[e.g.][]{Smith_2014,Beasor_2020,Massey_2023}. 
Modeling early SN emission can reveal interaction of the SN ejecta or radiation with circumstellar material (CSM) ejected from the progenitor star before explosion, and can thus constrain the properties and type of mass-loss experienced by the star during the final stages of its evolution.
 
Signatures of CSM interaction can be seen, for example, in SN spectra. The spectra of Type IIn SNe are characterized by narrow ($\sim$100--1000\,\kms) hydrogen emission lines, which indicate the presence of slower-moving material (compared to the SN ejecta which are typically moving at several 1000 to 10,000\,\kms). The narrow lines in SNe IIn are detected for months or even years after explosion, implying an extended CSM \citep[e.g.][]{Schlegel_1990, Chugai_1991, Kiewe2012, Shivvers_2015, Smith_2015}. 

Similar signatures, but much shorter lived (hours to days) have been discovered in many SNe that are not of Type IIn (i.e. not showing these signatures at later times). Such so-called ``flash features'' \citep{Gal-Yam_2014}, are attributed to confined CSM being ``flash ionized'' by the initial shock breakout flash from the SN. 
\cite{Yaron_2017} used spectra obtained $\sim$6 hours to a few days after explosion of the Type IIP SN\,2013fs to infer the presence of dense CSM confined to a distance of roughly ${10}^{15}$\,cm from the progenitor star at explosion, ejected at a rate of $\sim {10}^{−3}$\,\Msun\,yr$^{-1}$ during the final year before explosion. More recently, the prompt discovery of the very nearby Type IIP SN\,2023ixf \citep{2023ixf_discovery,2023ixf_classification} enabled exceptionally early-time observations, providing a unique opportunity to study its early emission, progenitor and CSM properties. These observations showed that SN\,2023ixf had a dense, solar-metallicity CSM confined to a distance of roughly $(0.5-1) \times {10}^{15}$\,cm, ejected by the progenitor at a mass-loss rate of $\sim {10}^{−3} - {10}^{−2}$\,\Msun\,yr$^{-1}$ shortly before explosion \citep[see for e.g.][and references therein]{Bostroem_2023,Hosseinzadeh_2023, Jacobson_2023, Teja_2023, Zimmerman_2024}.

``Flash features'' are common in early spectra of Type II SNe, indicating that many SN progenitors experience some form of enhanced mass-loss in the months or years prior to explosion \citep[e.g.][]{Gal-Yam_2014, Smith_2015, Yaron_2017, Hosseinzadeh_2018, Maayane}.
\cite{Khazov_2016} found that 12 out of 84 SNe II ($\sim$14$\%$) discovered by the Palomar Transient Factory \citep[PTF;][]{Law_2009,Rau_2009}, show flash-ionized features, detected in their first spectra obtained within 10 days from explosion. More recently, \cite{Bruch_2021,Bruch_2023} analyzed a sample of 30 SNe II, detected and classified by the Zwicky Transient Facility \citep[ZTF;][]{Bellm_2019}, with the first spectrum obtained $<$ 2 days from explosion, and found that $\gtrsim$ 36$\%$ (at the 95$\%$ confidence level) of them showed flash-ionization features, pointing to the presence of dense CSM close to the progenitor.

A complementary method for probing confined CSM in SNe is through their early light curves. \cite{Morozova_2017,Morozova_2018} use radiation-hydrodynamical modeling of Type IIP SNe and find that $\sim 0.18-0.83$\,\Msun\ of CSM, extending out to $\sim 5\times10^{13}$ -- $2\times10^{14}$\,cm is required to explain all 20 well-observed SN IIP light curves in their sample (without adding this CSM to the models, the light curve rise is significantly slower than observed). \cite{Hinds_2025} apply similar methods to a larger sample of Type II SNe and also find that substantial ($\geq10^{-2.5}$\,\Msun) CSM within $10^{15}$\,cm is common at the time of explosion. These results further strengthen the conclusion that significant mass-loss episodes are ubiquitous in RSGs shortly \citep[months to years;][]{Moriya_2011, Morozova_2018} before explosion.

Here we present a counterexample: the Type IIP SN\,2020bij, which is characterized by a relatively long rise of 14 days to the plateau in the $r$-band, as predicted by \cite{Morozova_2018} for CSM-free explosions. Most Type IIP SNe exhibit rise times of $\lesssim$10 days in the $r$-band \citep[e.g.][]{Gall_2015, Rubin_2016, Rubin_GalYam_2016} and even shorter rise times in the $g$-band \citep{Gonzalez_2015}. 
Longer $r$-band rise times of $\gtrsim$14--15 days are also observed, although they are less common and tend to lie toward the long rise end of the distribution of Type II SNe \citep{Gonzalez_2015, Pessi_2019}.

We present the discovery and observations of SN\,2020bij in Sections \ref{sec:Discovery} and \ref{sec:Observations}, respectively. In Section \ref{sec:Sample} we identify four additional (already published) Type IIP SNe with a similarly slow rise to plateau. We analyze our photometry and spectroscopy in Sections \ref{sec:Photometric_analysis} and \ref{sec:Spectroscopic_analysis} in the context of this broader sample, and compare their properties to those of typical SNe IIP. We discuss our main results in Section \ref{sec:Discussion} and summarize our conclusions in Section \ref{sec:Summary}.

\section{Discovery and Classification} 
\label{sec:Discovery}

\begin{figure}
\includegraphics[width=0.47\textwidth]{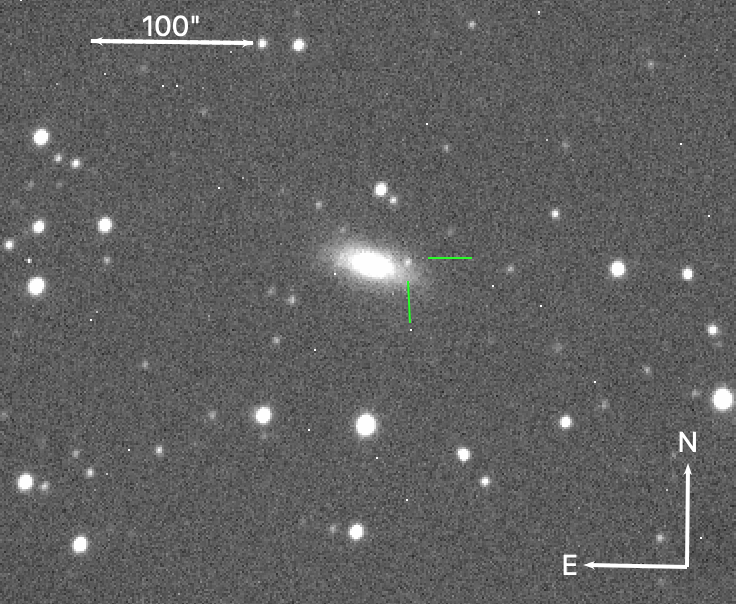}
\centering
\caption{The unfiltered discovery image of SN\,2020bij in the spiral galaxy NGC 3463, obtained with the KAF-1001E CCD mounted on the 0.35m reflector at the Itagaki Astronomical Observatory in Kochi, Japan \citep{Itagaki_2020bij_discovery}, taken on MJD 58877.68 (2020-01-29 16:22:49). The position of SN\,2020bij is indicated by green tick marks.}
\label{fig:Host_galaxy_NGC3463}
\end{figure}

We discovered SN\,2020bij in the spiral galaxy NGC 3463 \citep[which has a redshift of $z = 0.013276\pm0.000067$;][]{redshift_NED} on MJD 58877.68 (2020-01-29 16:22:49, UT used throughout) at a discovery magnitude of $\sim$17.5 in an unfiltered image, at Right Ascension ${10}^h{55}^m{11}^s.690$ and Declination $-{26}^d{08}^m{25}^s.80$ (J2000) using the KAF-1001E CCD mounted on the 0.35m reflector at the Itagaki Astronomical Observatory in Kochi, Japan \cite[Figure \ref{fig:Host_galaxy_NGC3463};][]{Itagaki_2020bij_discovery}. The last pre-explosion non-detection was on MJD 58875.58 (2020-01-27 13:52:36) by the Asteroid Terrestrial-impact Last Alert System \citep[ATLAS;][]{ATLAS_data1_Tonry2018} transient survey \citep{ATLAS_data2_Smith2020}. Using the {ATLAS Public Forced Photometry Server}\footnote{\url{https://fallingstar-data.com/forcedphot/}} \citep{ATLAS_data1_Tonry2018,ATLAS_data2_Smith2020, Shingles_2021} and co-addition of the individual exposures obtained during the same epoch, we find a $5\sigma$ non-detection limiting magnitude of 20.27 in the $o$-band (additional details about photometry extraction and the data processing of ATLAS can be found in \citealt{ATLAS_data1_Tonry2018} and \citealt{ATLAS_data2_Smith2020}). The first detection was on MJD 58877.49 (2020-01-29 11:49:50) by ATLAS at a magnitude of $18.27 \pm 0.07$ in the $c$-band. We adopt the midpoint between the last non-detection and the first detection as the explosion epoch of SN\,2020bij, and take its uncertainty as half of the difference, i.e. MJD 58876.54 (2020-01-28 12:51:13) $\pm$ 0.96.

The first spectrum of SN\,2020bij was obtained by the advanced Extended Public European Southern Observatory (ESO) Spectroscopic Survey of Transient Objects \citep[ePESSTO+;][]{PESSTO_Smartt_2015}\footnote{\url{https://www.pessto.org}} collaboration 3.57 days after discovery (i.e. 4.71 days from explosion) and was used to classify the transient as a Type II SN \citep{Irani_SN2020bij_classification,Zimmerman_SN2020bij_classification} based on the presence of broad hydrogen emission lines. 

We adopt the cosmological parameters from the Nine-Year Wilkinson Microwave Anisotropy Probe (WMAP9) data \citep[$H_0 = 69.32\ \mathrm{\kms\ Mpc^{-1}}$, and $\Omega_m = 0.286$;][]{WMAP9}, which imply a distance modulus of $\mu$ = 33.80 $\pm$ 0.10\,mag to SN\,2020bij from the host-galaxy redshift, corresponding to a distance of $57.54 \pm 2.65$\,Mpc. This is consistent with the redshift-independent Tully-Fisher-based distance modulus estimates of \cite{Willick1997}, retrieved via the NASA/IPAC Extra-galactic Database (NED)\footnote{\url{https://ned.ipac.caltech.edu/}}.

\section{Observations and Data Reduction} 
\label{sec:Observations}

\begin{figure*}
\centering
\includegraphics[width=1\textwidth]{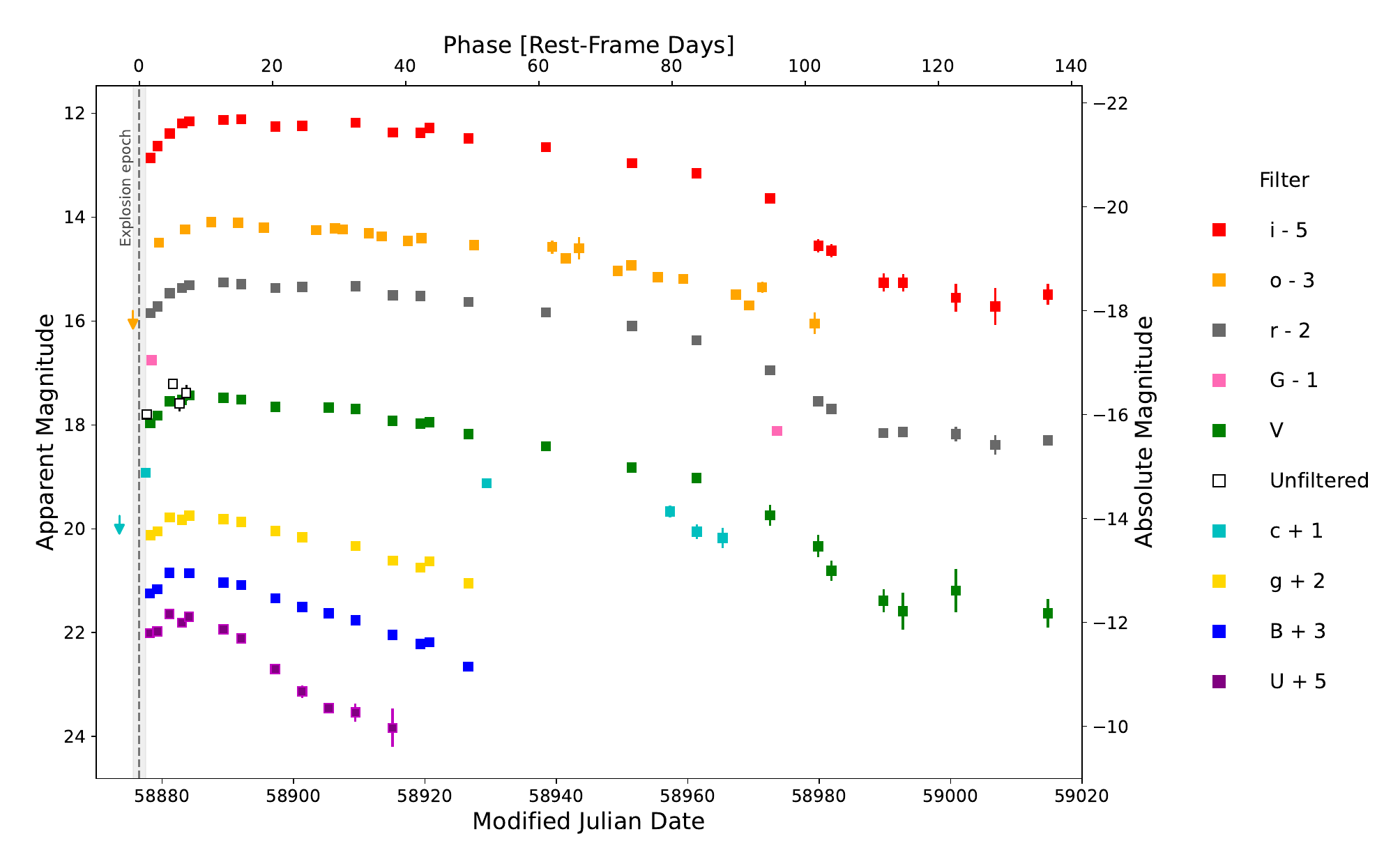}
\caption{Milky-Way extinction-corrected light curves of SN\,2020bij from observations obtained with Las Cumbres Observatory ($UBVgri$-band), Gaia ($G$-band), ATLAS ($oc$-band) and from the Itagaki Astronomical Observatory (unfiltered). Error bars represent 1$\sigma$ uncertainties and are sometimes smaller than the marker size. Arrows indicate 5$\sigma$ non-detection limits. The vertical dashed line and surrounding shaded region mark the adopted explosion epoch and its uncertainty, respectively.}
\label{fig:LC_2020bij}
\end{figure*}

\subsection{Photometry}

We obtained $UBVgri$-band imaging of SN\,2020bij through the Global Supernova Project (GSP; PI: D.A. Howell) with the 1m telescopes of the Las Cumbres Observatory global network \citep{Brown_2013} at the South African Astronomical Observatory (South Africa), the Cerro Tololo Inter-American Observatory (Chile) and the Siding Spring Observatory (Australia), from 2020-01-29 to more than 130 days after discovery. Reference images for Las Cumbres data were obtained on 2020-11-21 (approximately 300 days after explosion) in the $BVgri$-bands and on 2021-06-16 (approximately 500 days after explosion) in the $U$-band, once the SN faded significantly. 
Basic image reduction, including flat-field correction and bias subtraction, was performed using the Las Cumbres Observatory Beautiful Algorithms to Normalize Zillions of Astronomical Images \citep[\texttt{BANZAI};][]{McCully_2018} pipeline\footnote{\url{https://github.com/LCOGT/banzai}} for all Las Cumbres images. We then performed image subtraction and Point-Spread Function (PSF) fitting using the PyRAF-based \texttt{lcogtsnpipe}\footnote{\url{https://github.com/LCOGT/lcogtsnpipe}} pipeline \citep{Valenti_2016}, which uses the \texttt{PyZOGY} \citep{david_guevel_pyzogy} Python implementation of the image subtraction algorithm developed by \cite{Zackay_2016}, to remove host-galaxy contributions to the photometry of the SN. 
$UBV$-band magnitudes were calibrated to the Vega system from \cite{Stetson} using stars in \cite{Landolt} standard fields observed on the same night with the same telescope as SN\,2020bij, and $gri$-band magnitudes were calibrated to the AB system using the Sloan Digital Sky Survey Data Release 14 \citep[SDSS DR14;][]{SDSS-survey, data_release14}. 

We obtained unfiltered images during 14 epochs with the 0.35m reflector mounted at the Itagaki Astronomical Observatory using the KAF-1001E CCD. Image subtraction was performed using custom software with previous images of the field taken with the same camera as references. Photometry was extracted from the difference images and calibrated to the Fourth US Naval Observatory CCD Astrograph Catalog \citep[UCAC4;][]{Zacharias2013}.

Additional photometry was retrieved from the Gaia Photometric Science Alerts\footnote{Gaia Alerts Team \url{http://gsaweb.ast.cam.ac.uk/alerts/alert/Gaia20anq/}} \citep[$G$-band, 4 epochs;][]{Gaia_alerts}, through their event webpage\footnote{\url{http://gsaweb.ast.cam.ac.uk/alerts/alert/Gaia20anq/}}. 
These magnitudes are derived from a preliminary calibration of the photometry \citep{Lukasz_2012,Hodgkin_2021} and are not host subtracted. 
We also obtained $c$- and $o$-band host-subtracted photometry (94 epochs) from the ATLAS Public Forced Photometry Server \citep{ATLAS_data1_Tonry2018,ATLAS_data2_Smith2020, Shingles_2021}.

We correct all photometry for Milky Way extinction using the \cite{Schlafly} recalibrations of the \cite{Schlegel_extinction} infrared-based dust maps (which assume a \citealt{Fitzpatrick_1999} reddening law with \rv\ = 3.1), retrieved via NED for the $UBVgri$-bands (\ebv=0.064\,mag).
The extinction values for the $r$-band were used to correct the ATLAS $o$-band data and the Gaia data, the $g$-band extinction correction was used for the unfiltered data, and the $B$-band extinction correction was used for the ATLAS $c$-band data. We do not correct our photometry for the unknown host-galaxy extinction (our spectra show no evidence for strong host \NaiD\ absorption, see Section \ref{sec:Spectroscopic_analysis}). All observed photometry is listed in Table \ref{tab:photometry} and Milky Way extinction-corrected photometry is presented in Figure \ref{fig:LC_2020bij}.

\begin{deluxetable}{cccc}
\caption{Photometric Observations, and 5\,$\sigma$ pre-explosion non-detection upper limits of SN\,2020bij.}
\label{tab:photometry}
\centering
\tablehead{
\colhead{MJD} & \colhead{Band} & \colhead{Magnitude} & \colhead{Source} 
}
\startdata
58873.5011 & $c$ & $>$20.35 & ATLAS \\ 
58875.5782 & $o$ & $>$20.27 & ATLAS \\ 
58878.0864 & $V$ & 18.181 $\pm$ 0.025 & Las Cumbres \\ 
58878.0731 & $U$ & 17.314 $\pm$ 0.041 & Las Cumbres \\
58878.0770 & $U$ & 17.272 $\pm$ 0.041 & Las Cumbres \\
58878.2653 & $r$ & 18.013 $\pm$ 0.016 & Las Cumbres \\
58878.0809 & $B$ & 18.541 $\pm$ 0.019 & Las Cumbres \\
58878.2670 & $r$ & 18.011 $\pm$ 0.016 & Las Cumbres \\
58878.2508 & $B$ & 18.452 $\pm$ 0.016 & Las Cumbres \\ 
58878.2431 & $U$ & 17.367 $\pm$ 0.038 & Las Cumbres \\
58878.2624 & $g$ & 18.379 $\pm$ 0.014 & Las Cumbres \\
\hline
\enddata
\tablecomments{This table is published in its entirety in the machine-readable format. A portion is shown here for guidance regarding its form and content.}
\end{deluxetable}

\subsection{Spectroscopy}

In addition to the classification spectrum, we obtained a second optical spectrum through the ePESSTO+ collaboration \citep{PESSTO_Smartt_2015} using the ESO Faint Object Spectrograph and Camera version 2 (EFOSC2), mounted at the 3.58\,m New Technology Telescope \citep[NTT;][]{EFOSC2}. Grism 13 was used with a spectral resolution of $R \sim 200$--$550$.
We obtained seven more optical spectra of SN\,2020bij through the GSP with the Las Cumbres Observatory FLOYDS spectrographs mounted on the 2-meter Faulkes Telescope North (FTN) at Haleakalā (HI, USA) and Faulkes Telescope South (FTS) at Siding Spring (Australia) observatories \citep{Brown_2013}.
The FLOYDS spectra have a spectral resolution of $R \sim 300$--$600$.

\begin{deluxetable}{ccccc}
\caption{Log of the spectroscopic observations of SN\,2020bij.}
\label{tab:spectroscopic_measurements}
\centering
\tablehead{
 \colhead{MJD} & \colhead{Phase} & \colhead{Telescope} & \colhead{Slit Width} & \colhead{Exposure Time}
 \\ & {(days)} & & (\arcsec) & (s)
}
\startdata 
58881.25 & 4.65 & ESO-NTT & 1.0 & 600 \\
58882.42 & 5.81 & OGG 2m & 2.0 & 2700 \\
58884.44 & 7.80 & OGG 2m & 2.0 & 3600 \\
58893.67 & 16.91 & COJ 2m & 2.0 & 2700 \\
58906.52 & 29.59 & COJ 2m & 2.0 & 2700 \\
58917.47 & 40.41 & OGG 2m & 2.0 & 2700 \\
58920.53 & 43.43 & COJ 2m & 2.0 & 3600 \\
58923.25 & 46.11 & ESO-NTT & 1.0 & 1800 \\
58928.44 & 51.23 & COJ 2m & 2.0 & 3600 \\ 
\enddata
\centering
\tablecomments{The phase of each spectrum is listed in rest-frame days relative to explosion. OGG denotes the Haleakalā site and COJ denotes the Siding Spring site.}
\end{deluxetable}

\begin{figure*}
\includegraphics[width=0.7\textwidth]{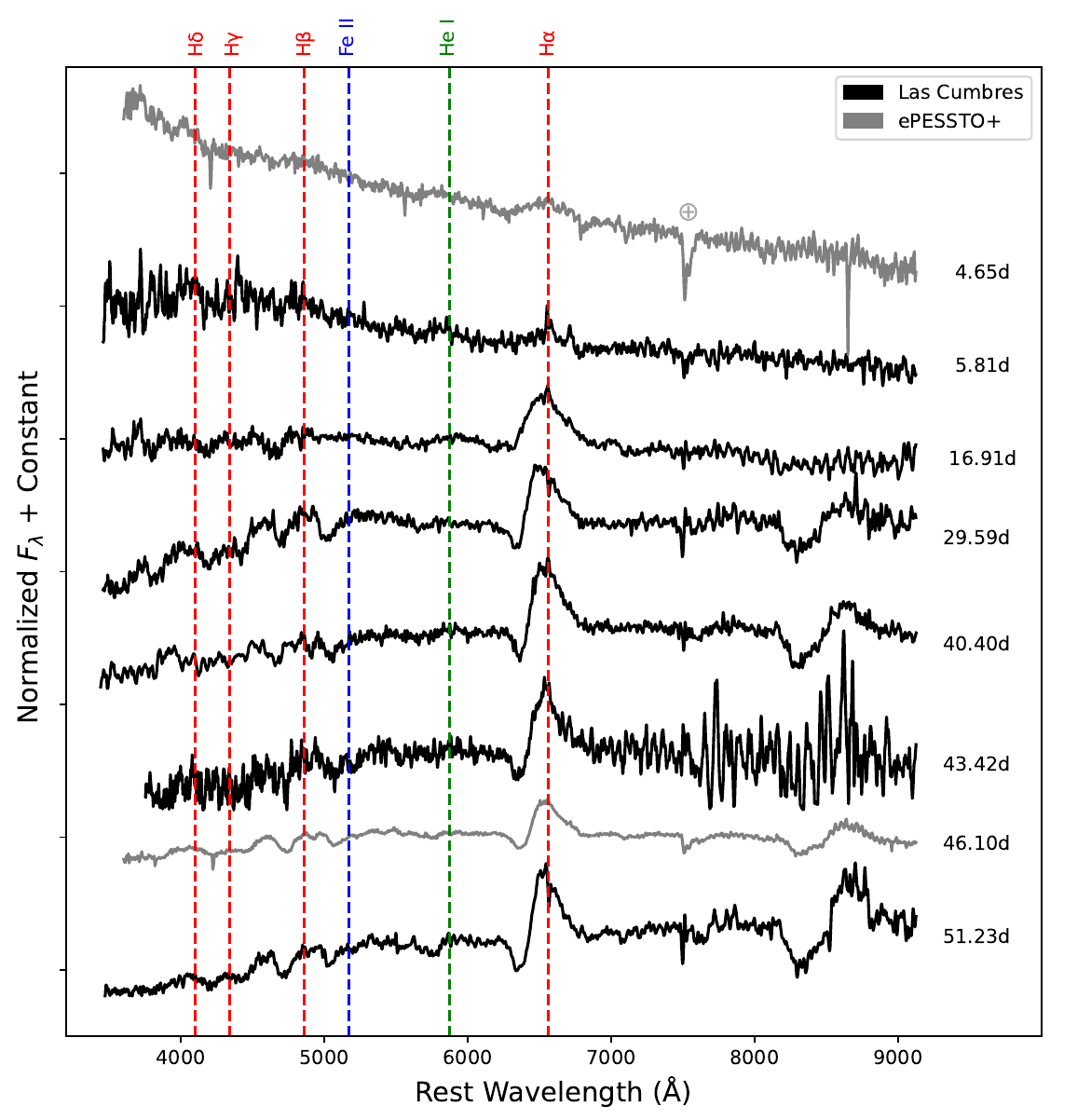}
\centering
\caption{The spectroscopic evolution of SN\,2020bij. The phase of each spectrum in rest-frame days relative to explosion is shown on the right. All spectra have been calibrated to the Milky-Way extinction-corrected photometry and are shifted in flux for clarity. The vertical colored lines at the rest wavelengths of hydrogen (\hd\ 4101, \hg\ 4340, \hb\ 4861, and \ha\ 6563\,\AA), helium (\he\ 5875\,\AA) and iron (\Feii\ 5169\,\AA), seen in all spectra, denote spectral features common in Type II SNe. The main telluric features are marked.}
\label{fig:Spectra}
\end{figure*}

The FLOYDS and EFOSC2 spectra of SN\,2020bij were reduced using the {FLOYDS pipeline}\footnote{\url{https://github.com/svalenti/FLOYDS_pipeline/}} \citep[based on PyRAF routines;][]{FLOYDS_REDUCTION} and the PESSTO pipeline \citep{PESSTO_Smartt_2015}, respectively. We calibrate all spectra of SN\,2020bij to the Milky-Way extinction-corrected photometry using the Light Curve Fitting package \citep{hosseinzadeh_2024_11405219}.
A log of our spectroscopic observations is available in Table \ref{tab:spectroscopic_measurements} and the spectra are displayed in Figure \ref{fig:Spectra}. All spectra will be made available through the Weizmann Interactive Supernova Data Repository \citep[WISeREP;][]{yaron_2012}\footnote{\url{https://www.wiserep.org}}.

\section{Sample}
\label{sec:Sample}

To analyze SN\,2020bij in context, we searched the literature for other well-observed SNe IIP with slow light curve rises. Here we define a ``slowly rising SN IIP'' as a SN having a rise lasting $\gtrsim$10 days directly followed by a plateau. In this definition we do not include events showing a slow rise to a peak followed by a decline to the plateau, such as SN\,2023ixf \citep{Bostroem_2023, Hosseinzadeh_2023, 2023ixf_discovery, Jacobson_2023, 2023ixf_classification, Teja_2023, Zimmerman_2024}, nor Type IIL SNe which show a slow rise followed by a prolonged decline such as SN\,2013ai \citep{Davis_2021}. We also exclude
SN\,1987A-like events (see \citealt{Sit_2023} for a recent compilation) which do have a very long rise, but do not have a plateau at all. 

\begin{deluxetable*}{cccccccc}
\tablecaption{Slowly rising Type IIP SNe included in this study. \label{tab:SN_list}}
\tablehead{
\colhead{SN} & \colhead{R.A.} & \colhead{Decl.} & \colhead{Host-Galaxy} & \colhead{Redshift} & \colhead{Explosion Epoch$^a$} & \colhead{Distance} & \colhead{Reference} \\
\colhead{Name} & (J2000) & (J2000) & & & & (Mpc) &
}
\startdata
ASASSN-14kg & ${01}^h{44}^m{38.^s3808}$ & $+{35}^d{48}^m{20^s.448}$ & CGCG 521-075 & 0.0145 & 56969.50 $\pm$ 3.00 & 58.05 $\pm$ 4.07 & 1 \\
SN\,2018fif & ${00}^h{09}^m{26.^s5500}$ & $+{47}^d{21}^m{14^s.700}$ & UGC 85 & 0.0172 & 58350.88 $\pm$ 0.48 & 76.50 & 2,3 \\
SN\,2020bij & ${10}^h{55}^m{11.^s690}$ & $-{26}^d{08}^m{25^s.80}$ & NGC 3463 & 0.0133 & 58876.54 $\pm$ 0.96 & 57.54 $\pm$ 2.65 &This work \\
SN\,2021yja & ${03}^h{24}^m{21.^s1790}$ & $-{21}^d{33}^m{56^s.090}$ & NGC 1325 & 0.0053 & 59464.46 $\pm$ 0.17 & 23.40 $\pm$ 4.50 & 4 \\
SN\,2023axu & ${06}^h{45}^m{55.^s3210}$ & $-{18}^d{13}^m{53^s.480}$ & NGC 2283 & 0.0028 & 59971.82 $\pm$ 0.30 & 13.68 $\pm$ 2.05 & 5 \\
\enddata
\tablenotetext{a}{The explosion epochs are estimated as the midpoint between the last non-detection and first detection.}
\tablecomments{
\textbf{References.} (1) \cite{Valenti_2016}; (2) \cite{Maayane}; (3) \cite{Bruch_2021}; (4) \cite{Hosseinzadeh_2022}; (5) \cite{Shrestha_2023}
}
\end{deluxetable*}

We identified ASASSN-14kg \citep{Valenti_2016, Davis_2021}, SN\,2018fif \citep{Maayane}, SN\,2021yja \citep{Hosseinzadeh_2022} and SN\,2023axu \citep{Shrestha_2023}. The basic parameters of these events are listed in Table \ref{tab:SN_list}. This is likely not a comprehensive list. Indeed, \citet{Irani_2024} recently presented a large sample of Type II SNe discovered early, including a few events that could fit our definition of slowly rising Type IIP SNe. 
However, their explosion epochs (critical for determining the duration of the rise to plateau) are estimated using early-time light curve fits to various power laws, rather than the more model-agnostic approach used here which relies on recent pre-explosion non-detections. Therefore, the events in \cite{Irani_2024} can not be directly compared to our selection in a systematic way.
We leave a comparison to their and other possible events for future work aimed at analyzing a more systematically-collected sample. Our goal here is only to provide some context for interpreting SN\,2020bij.

Consistent with the expectations of \cite{Morozova_2018} for slowly rising Type IIP light curves, \cite{Maayane} find that shock cooling alone can explain the early light curve of SN\,2018fif, indicating a null to small CSM-interaction contribution to its early emission. 
For SN\,2021yja, hydrodynamical light curve modeling by \citet{Kozyreva_2022} suggests that CSM is required to reproduce the early-time light curve. However, light curve modeling by \citet{Hosseinzadeh_2022} suggests only weak CSM interaction. Consistent with this, radiative-transfer modeling of early UV and optical spectroscopy by \citet{Vasylyev_2022} does not find evidence for a significant CSM contribution to the observed UV flux. Similarly, for SN\,2023axu,
\citet{Shrestha_2023} find evidence for only weak CSM interaction.

In the following sections, we analyze and compare the light curve, color evolution, temperature evolution, spectra and expansion velocities of SN\,2020bij to this sample and to typical Type IIP SNe.
For typical Type IIP SNe, we choose \cite{Faran_2018} as a comprehensive uniform photometric comparison sample and \cite{Gutierrez} as a comprehensive uniform spectroscopic comparison sample.

\section{Photometric Analysis} \label{sec:Photometric_analysis}

\subsection{Light Curve Parameters \& Color Evolution}
\label{subsec:LC} 

\begin{figure}
\centering
\includegraphics
[width=0.5\textwidth]{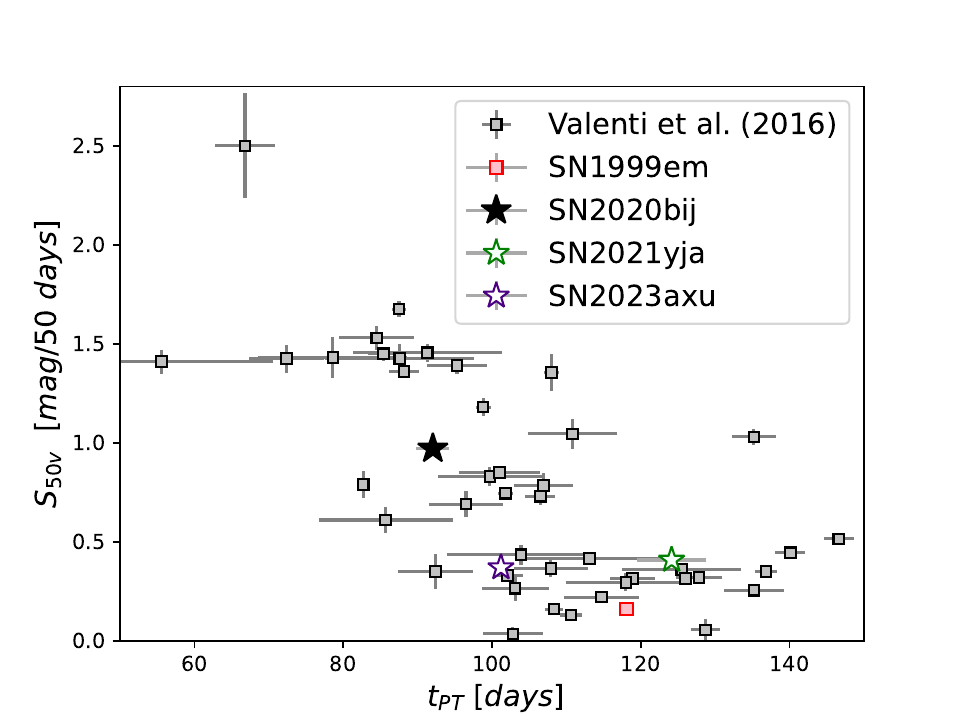}
\caption{The $V$-band decline rate ($s_{50V}$) and plateau duration ($t_{PT}$) for Type IIP SNe, as defined by \cite{Valenti_2016} for their sample. The events in our slow-rising sample for which these parameters can be measured (SN\,2020bij and SN\,2021yja measured here, SN\,2023axu from \citealt{Shrestha_2023}) are shown in stars. The prototypical Type IIP SN\,1999em is also shown, with values taken from \cite{Valenti_2016}. The slow-rising events are typical among the broader sample of Type IIP SNe.}
\label{fig:lc_parameters}
\end{figure}

The light curve of SN\,2020bij shows a plateau of approximately 100 days (Fig. \ref{fig:LC_2020bij}), which is typical for Type IIP SNe. More precisely, we calculate the plateau duration, $t_{PT}$, as well as the plateau decline rate in the $V$-band, $s_{50V}$, of SN\,2020bij as defined by \cite{Valenti_2016} and find $t_{PT} = 92.07 \pm 2.35$ days and $s_{50V} = 0.9709 \pm 0.0011$\,mag/50 days. As can be seen in Figure \ref{fig:lc_parameters}, these values are typical of SNe IIP, and SN\,2020bij falls in the center of the Type IIP SN sample from \cite{Valenti_2016}. 

We also calculate these parameters for SN\,2021yja and find $t_{PT} = 124.12 \pm 4.65$ days and $s_{50V} = 0.40822 \pm 0.00027$\,mag/50 days. For SN\,2023axu, these parameters were measured by \citet{Shrestha_2023}, and we present them too in Figure \ref{fig:lc_parameters}. SN\,2021yja and SN\,2023axu also lie within the typical range of Type IIP SNe in this phase space.
We do not calculate these light curve parameters for ASASSN-14kg and SN\,2018fif, as their available $V$-band photometry does not extend sufficiently beyond the plateau phase to allow a reliable determination.

\begin{figure}
\includegraphics[width=0.5\textwidth]{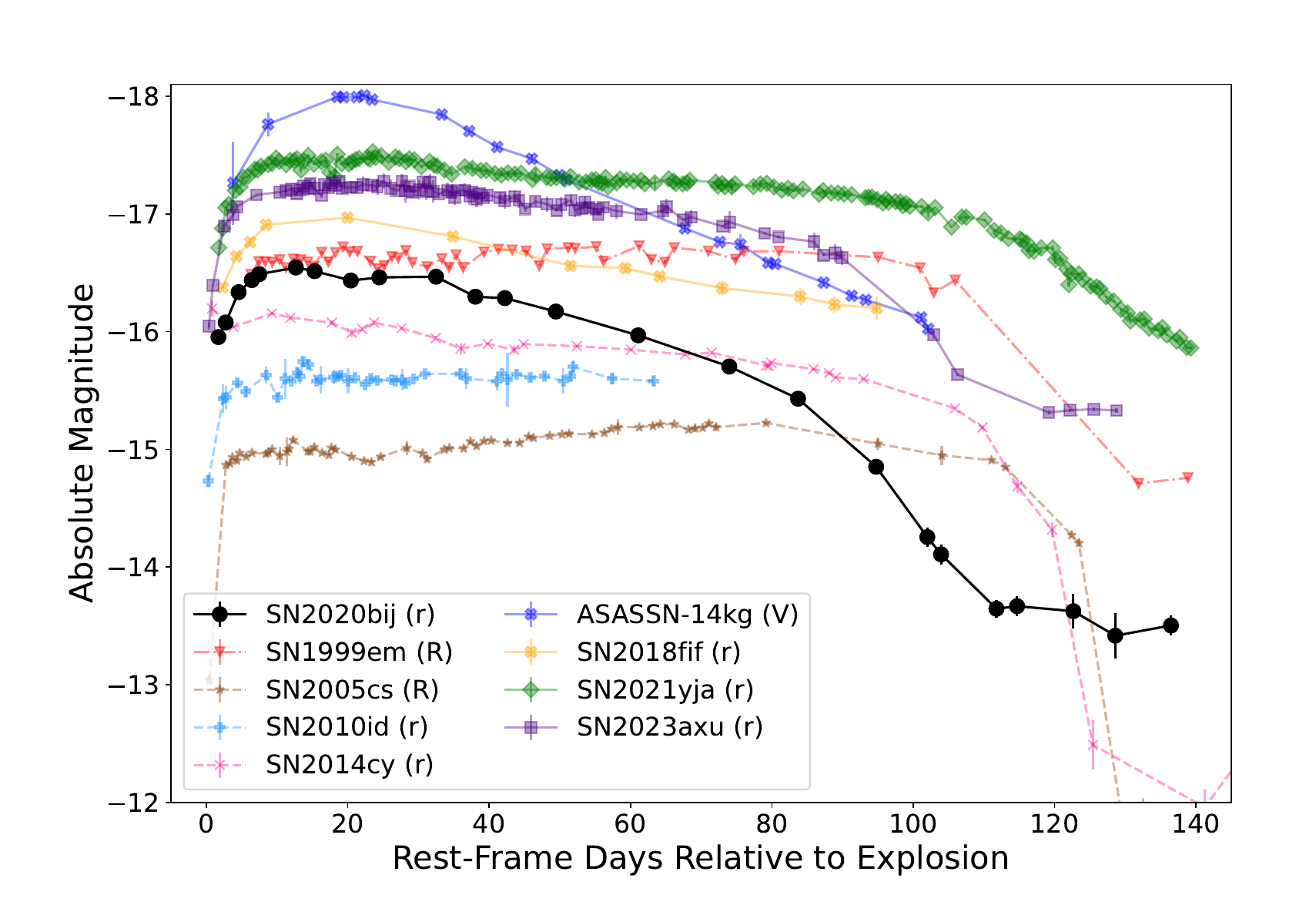}
\centering
\caption{The $r$-band light curve of SN\,2020bij compared to light curves of the prototypical Type IIP SN\,1999em (dot-dashed line), the slowly rising Type IIP ASASSN-14kg, SN\,2018fif, SN\,2021yja and SN\,2023axu (solid lines), and the well-sampled SN\,2005cs, SN\,2010id and SN\,2014cy (dashed lines). All events are shown in $r$- or $R$-band, except ASASSN-14kg for which $V$-band data are shown, as no early-time $r$-band observations are available for it. See text for data sources.}
\label{fig:abs_mag}
\end{figure}

\begin{figure*}
\includegraphics[width=1\textwidth]{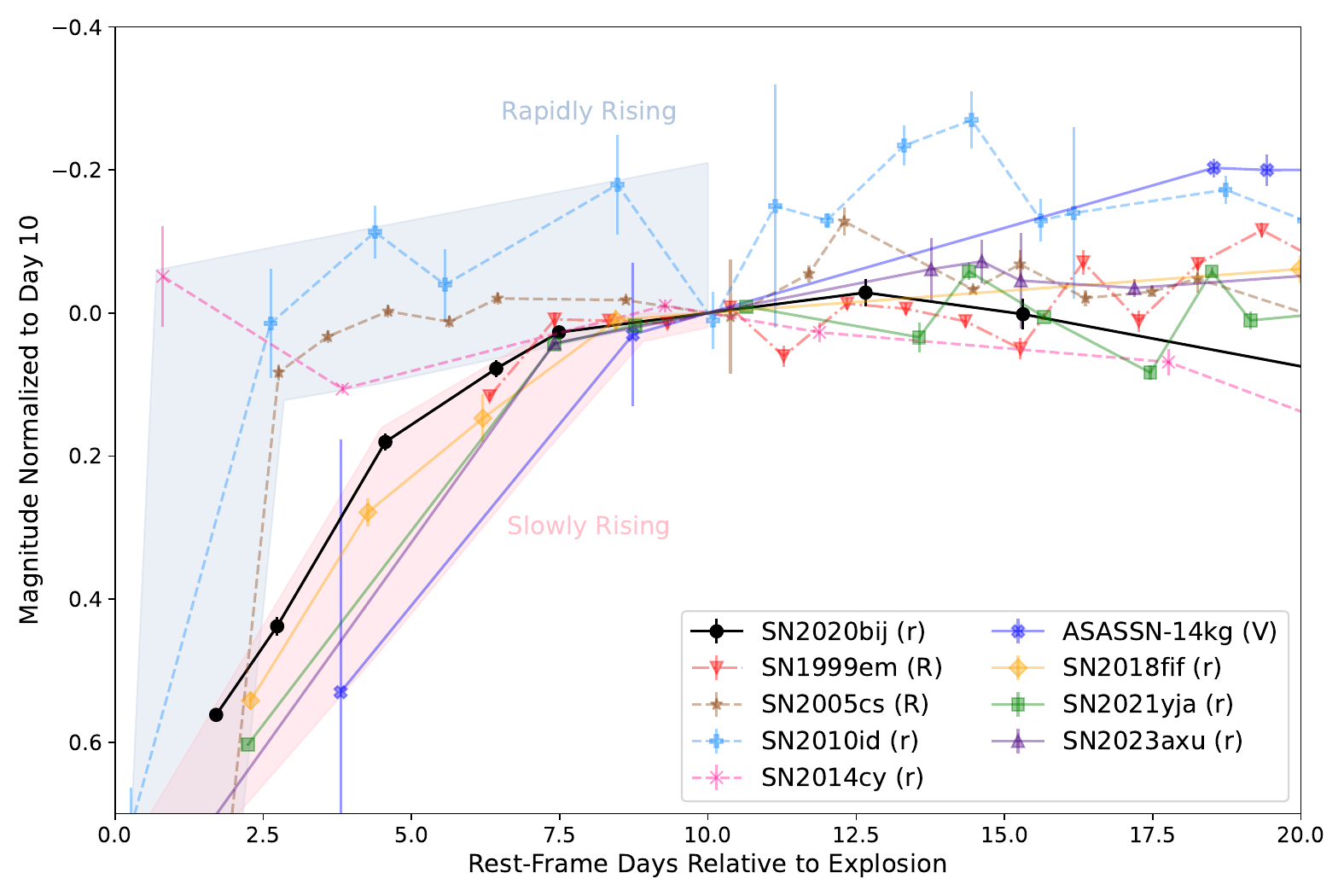}
\centering
\caption{Same as Figure \ref{fig:abs_mag}, but normalized to the magnitude on day 10, and focusing only on the first 20 days to show the difference in light curve rise rates. Shaded regions are added to highlight the difference in the light curve rise between the slowly rising and rapidly rising Type IIP SNe.}
\label{fig:inset_abs_mag}
\end{figure*}

What sets SN\,2020bij apart from the majority of the SN IIP population, as mentioned above, is its relatively long rise to the plateau, lasting roughly 14 days, compared to $\lesssim$10 days in a typical SN IIP \citep[e.g.][]{Gall_2015, Gonzalez_2015, Rubin_2016, Rubin_GalYam_2016}. ASASSN-14kg \citep{Valenti_2016}, SN\,2018fif \citep{Maayane}, SN\,2021yja \citep{Hosseinzadeh_2022} and SN\,2023axu \citep{Shrestha_2023} show a similarly slow rise to the plateau despite having different plateau luminosities and decline rates. This can be seen in Figure \ref{fig:abs_mag}, where we also plot the prototypical Type IIP SN\,1999em \citep[][retrieved via the Open Supernova Catalog; \citealt{Guillochon_2017}]{Hamuy_2001, Faran_2014, Galbany_2016} and the well-sampled SN\,2010id \citep{Gal_Yam_2010id}, SN\,2005cs \citep[data from][]{Pastorello_2009} and SN\,2014cy \citep{Valenti_2016} as comparison events having the highest-cadence SN IIP light curves and best constrained ($<$1\,day) explosion dates from the \cite{Morozova_2018} sample\footnote{SN\,2013ab \citep{Bose_2015} is often classified as rapidly rising. However, its rise is non-uniform, with a sharp initial rise followed by slower evolution. Thus, its $\sim$1-day explosion epoch uncertainty introduces a large uncertainty in its rise-time classification. We therefore do not include it here.} (distance moduli and explosion dates were taken from \citealt{Valenti_2016}).
In Figure \ref{fig:inset_abs_mag} we plot the first 20 days of each light curve from Figure \ref{fig:abs_mag}, normalizing the magnitude to its value on day 10 to emphasize the differences in light curve rise time.

\begin{figure}
\includegraphics[width=0.5\textwidth]{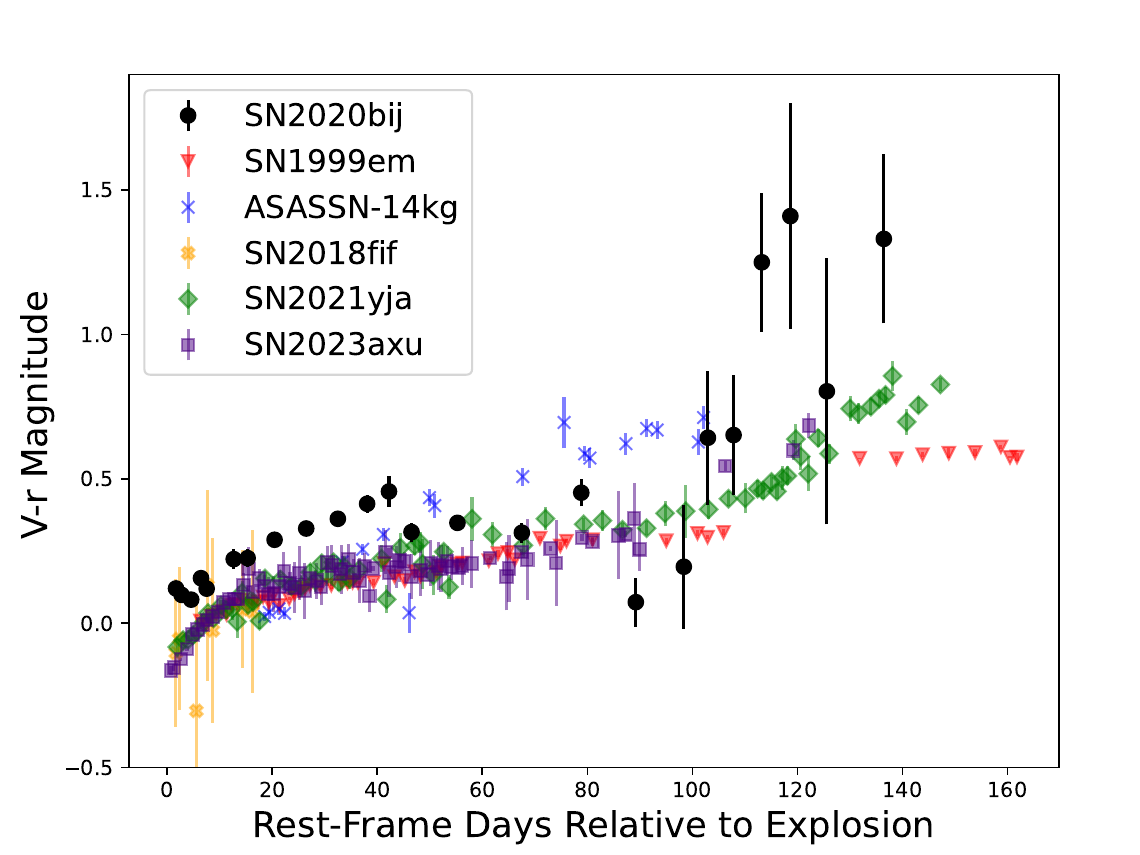}
\centering
\caption{$V$-$r$ color evolution of SN\,2020bij (black circles) and of other slowly rising SNe in our sample compared to the color evolution of the prototypical Type IIP SN\,1999em. For SN\,2018fif, the $V$ data are from \textit{Swift}/UVOT, while the $r$-band combines measurements from $r_{\rm SDSS}$ and $r_{\rm P48}$ \citep{Bellm_2019,Masci_2019}.}
\label{fig:color_evolution_comp}
\end{figure}

The $V - r$ color evolution of SN\,2020bij, together with those of the other slowly rising SNe in our sample and of SN\,1999em\footnote{We convert its $R$ band data to $r$ band using the empirical $R$--$r$ color relations of \citet{Jordi_2006}.}, is shown in Figure \ref{fig:color_evolution_comp}. All SNe show similar color evolution, with SN\,2020bij on the redder side of the distribution at early-times.

\subsection{Blackbody Fits \& Bolometric Luminosity} \label{subsec:bb} 

We fit the spectral energy distribution (SED) of each photometric epoch (spanning up to 1 day) of the slowly rising SNe ASASSN-14kg, SN\,2020bij and SN\,2023axu with a blackbody spectrum using the {Light Curve Fitting}\footnote{\url{https://github.com/griffin-h/lightcurve_fitting}} Python package \citep{hosseinzadeh_2024_11405219} to obtain the blackbody temperature and radius evolution. This package uses two methods for the fit, Markov Chain Monte Carlo (MCMC), and non-linear least squares minimization as implemented through {Scipy}\footnote{\url{https://github.com/scipy/scipy/tree/v1.9.3}}. Both methods yield similar results. Here we adopt the MCMC method. We used the default MCMC parameters set by the package (namely, 10 walkers, at least 3 filters per epoch, and 200 burn-in steps followed by an additional 100 steps). 

\begin{figure}
\includegraphics[width=0.5\textwidth]{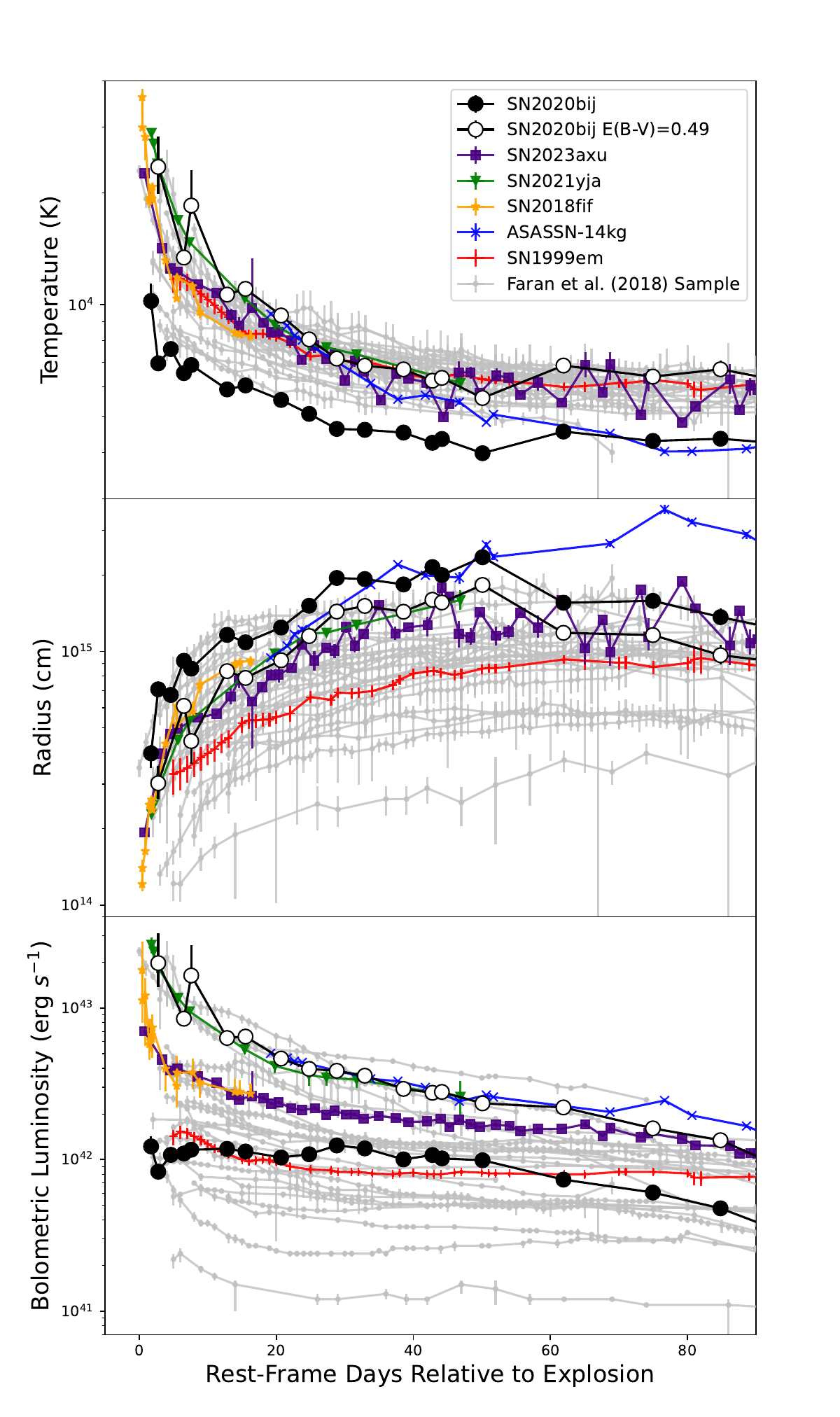}
\centering
\caption{Best-fit blackbody temperature (top), radius (middle) and resulting bolometric luminosity (bottom) of SN\,2020bij (without correction for host-galaxy extinction in filled black circles and with correction to the best-fit host-galaxy extinction from the \citet{Sapir_and_Waxman} shock cooling model fits in empty circles), compared to those from the \cite{Faran_2018} sample of SNe IIP, and to the slowly rising Type IIP ASASSN-14kg, SN\,2018fif \citep{Maayane}, SN\,2021yja \citep{Hosseinzadeh_2022} and SN\,2023axu. All SNe from the \cite{Faran_2018} sample are shown in gray, except for the prototypical IIP SN\,1999em which is highlighted in red. No host-galaxy extinction corrections are applied to the comparison sample, as they are not known.}
\label{fig:BB_Evo}
\end{figure}

We use the best-fit blackbody temperature and radius as a function of time to derive the bolometric luminosity using the Stefan–Boltzmann law, $L = 4 \pi R^2 \sigma_{\rm{SB}} T^4$, where $\sigma_{\rm{SB}}$ is the Stefan-Boltzmann constant.
The results are presented in Table \ref{tab:BB_data} and Figure \ref{fig:BB_Evo}, where we also compare them to those of other SNe IIP taken from the \cite{Faran_2018} sample and to the slowly rising Type IIP SN\,2018fif \citep{Maayane} and SN\,2021yja \citep{Hosseinzadeh_2022} using their published blackbody best-fit parameters.

SN\,2020bij exhibits the lowest blackbody temperatures in the sample (filled black circles in Figure \ref{fig:BB_Evo}), when neglecting host-galaxy extinction (as done for the rest of the sample). If we assume the host-galaxy extinction values derived from the 
\cite{Sapir_and_Waxman} and \cite{Morag_2023} shock cooling fits (\ebv\,$ = 0.49--0.58$\,mag; see Section \ref{subsec:Shock_cooling}), we find that SN\,2020bij has more typical temperatures (empty circles in Figure \ref{fig:BB_Evo}) compared to other Type IIP SNe. 
However, such high extinction would result in strong \NaiD\ absorption \citep[e.g.][]{Poznanski_2012}, which is not seen in the spectra of SN\,2020bij. The corrected temperatures shown in Figure \ref{fig:BB_Evo} correspond to the \cite{Sapir_and_Waxman} extinction estimate (\ebv\,$ = 0.49 \pm 0.04$\,mag), adopting the \cite{Morag_2023} value leads to a similar result.

\begin{deluxetable*}{cccccc}
\setlength{\tabcolsep}{12pt}
\caption{Best-fit blackbody temperatures, radii and the resulting bolometric luminosities of ASASSN-14kg, SN\,2020bij, and SN\,2023axu. We report the median of the posterior distribution with the 16th and 84th percentiles as uncertainties.}
\label{tab:BB_data}
\centering
\tablehead{
\colhead{SN} & \colhead{Phase} & \colhead{Temperature} & \colhead{Radius} & \colhead{Luminosity} & \colhead{Bands} \\
 {Name} & {(days)} & {($10^3$\,K)} & {($10^{14}$\,cm)} & {($10^{42}$\,erg\,s$^{-1}$)} & 
}
\startdata
ASASSN-14kg & $18.809_{-0.008}^{+0.014}$ & $9.468_{-0.086}^{+0.058}$ & $13.467_{-0.127}^{+0.190}$ & $5.029_{-0.048}^{+0.040}$ & $BgVri$ \\
ASASSN-14kg & $19.711_{-0.008}^{+0.009}$ & $9.371_{-0.091}^{+0.071}$ & $13.709_{-0.157}^{+0.217}$ & $4.997_{-0.053}^{+0.056}$ & $BgVri$ \\
ASASSN-14kg & $21.588_{-0.008}^{+0.009}$ & $8.779_{-0.065}^{+0.089}$ & $15.102_{-0.257}^{+0.201}$ & $4.670_{-0.040}^{+0.035}$ & $BgVri$ \\
\hline
SN\,2020bij & $1.726_{-0.101} ^{+0.189}$ & $10.221_{-0.547} ^{+1.150}$ & $5.697_{-0.437} ^{+0.723}$ & $1.222_{-0.085} ^{+0.205}$ & $UBgVGri$ \\
SN\,2020bij & $2.767_{-0.016} ^{+0.224}$ & $6.946_{-0.053} ^{+0.075}$ & $10.187_{-0.236} ^{+0.146}$ & $0.832_{-0.004} ^{+0.003}$ & $UBgVori$ \\
SN\,2020bij & $4.614_{-0.014} ^{+0.011}$ & $7.589_{-0.088} ^{+0.072}$ & $9.683_{-0.168} ^{+0.234}$ & $1.075_{-0.005} ^{+0.007}$ & $UBgVcri$ \\
\enddata
\centering
\tablecomments{The phase is listed in rest-frame days relative to explosion. The values and errors denote the time bins used (uncertainties in phase due to uncertainties in explosion date are not noted). This table is published in its entirety in the machine-readable format. A portion is shown here for guidance regarding its form and content.}
\end{deluxetable*}

\begin{figure}
\includegraphics[width=0.5\textwidth]{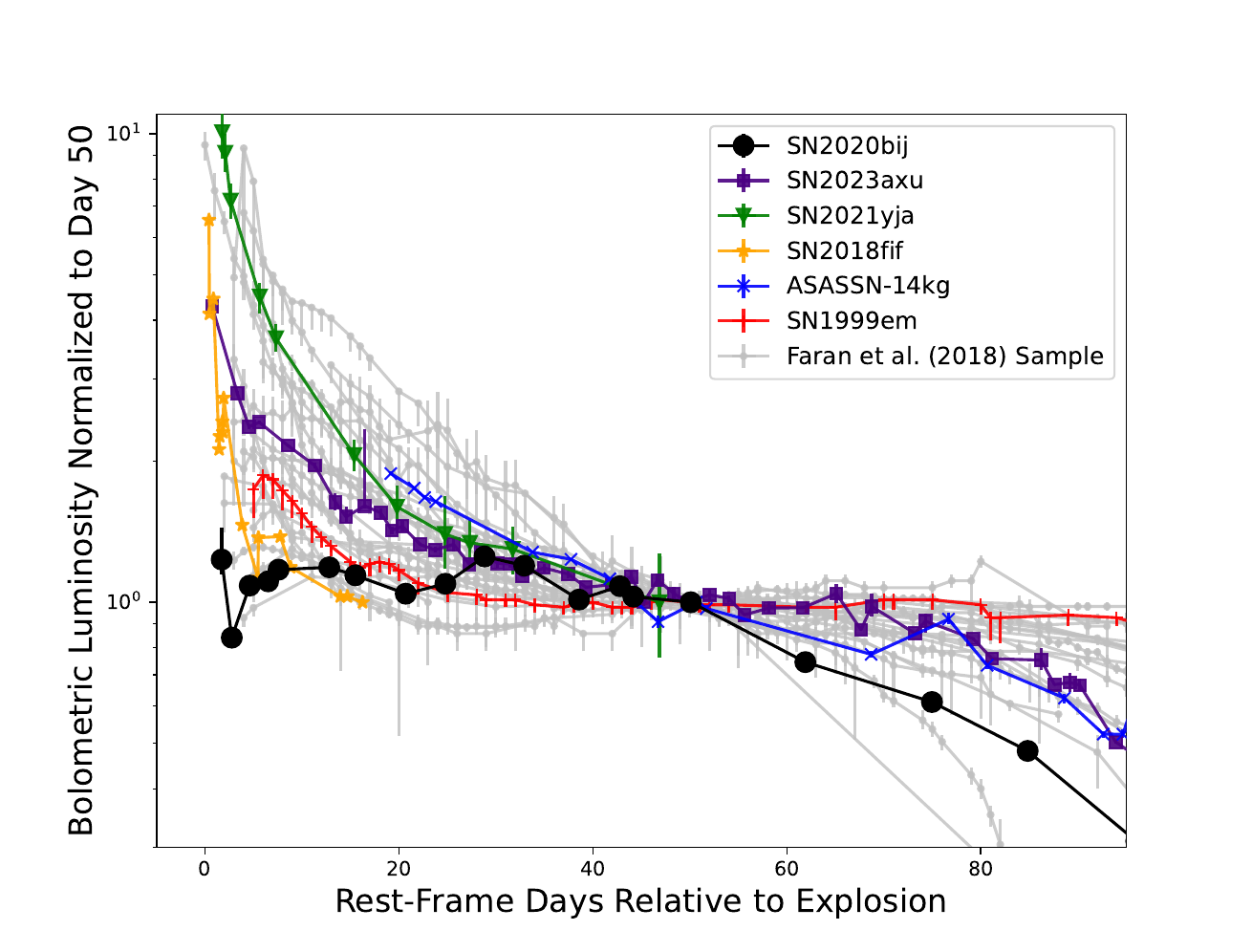}
\centering
\caption{Same as the bottom panel of Figure \ref{fig:BB_Evo} but normalized to day 50. SN\,2020bij lacks the early-time decline seen in many other Type IIP SNe.}
\label{fig:Nor_BB}
\end{figure}

The bolometric luminosity of SN\,2020bij during the plateau is consistent with that of the comparison SN IIP sample and the slowly rising SNe, although the shape of the bolometric light curve of SN\,2020bij is different at early-times. In Figure \ref{fig:Nor_BB} we present the bolometric light curves normalized to day 50. SN\,2020bij exhibits a nearly constant early-time evolution, and does not show the early-time decline seen in many of the other events.

\subsection{Shock Cooling Model Fit} \label{subsec:Shock_cooling} 

We fit the analytic shock cooling model from \cite{Sapir_and_Waxman} and its extended formulation developed by \cite{Morag_2023} to the early light curve of SN\,2020bij using the MCMC routine implemented in the Light Curve Fitting package \citep{hosseinzadeh_2024_11405219}. This model assumes no emission from CSM interaction but rather that the early emission is powered entirely by energy deposited in the stellar envelope by the supernova shock wave \citep[e.g.][]{Nakar2010,RabinakWaxman_2011,Sapir_and_Waxman,Morag_2023}. 
The \citet{Morag_2023} model is based on the same general shock cooling framework as \citet{Sapir_and_Waxman}, but includes an updated treatment of the earliest phases of the emission and incorporates an approximate correction for UV line blanketing, leading to a modified SED.

For the \citet{Morag_2023} model, we modified the Light Curve Fitting package to include the distance and host-galaxy extinction as free parameters, as is done for the \cite{Sapir_and_Waxman} model there.
Here we take the $n = 1.5$ models, where $n$ is the power-law index of the progenitor envelope density profile, which is relevant for efficiently convective envelopes, as expected for an RSG progenitor.
The model parameters and fit priors are presented in Table \ref{tab:shockcooling_parameters}. The intrinsic scatter term represents a scaling factor of the photometric uncertainty, and is added in quadrature to each point (see \citealt{hosseinzadeh_2024_11405219} for more details). The prior on the distance, \DL\, is assumed to follow a Gaussian distribution based on the Tully-Fisher distance of the host-galaxy \citep{Tully_2016} retrieved via NED. The prior for the explosion time, \too, is uniform within the explosion window discussed in Section \ref{sec:Discovery}. Priors for all physical parameters are uniform (see Table \ref{tab:shockcooling_parameters}).

\begin{figure*}
\centering
\includegraphics[width=1\textwidth]{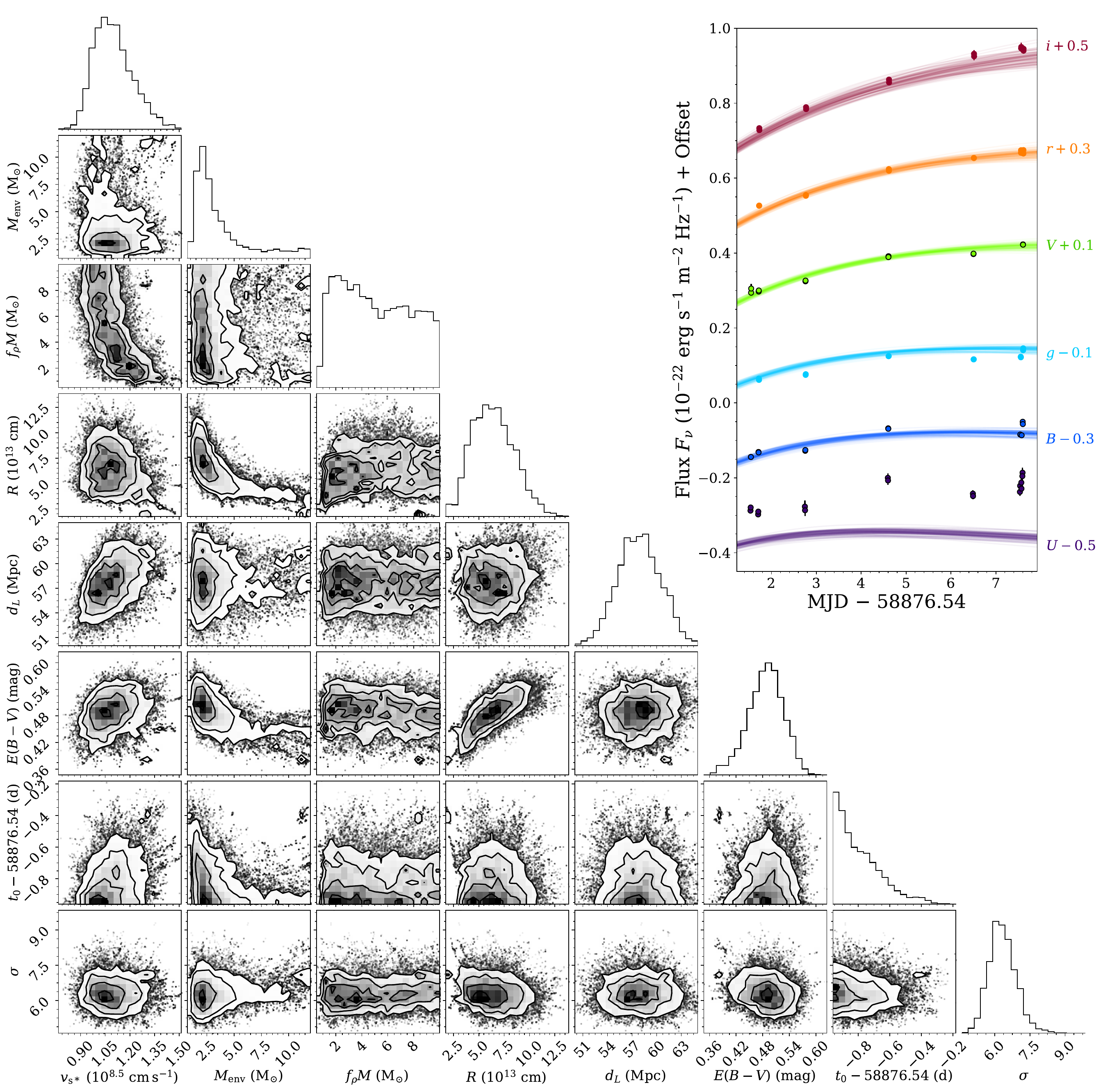}
\caption{Posterior probability distributions of fitting the \cite{Sapir_and_Waxman} shock cooling model to the early light curve of SN\,2020bij. The top-right panel shows 100 random shock cooling models (lines) drawn from the MCMC posteriors on top of our data (points). Error bars represent 1$\sigma$ uncertainties and are sometimes smaller than the marker size. The model fits the early light curve of SN\,2020bij well in all the optical bands using reasonable physical parameters, but it underestimates the observed $U$-band flux.}
\label{fig:cornerplot_shockcooling}
\end{figure*}

\begin{figure*}
\centering
\includegraphics[width=1\textwidth]{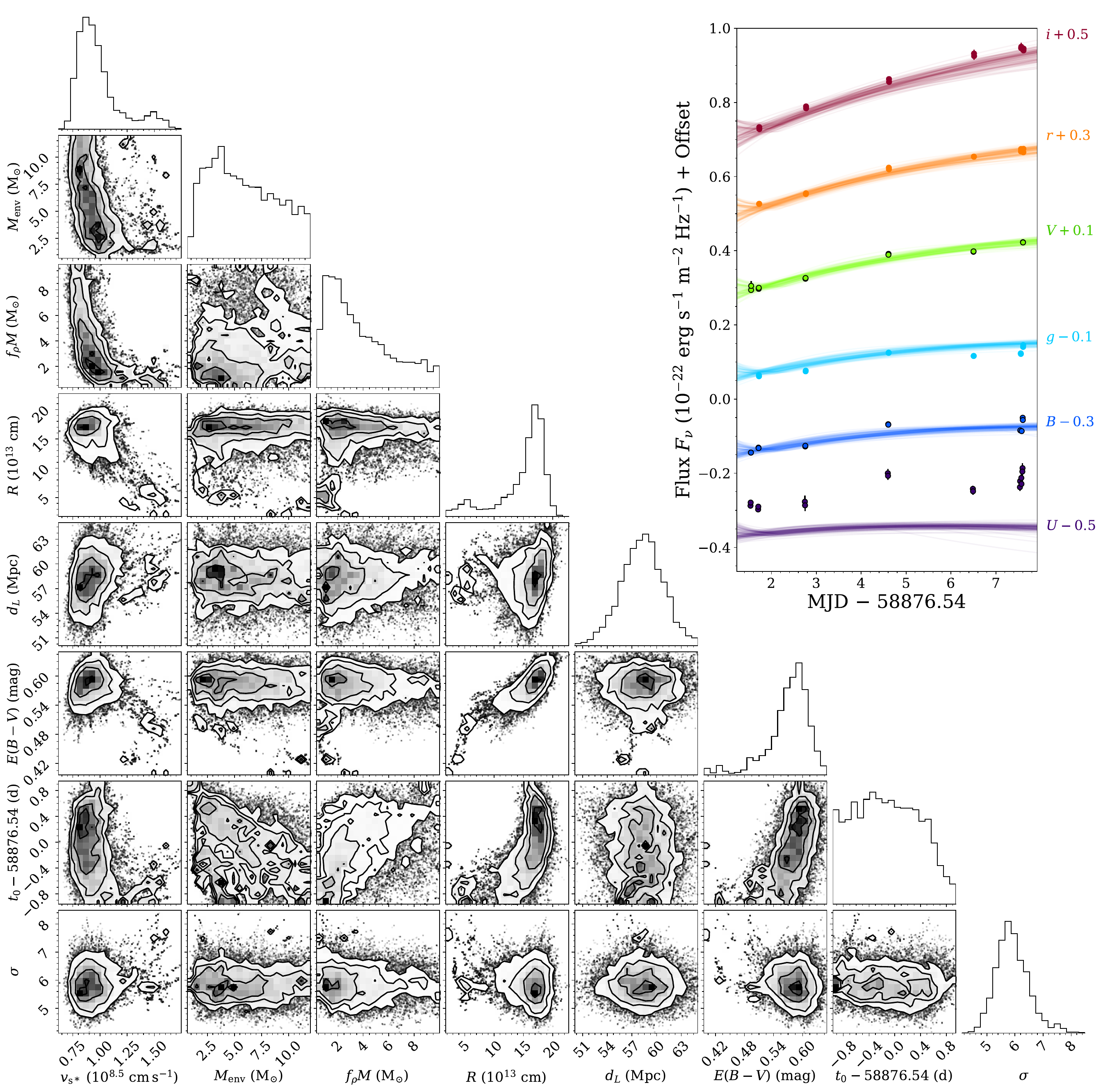}
\caption{Same as Figure \ref{fig:cornerplot_shockcooling}, but using the shock cooling model of \citet{Morag_2023}.}
\label{fig:cornerplot_morag}
\end{figure*}

\begin{deluxetable*}{ccccc}
\caption{Shock cooling model parameters, units, and their priors for SN\,2020bij.}
\label{tab:shockcooling_parameters}
\centering
 \tablehead{
\colhead{Parameter} & \colhead{Notation} & \colhead{Units} & \colhead{Prior Shape} & \colhead{Prior Parameters$^c$}
 }
\startdata
Shock velocity$^a$ & \Vs & $10^{8.5}$\,\cms & Uniform & 0--10\\
Envelope mass$^a$ & \Menv & \Msun & Uniform & 0--10 \\
Ejecta mass $\times$ numerical factor$^a$ & \frho M & \Msun & Uniform & 0--10\\
Progenitor radius$^a$ & \Rad & $10^{13}$\,cm & Uniform & 0--30\\
Distance$^b$ & \DL & Mpc & Gaussian & $57.54 \pm 2.65$\\
Extinction$^b$ & \ebv & mag & Uniform & 0.01--1\\
Explosion time$^b$ & \too & MJD & Uniform & 58875.6--58877.5\\
Intrinsic scatter$^b$ & $\sigma$ & Unitless & Uniform & 0--11\\ 
\enddata
\tablenotetext{a}{Physical parameters from the analytical models of 
\cite{Sapir_and_Waxman} and \cite{Morag_2023}.} \tablenotetext{b}{Observational parameters from the Light Curve Fitting Python package by \cite{hosseinzadeh_2024_11405219}.}
\tablenotetext{c}{For Gaussian priors, the listed values correspond to the mean and standard deviation, and for uniform priors, the bounds indicate the minimum and maximum of the parameter range.}
\end{deluxetable*}

The \cite{Sapir_and_Waxman} model is only valid for temperatures above 0.7\,eV ($\sim$8000\,K), since it does not take into account recombination.
We fit the model self-consistently in this range and find that it is valid for the first $\sim$8 days from explosion. For the \cite{Morag_2023} model, we verified that the fitted data are within the model validity range using the \texttt{t\_min} and \texttt{t\_max} limits implemented in the Light Curve Fitting package, and find that it is also valid over the fitted time range.
For both models, we use the default MCMC parameters set by the package, namely 100 walkers with 1000 burn-in steps followed by an additional 1000 steps. 

In Figures \ref{fig:cornerplot_shockcooling} and \ref{fig:cornerplot_morag} we plot the early light curve of SN\,2020bij, together with the shock cooling model fits of \cite{Sapir_and_Waxman} and \cite{Morag_2023}, respectively, and the posterior probability distributions for the model parameters.
The shock cooling fits from both models reproduce the optical data, but underestimate the luminosity of SN\,2020bij in the $U$-band. This could be an indication that the host-galaxy extinction is over-estimated, or that an additional hot power source is present, such as weak CSM interaction. 
Similar discrepancies at UV wavelengths have been reported for SN\,2023axu \citep{Shrestha_2023} and SN\,2024ggi \citep{Jacobson_2024}, where they have been attributed to uncertainties in the treatment of UV line blanketing in the models.

In addition, both models prefer a high host-galaxy extinction value, which is inconsistent with the lack of strong \NaiD\ absorption in the spectra. However, the correlation between Na absorption and extinction is strongest when using high-resolution spectra \citep{Poznanski_2012}, which are not available in this case.

The best-fit parameters from both models are provided in Table \ref{tab:shockcooling_sample}. Most inferred parameters are consistent between the models within their uncertainties. However, the \citet{Morag_2023} model prefers a larger progenitor radius and larger extinction compared to the \citet{Sapir_and_Waxman} model.
The best-fit progenitor radius of $860 \pm 290$\,\Rsun\ from the \citet{Sapir_and_Waxman} model is consistent with that of RSGs, which have radii in the range of 100--1500\,\Rsun\ \citep[e.g.][and references therein]{Levesque_2017}, as are the best-fit envelope masses of $3^{+4}_{-1}$\,\Msun\ and $5^{+4}_{-3}$\,\Msun\ from the \cite{Sapir_and_Waxman} and \cite{Morag_2023} models, respectively \citep[e.g.][]{Sukhbold_2016}.
This is in contrast to other SNe IIP, where shock cooling alone does not provide a good fit to the early light curve \citep[e.g.][]{Hosseinzadeh_2018,Dong_2020, Tartaglia_2021,Ailawadhi_2023}, possibly due to a significant contribution from CSM interaction in those cases. Overall, both models reproduce the early light curve of SN\,2020bij and yield broadly consistent progenitor and explosion properties.

\begin{deluxetable*}{cccccccccc}
\caption{Best-fit parameters from the shock cooling model for the slowly rising SNe studied here. We report the median of the posterior distribution with the 16th and 84th percentiles as uncertainties.}
\label{tab:shockcooling_sample}
\centering
 \tablehead{SN & $v_{\rm s}^{*}$ & $M_{\rm env}$ & $f_\rho M$ & $R_*$ & $t_0$ & $E(B{-}V)$ & $d_L$ & Reference & Model$^a$ \\
Name & ($10^{8.5}$\,\cms) & ($M_\odot$) & ($M_\odot$) & ($R_\odot$) & (MJD) & (mag) & (Mpc) & & }
\startdata
SN\,2018fif & $0.83^{+0.05}_{-0.12}$ & $9.3^{+5.8}_{-0.4}$ & --- & $744^{+183}_{-128}$ & $58350.95^{+0.20}_{-0.13}$ & $0.20^{+0.04}_{-0.02}$ & --- & 1 & SW17\\
SN\,2020bij & $1.1 \pm 0.1$ & $3^{+4}_{-1}$ & $5^{+4}_{-3}$ & $860 \pm 290$ & $58875.70^{+0.20}_{-0.10}$ & $0.49 \pm 0.04$ & $58 \pm 3$ & 4 & SW17\\
SN\,2020bij & $0.9^{+0.2}_{-0.1}$ & $5^{+4}_{-3}$ & $3^{+4}_{-2}$ & $2300^{+290}_{-720}$ & $58876.40^{+0.60}_{-0.50}$ & $0.58^{+0.03}_{-0.04}$ & $58 \pm 3$ & 4 & MSW23\\
SN\,2021yja & $1.0 \pm 0.2$ & $0.6 \pm 0.1$ & $60 \pm 30$ & $2010 \pm 290$ & $59464.40 \pm 0.06$ & $0.11^{+0.01}_{-0.01}$ & $25 \pm 5$ & 2 & SW17\\
SN\,2023axu & $0.87^{+0.04}_{-0.03}$ & $1.2 \pm 0.1$ & $0.7 \pm 0.2$ & $560 \pm 43$ & $59971.26^{+0.01}_{-0.02}$ & --- & --- & 3 & MSW23\\ \hline
\enddata
\tablenotetext{a}{SW17 and MSW23 refer to
the models from \cite{Sapir_and_Waxman} and \cite{Morag_2023}, respectively.}
\tablecomments{
\textbf{References.} (1) \cite{Maayane}; (2) \cite{Hosseinzadeh_2022}; (3) \cite{Shrestha_2023}; (4) This work
}
\end{deluxetable*}

The early light curves of the slowly rising SN\,2018fif, SN\,2021yja and SN\,2023axu were modeled by \cite{Maayane}, \cite{Hosseinzadeh_2022} and \cite{Shrestha_2023}, respectively, using the analytic framework of \cite{Sapir_and_Waxman}, or using its extended formulation developed in \cite{Morag_2023}. We summarize their results in Table \ref{tab:shockcooling_sample}. As for SN\,2020bij, the model successfully reproduced the early photometric data and yielded reasonable progenitor parameters in all three cases \citep{Maayane, Hosseinzadeh_2022,Shrestha_2023}. For SN\,2023axu the model also under-predicted the $UV$ flux (as in SN\,2020bij). There, \cite{Shrestha_2023} interpreted the discrepancy as over-correction of UV line blanketing in the models. Together, these results suggest that, similarly to SN\,2020bij, the early emission of SN\,2018fif, SN\,2021yja and SN\,2023axu was likely dominated by shock cooling. Unfortunately, not enough early data is available for ASASSN-14kg to reliably fit the shock cooling model to that event.

\subsection{SNEmcee Fit} \label{subsec:SNEmcee_Fit} 

\begin{deluxetable*}{cccc}
\caption{SNEmcee model parameters, their units and the grid values.}
\label{tab:Snemcee_parameters}
\centering
 \tablehead{
\colhead{Parameter} & \colhead{Notation} & \colhead{Grid values} & \colhead{Units}
}
\startdata 
Progenitor ZAMS mass & \Mzams & 9,10,11,13,15,17 & \Msun \\ 
CSM density & $K$ & 0,10,30,60 & $10^{17}\,g\,{cm}^{-1}$ \\
CSM thickness & $\Delta\Rcsm$ & 0,500,1000,2000 & \Rsun \\
Explosion energy & $E$ & 0.1,0.3,0.5,0.7,0.9,1.3,1.7 & $10^{51}$\,erg \\
$^{56}$\Ni\ mass & \Ni\ & 0.001,0.02,0.07,0.12 & \Msun \\
$^{56}$\Ni\ mixing & $Mix$ & 2,8 & \Msun \\
Explosion time & $t$ & - & days \\
Luminosity scaling & $S$ & - & unitless \\\hline
\enddata
\tablecomments{The two last parameters are not part of the grid but part of the SNEmcee output.}
\end{deluxetable*}

We also fit numerical stellar evolution and explosion models to our data of SN\,2020bij using SNEmcee (Mohsen et al. in preparation)\footnote{\url{https://github.com/sondosmohsen/SNemcee}}. SNEmcee, a new open-source tool, performs MCMC fitting to an interpolated grid of numerical SN progenitor models from KEPLER \citep{Weaver_1978,Woosley_Heger_2007,Woosley_Heger_2015,Sukhbold_2014,Sukhbold_2016} exploded with the Supernova Explosion Code \citep[SNEC;][]{Morozova_2015,Morozova_2016,Morozova_2017,Morozova_2018} with a parametrized CSM to H-rich SN light curves and velocity measurements. The CSM is assumed to be a constant-velocity mass-loss wind attached to the surface of the progenitor out to some outer radius, \Rext, with a density profile:
\begin{equation} \label{eq:density}
\rho (r) = \frac {\dot{M}}{4 \pi r^2 {v}_{wind}} \equiv \frac{K} {r^2}
\end{equation}
where $\dot{M}$ is the mass-loss rate and $v_{wind}$ is the wind velocity.
The CSM is thus parametrized by two quantities: $K$, the overall scaling of the CSM density, and a cutoff radius where the CSM is assumed to abruptly terminate. \cite{Morozova_2018} parametrize this cutoff radius as \Rext, which is measured from the center of the star (i.e. it includes the pre-explosion progenitor radius \Rstar). SNEmcee, on the other hand, uses:
\begin{equation} \label{eq:thickness}
\Delta\Rcsm = \Rext - \Rstar
\end{equation}
All CSM-related quantities presented hereafter are expressed in terms of this CSM thickness ($\Delta\Rcsm$) rather than the total CSM outer radius (\Rext).

The SNEmcee fit parameters are the progenitor mass, the explosion energy, the $^{56}$\Ni\ mass and mixing, and the CSM scaling $K$ and thickness $\Delta\Rcsm$. Two additional parameters are the time between explosion and discovery (hereafter referred to as ``explosion time'') and an overall luminosity scaling parameter to account for uncertainties in the distance estimation. The $^{56}$\Ni\ mixing parameter is the mass coordinate (in solar masses) up to which the $^{56}$\Ni\ is mixed in the ejecta. The SNEmcee parameters are summarized in Table \ref{tab:Snemcee_parameters}.
We assume uniform priors for all physical parameters within the bounds of the pre-computed model grid. The luminosity scaling parameter, $S$, is assigned a Gaussian prior centered at $S=1$, with a width set by the distance uncertainty.

\begin{figure*}
\includegraphics[width=1\textwidth]{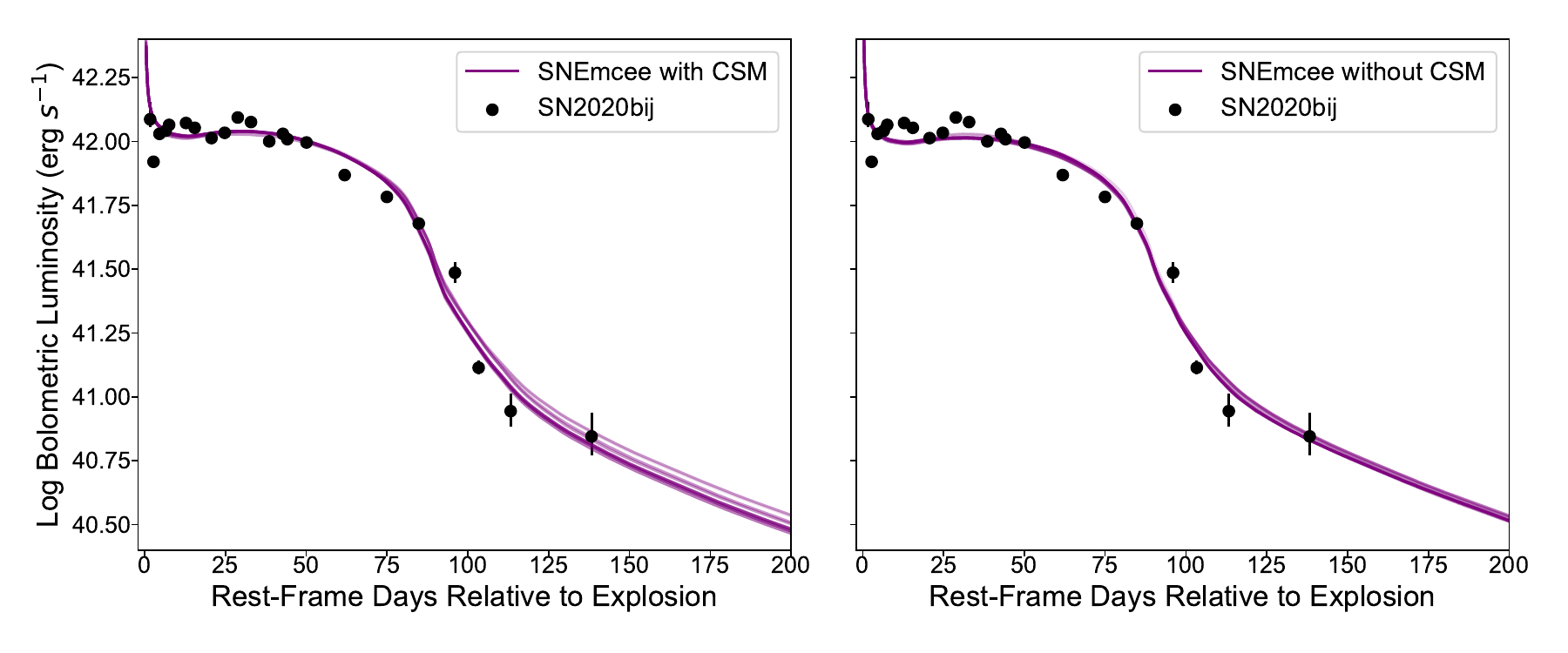}
\centering
\caption{SNEmcee models (purple lines) drawn from the MCMC posterior compared to the observed bolometric luminosity of SN\,2020bij (black circles) with (left) and without (right) CSM interaction. Both cases fit the data well.}
\label{fig:comparison_snemcee}
\end{figure*}

\begin{deluxetable*}{cccccccccc}
\setlength{\tabcolsep}{3pt}
\tablecaption{Best-fit SNEmcee parameters to the bolometric light curves of the slowly rising SNe IIP in our sample. We report the median of the posterior distribution with the 16th and 84th percentiles as uncertainties.}
\label{tab:parameters_SNEmcee_sample}
\tablehead{
\colhead{SN} & \colhead{t} & \colhead{S} & \colhead{Mix} & \colhead{E} & \colhead{\Ni} & \colhead{\Mzams} & \colhead{K} & \colhead{$\Delta\Rcsm$} & \colhead{\Mcsm} \\ 
\colhead{Name}
& (days) & & (\Msun) & ($10^{51}$\,erg) & (\Msun) & (\Msun) & ($10^{17}\,g\,{cm}^{-1}$) & (\Rsun) & (\Msun)
}
\startdata
\multicolumn{10}{c}{Fits with CSM} \\
\hline
ASASSN-14kg & $1.41_{-0.43}^{+0.47}$ & $1.09_{-0.043}^{+0.021}$ & $2.31_{-0.11}^{+0.29}$ & $1.20_{-0.026}^{+0.049}$ & $0.0759_{-0.0029}^{+0.0049}$ & $12.90_{-0.19}^{+0.22}$ & $20.6_{-3.17}^{+3.12}$ & $931_{-62.2}^{+64.6}$ & $0.84_{-0.14}^{+0.14}$\\
SN\,2018fif & $-0.51_{-0.33}^{+0.35}$ & $0.99_{-0.052}^{+0.051}$ & $5.12_{-1.99}^{+2.10}$ & $0.95_{-0.39}^{+0.33}$ & $0.0481_{-0.015}^{+0.017}$ & $13.1_{-2.7}^{+2.8}$ & $26.9_{-19.6}^{+21.1}$ & $413_{-318}^{+282}$ & $0.48_{-0.517}^{+0.507}$\\
SN\,2020bij & $-0.511_{-0.082}^{+0.069}$ & $0.604_{-0.012} ^{+0.164}$ & $4.7_{-1.6} ^{+1.6}$ & $0.799_{-0.102}^{+0.085}$ & $0.025_{-0.0139} ^{+0.014}$ & $14.0_{-0.97} ^{+0.96}$ & $35.2_{-19.9}^{+22.7}$ & $75.7_{-31.4} ^{+73.8}$ & $0.117_{-0.136}^{+0.082}$ \\
SN\,2021yja & $0.0055_{-0.0030}^{+0.0029}$ & $1.45_{-0.41}^{+0.35}$ & $4.7_{-1.7}^{+1.9}$ & $1.11_{-0.32}^{+0.33}$ & $0.067_{-0.033}^{+0.032}$ & $12.9_{-2.9}^{+2.4}$ & $12.1_{-2.4}^{+2.6}$ & $780_{-136}^{+142}$ & $0.41_{-0.11}^{+0.12}$ \\
SN\,2023axu & $0.046_{-0.026}^{+0.027}$ & $0.62_{-0.12}^{+0.12}$ & $6.6_{-1.0}^{+1.2}$ & $1.26_{-0.19}^{+0.20}$ & $0.094_{-0.025}^{+0.027}$ & $12.0_{-0.24}^{+0.65}$ & $40.8_{-6.1}^{+7.4}$ & $491_{-63}^{+75}$ & $0.88_{-0.17}^{+0.21}$\\
\hline
\multicolumn{10}{c}{Fits without CSM} \\
\hline
ASASSN-14kg & $0.139_{-0.013}^{+0.14}$ & $0.99_{-0.012}^{+0.033}$ & $2.16_{-0.046}^{+0.16}$ & $1.61_{-0.061}^{+0.024}$ & $0.1140_{-0.0059}^{+0.0022}$ & $14.8_{-0.18}^{+0.14}$ & -- & -- & --\\
SN\,2018fif & $-0.32_{-0.27}^{+0.33}$ & $1.0_{-0.055}^{+0.053}$ & $4.81_{-2.26}^{+2.09}$ & $1.45_{-0.17}^{+0.16}$ & $0.0458_{-0.015}^{+0.015}$ & $13.9_{-2.2}^{+2.1}$ & -- & -- & --\\
SN\,2020bij & $-0.69_{-0.44}^{+0.31}$ & $0.478_{-0.037}^{+0.066}$ & $4.38_{-1.66}^{+1.72}$ & $0.92_{-0.093}^{+0.083}$ & $0.030_{-0.013}^{+0.014}$ & $14.3_{-1.14}^{+1.18}$ & -- & -- & --\\
SN\,2021yja & $0.0052_{-0.0034}^{+0.0032}$ & $1.49_{-0.087}^{+0.10}$ & $3.5_{-1.2}^{+1.4}$ & $1.59_{-0.077}^{+0.089}$ & $0.054_{-0.027}^{+0.027}$ & $13.1_{-0.36}^{+0.37}$ & -- & -- & -- \\
SN\,2023axu & $0.085_{-0.019}^{+0.014}$ & $0.61_{-0.042}^{+0.057}$ & $2.2_{-0.18}^{+0.11}$ & $1.59_{-0.17}^{+0.11}$ & $0.107_{-0.017}^{+0.0095}$ & $12.8_{-0.068}^{+0.022}$ & -- & -- & -- \\
\enddata
\end{deluxetable*}

\begin{figure*}
\includegraphics[width=1\textwidth]{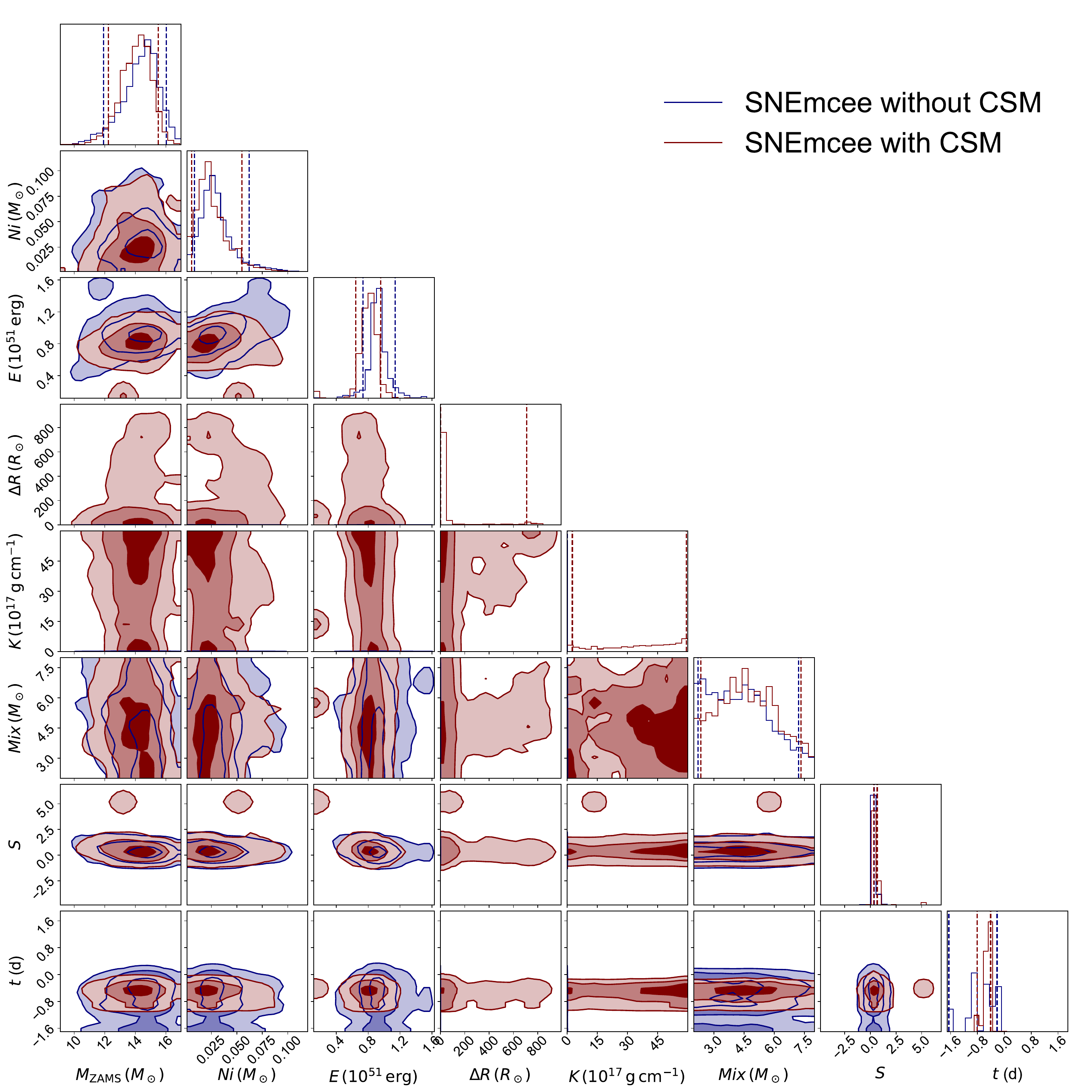}
\centering
\caption{Posterior probability distributions of fitting the SNEmcee models to the bolometric light curve of SN\,2020bij with (red) and without (blue) CSM. Both fits converge on similar parameters, with the CSM fit consistent with a cutoff radius thickness of 0 at 1.35$\sigma$ (i.e. consistent with no CSM).}
\label{fig:Snemcee_with_and_without_CSM}
\end{figure*}

SNEmcee has two fitting modes, one with all eight parameters free, and one with no contribution from the CSM, where the two CSM parameters are fixed to zero.
We fit the model to the bolometric light curve of SN\,2020bij calculated in Section \ref{subsec:bb} using each fitting mode (i.e. with and without CSM). We perform the MCMC fitting using 100 walkers with 500 burn-in steps followed by an additional 1000 steps.
The fit results are shown in Figure \ref{fig:comparison_snemcee} and Table \ref{tab:parameters_SNEmcee_sample}, where we also calculate the total CSM mass by integrating Equation \ref{eq:density} from \Rstar to \Rext. The respective corner plot is shown in Figure \ref{fig:Snemcee_with_and_without_CSM}. 
The inferred explosion time is consistent with the explosion window defined by the last non-detection and first detection in both fits.
SNEmcee provides a good fit to the data of SN\,2020bij without CSM. 
Adding CSM to the fit does not significantly affect the results. In fact, despite the added degrees of freedom, the fit with CSM is marginally consistent with a CSM thickness of 0. We compared the two SNEmcee fits using the Bayesian Information Criterion (BIC)\footnote{ The BIC penalizes models with additional free parameters. Differences of $\Delta{\rm BIC}\gtrsim 2$ are commonly interpreted as evidence against the model with the higher BIC, while $\Delta{\rm BIC}\gtrsim 6$ indicate strong evidence \citep[e.g.,][]{Kass_1995}.}, finding that the model without CSM is preferred ($\Delta{\rm BIC}\simeq 5.91$), suggesting that the data do not require a CSM component.

We repeat the fits for the other slowly rising SNe in our sample and present the results in Figures \ref{fig:snemcee_sample}--\ref{fig:cornerplot_SN2023axu} in Appendix \ref{sec:snemcee_posteriors}, as well as in Table \ref{tab:parameters_SNEmcee_sample}. For ASASSN-14kg, SN\,2021yja and SN\,2023axu, models with CSM provide better fits, as quantified by $\Delta{\rm BIC}\simeq 177.1$, $19.0$, and $349.5$, respectively. The data of SN\,2018fif are fit well both with and without CSM, with a preference for the model without CSM ($\Delta{\rm BIC}\simeq 5.4$), yet its $^{56}$\Ni\ mass can not be constrained due to the lack of late-time data. In all cases, the best-fit explosion energies and progenitor masses are typical of SNe IIP \citep[e.g.][]{kasen,Sukhbold_2016,Morozova_2018}.

\section{Spectroscopic Analysis} \label{sec:Spectroscopic_analysis}

The spectra of SN\,2020bij (Fig. \ref{fig:Spectra}) show common Type II SN features, such as strong emission lines of H (4101, 4340, 4861, and 6563\,\AA) as well as \Feii\,(5169\,\AA). No prominent flash-ionized features \citep{Gal-Yam_2014} are seen in the early spectra, though our earliest spectrum was obtained 4.71$\pm$0.96 days after explosion, so such lines, if they were present, could have disappeared by then \citep{Khazov_2016,Bruch_2023}.

\begin{figure}
\includegraphics[width=0.5\textwidth]{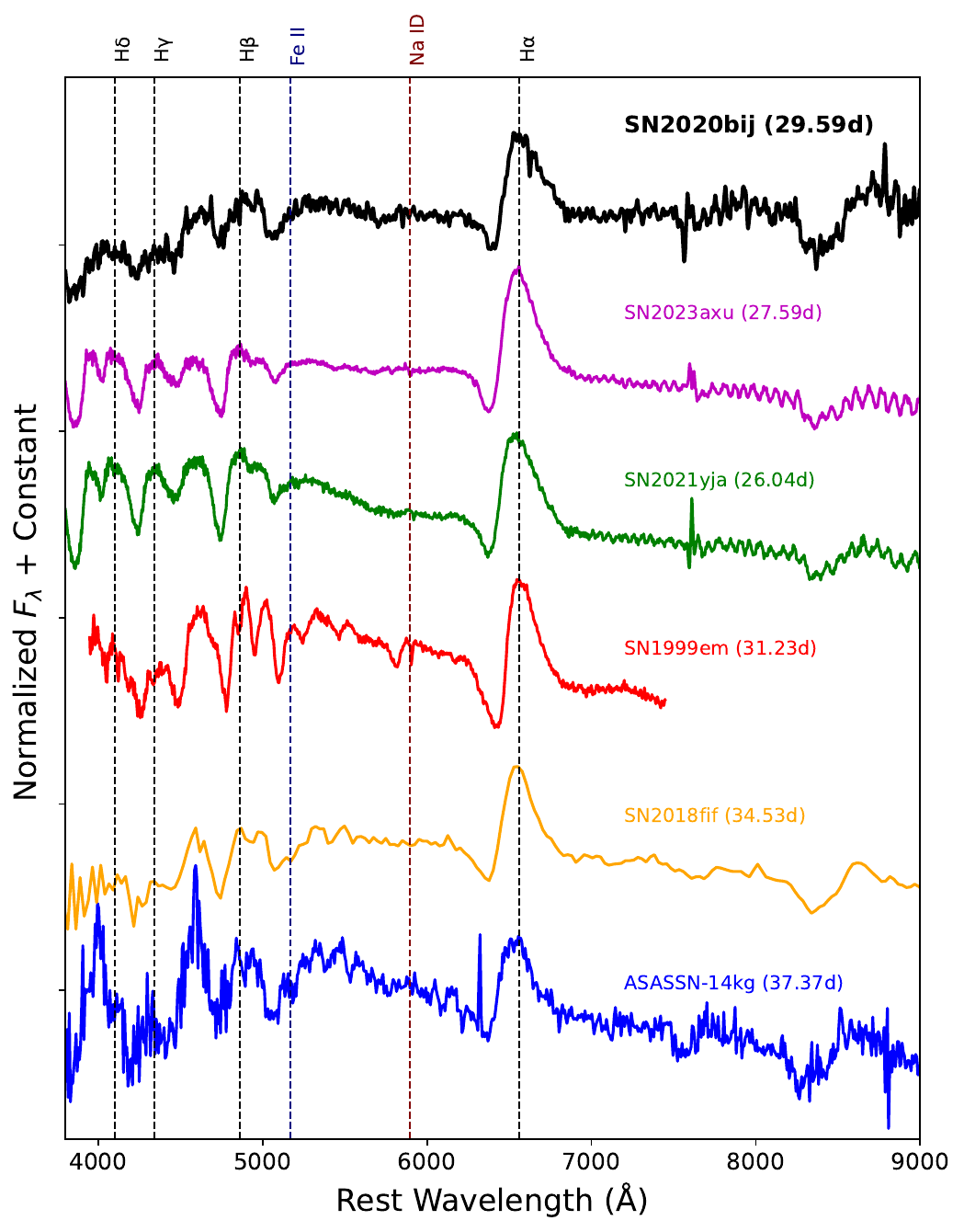}
\centering
\caption{The spectrum of SN\,2020bij (black) and of our slowly rising sample, compared to the prototypical Type IIP SN\,1999em, all at a phase of roughly 30 rest-frame days from explosion. All events show similar spectral features typical of Type IIP SNe (namely the Balmer series, marked in black vertical lines, and \Feii\ 5169\,\AA\, marked in blue). The red vertical line marks the \NaiD\ region. Notably, SN\,2020bij and the other slowly rising SNe lack strong narrow \NaiD\ absorption, indicative of low local extinction. Phases in rest-frame days relative to explosion are shown on the right. See text for data sources.}
\label{fig:comparison_spectra}
\end{figure}

Figure \ref{fig:comparison_spectra} shows a comparison between the spectra of SN\,2020bij to those of the slowly rising SN\,2018fif \citep{Maayane}, SN\,2021yja \citep{Hosseinzadeh_2022}, SN\,2023axu \citep{Shrestha_2023} and the prototypical Type IIP SN\,1999em \citep{Leonard_2002}, all at approximately 30 rest-frame days after explosion. We also plot a spectrum of ASASSN-14kg that we obtained with Las Cumbres and reduced in the same way that our SN\,2020bij spectra were obtained and reduced (see Section \ref{sec:Observations}).
The full set of ASASSN-14kg spectra is presented in Appendix \ref{sec:ASASSN-14kg_spectra}.
The data of SN\,1999em were retrieved via WISeREP \citep{Leonard_2002}. We find that SN\,2020bij and the other slowly rising SNe in our sample all lack strong narrow \NaiD\ absorption which could indicate low local extinction for these events.

SN\,2018fif, SN\,2021yja and SN\,2023axu exhibit a ‘ledge’ feature (a broad absorption-like structure around 4500–4800\,\AA\, first described by \cite{Maayane}, during the first $\sim$1–2 days after explosion, disappearing by day $\sim$3 \citep[see Figure 8 in][]{Shrestha_2023}. This feature has been attributed to interaction with low-density CSM \citep{Bruch_2021,Hosseinzadeh_2022}. Since no spectra exist for SN\,2020bij, SN\,1999em and ASASSN-14kg at those phases, we cannot confirm whether this feature was present also in those events.

\subsection{\ha\ and \Feii\ Velocities}

We calculate the expansion velocities associated with the \ha\ line for all the slowly rising events in our sample and with the \Feii\ 5169\,\AA\ line for SN\,2020bij, SN\,2021yja and SN\,2023axu by measuring the offset of the minimum of the P-Cygni profile (determined from second order polynomial fits) from the rest wavelength of each line.
For SN\,2018fif and ASASSN-14kg, \Feii\ 5169\,\AA\ line velocity measurements were not possible, as no data were available in the literature and we were unable to derive them for ASASSN-14kg from the existing spectra given their low signal to noise ratio. 

\begin{figure}
\includegraphics[width=0.5\textwidth]{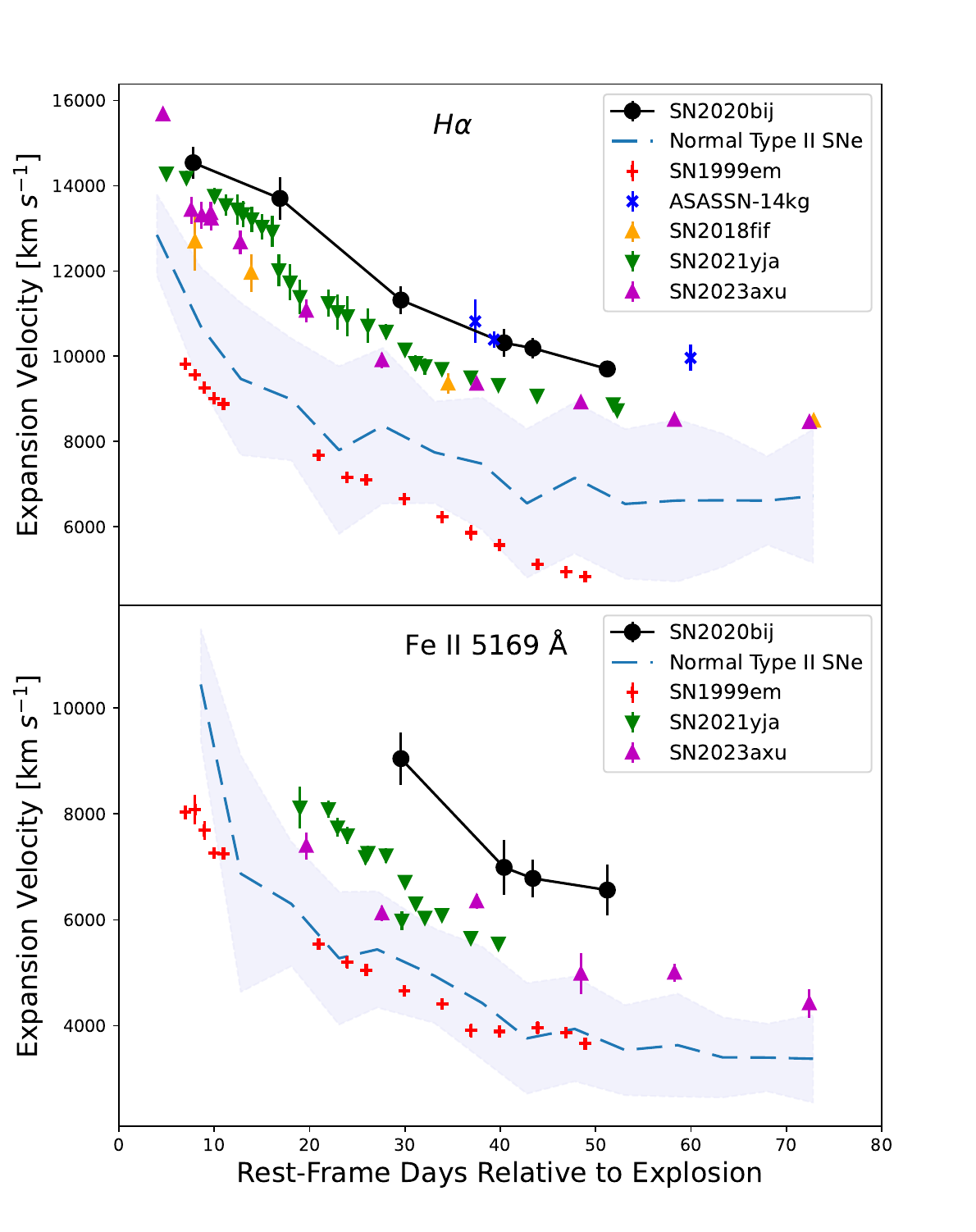}
\centering
\caption{Velocity evolution of \ha\ and \Feii\, for the slowly rising events in our sample compared to the prototypical Type IIP SN\,1999em \citep{Leonard_2002} and to the distribution of velocities of normal SNe II from \cite{Gutierrez}. The blue shaded region denotes $\pm1\sigma$ around the mean of their sample, which is marked by the dashed blue line.}
\label{fig:Velo_comparison}
\end{figure}

Our results are listed in Table \ref{tab:velocities} and plotted in Figure \ref{fig:Velo_comparison}, where we also compare them to the Type II SN velocity distribution from \cite{Gutierrez} and the velocities of the prototypical Type IIP SN\,1999em \citep{Leonard_2002}.
All slowly rising events exhibit velocities $>1\sigma$ above the \cite{Gutierrez} sample mean, with SN\,2020bij showing the highest \ha\ and \Feii\ velocities.

\begin{deluxetable}{cccc}
\caption{Expansion velocities measured from P-Cygni absorption minima of \ha\ and \Feii\ 5169\,\AA\ for the slowly rising events in our sample.}
\label{tab:velocities} 
\centering
 \tablehead{
\colhead{SN {Name}} & \colhead{Phase} & \colhead{\ha\ Velocity} & \colhead{\Feii\ Velocity} \\
 &{(days)} & {(km\,s$^{-1}$)} & {(km\,s$^{-1}$)} 
}
\startdata
SN\,2021yja & 4.99 & 14286.17 $\pm$ 148.85 & $-$ \\
SN\,2021yja & 7.09 & 14188.29 $\pm$ 162.48 & $-$ \\
SN\,2020bij & 29.59 & 11318.19 $\pm$ 322.32 & 9045.71 $\pm$ 501.16 \\
SN\,2020bij & 40.41 & 10314.35 $\pm$ 330.95 & 6990.28 $\pm$ 515.55 \\
SN\,2020bij & 43.43 & 10189.11 $\pm$ 245.66 & 6782.58 $\pm$ 350.76 \\
SN\,2020bij & 51.23 & 9701.21 $\pm$ 151.65 & 6561.93 $\pm$ 479.83 \\
\hline
\enddata
\tablecomments{Phases are in rest-frame days relative to explosion. This table is published in its entirety in the machine-readable format. A portion is shown here for guidance regarding its form and content.}
\end{deluxetable}

\section{Discussion} \label{sec:Discussion}

SN\,2020bij is generally similar both photometrically and spectroscopically to normal SNe IIP. Its global light curve shape (Fig. \ref{fig:LC_2020bij}), $s_{50V}$ and $t_{PT}$ values (Fig. \ref{fig:lc_parameters}), color evolution (Fig. \ref{fig:color_evolution_comp}), and spectral features (Figs. \ref{fig:Spectra} and \ref{fig:comparison_spectra}) are all typical of Type IIP SNe. However, a few key properties set SN\,2020bij apart from the majority of the Type IIP SN population.

\begin{figure}
\centering
\includegraphics[width=0.5\textwidth]{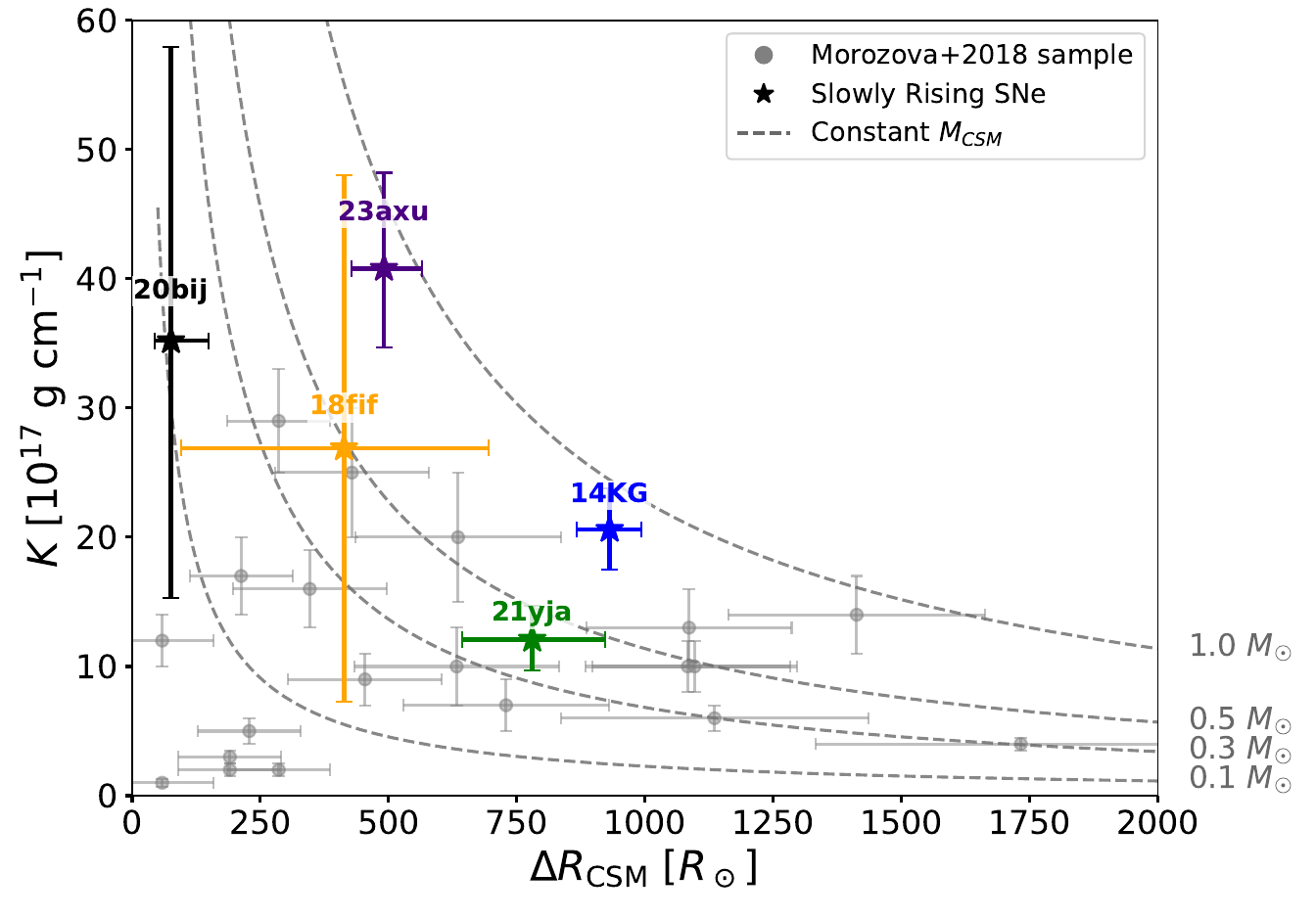}
\centering
\caption{The $K$ and $\Delta\Rcsm$ values for the slowly rising Type IIP SNe in our sample compared to the Type IIP SNe from \citet{Morozova_2018}. Dashed lines indicate constant \Mcsm.}
\label{fig:csm_space}
\end{figure}

The most notable difference between SN\,2020bij and typical SNe IIP is its slow rise to the plateau of about 14 days in the $r$-band, compared to the typical $\lesssim$10-day rise \citep{Gall_2015, Gonzalez_2015, Rubin_2016, Rubin_GalYam_2016} seen in many SNe IIP (Fig. \ref{fig:inset_abs_mag}). 
A slow rise to plateau is reproduced in radiation-hydrodynamical models of explosions with little or no CSM interaction \citep{Morozova_2018}, as well as by analytical shock cooling emission models, which include no CSM interaction powered emission \citep[Figure \ref{fig:cornerplot_shockcooling}; see also][]{Maayane,Hosseinzadeh_2022,Shrestha_2023}\footnote{Discrepancies in the $U$-band have been seen in shock cooling modeling of other Type II SNe \citep[e.g.][]{Hosseinzadeh_2022,Shrestha_2023} and were attributed to over- or under-correction for UV line blanketing.}. This suggests that the slow rise could indicate an explosion of a star that did not produce substantial CSM before core collapse (but see \citealt{Irani_2024}). Indeed, SNEmcee fits to SN\,2020bij prefer a very small value of $\Delta\Rcsm$ compared to other SNe IIP, indicating a confined CSM, while the rest of its progenitor and explosion best-fit parameters are typical of SNe IIP \citep{kasen,Sukhbold_2016,Morozova_2018}. We compare the CSM parameters of SN\,2020bij to those of \cite{Morozova_2018} in Figure \ref{fig:csm_space}. Here we convert the \cite{Morozova_2018} $R_{ext}$ values to $\Delta\Rcsm$ using Equation \ref{eq:thickness} and the stellar radii values of \cite{Sukhbold_2016}.

The second notable difference between SN\,2020bij and typical Type IIP SNe lies in its high expansion velocities ($\gtrsim\,$2$\sigma$ above the mean of the population) as measured in both the \ha\ and \Feii\ 5169\,\AA\ lines (Fig. \ref{fig:Velo_comparison}). The ``missing'' early bolometric luminosity due to the flat early light curve (compared to other Type IIP SNe where the bolometric luminosity declines at early-times; Fig. \ref{fig:BB_Evo}) is of order $\sim10^{49}$\,erg. In contrast, the added kinetic energy given the higher velocities is of order $\sim10^{50}$\,erg per solar mass of ejecta. Therefore, it is not clear if the absence of extended CSM is enough to account for the ejecta experiencing less deceleration and hence higher expansion velocities, or if the high expansion velocities are due to higher explosion energies or lower ejecta masses in such events compared to normal SNe IIP.
The velocities of SN\,2020bij are $\sim$20--60\% larger than typical SN IIP velocities. This implies either a $\sim$45--250\% higher explosion energy than normal (which is not obviously the case here; Table \ref{tab:parameters_SNEmcee_sample})\footnote{We do not see higher than normal nickel masses, which might have been an indication of a larger explosion energy \citep{Burrows_2024}.} or that the ejecta masses (not constrained here) are $\sim$20--60\% lower than normal. Either way, the observed high velocities in SN\,2020bij could point to a connection between explosion energy or ejecta mass and the amount of pre-explosion mass-loss. 

\begin{deluxetable*}{ccccc}
\tablecaption{Qualitative summary of properties of the slowly rising SNe IIP studied here.}
\label{tab:sample_summary}
\tablehead{
\colhead{Name} & \colhead{Slow-Rising} & \colhead{High Velocities} & \colhead{SNEmcee CSM Fit} & \colhead{Shock Cooling Model Fits}
}
\startdata
SN\,2020bij & yes & yes & Very confined & yes \\
SN\,2023axu & yes & yes & Confined & yes \\
SN\,2021yja & yes & yes & Confined & yes \\
SN\,2018fif & yes & yes & Confined & yes \\
ASASSN-14kg & yes & yes & Confined & n/a\textsuperscript{b} \\
\enddata
\tablenotetext{b}{Shock cooling model parameters could not be constrained for ASASSN-14kg due to the lack of early-time data.}
\end{deluxetable*}

A third distinctive property of SN\,2020bij is its best-fit blackbody temperature during the plateau. While initially SN\,2020bij exceeds 10,000\,K and later cools, as is typical of SNe IIP, its temperature flattens at $\sim$4000--5000\,K, which is cooler than the typical plateau temperature of SNe IIP of $\sim$6000--8000\,K (top panel of Figure \ref{fig:BB_Evo}). 
Unknown host-galaxy extinction could produce an apparent lower plateau temperature, however the luminous UV emission and lack of strong \NaiD\ absorption in SN\,2020bij argue against stronger than typical SN IIP host extinction (for which the rest of the sample is also not corrected). At early-times, the lack of data bluer than the $U$-band could lead to an underestimate of the temperature. However, once the temperature drops below $\sim$10,000\,K, optical data provide a reliable estimate of the temperature \citep{Arcavi_2022}.
We conclude that SN\,2020bij is likely among the lowest plateau-temperature Type IIP SNe known.

Slowly rising light curves have been seen in the $r$-band for a few other events (spanning different plateau luminosities and decline rates) such as SN\,2018fif \citep{Maayane}, SN\,2021yja \citep{Hosseinzadeh_2022}, and SN\,2023axu \citep{Shrestha_2023}, and in the $V$-band in ASASSN-14kg \citep{Valenti_2016,Davis_2021}. The first three, which have enough early data for shock cooling model fitting, are also well fit by shock cooling emission \citep{Maayane,Hosseinzadeh_2022,Shrestha_2023} indicating that, as in SN\,2020bij, CSM interaction is not a dominant source of emission for these events. 

Using SNEmcee, we find that while all four slow-rising events prefer CSM that is not as confined as that of SN\,2020bij, they still lie in the lower half of the $\Delta\Rcsm$ distribution (Fig. \ref{fig:csm_space}; though these are still small number statistics). The slow-rising sample does not seem to necessarily prefer low-mass CSM, but rather concentrates in a corner of phase space indicating high density but low radial-extent CSM.

Curiously, SN\,2018fif, SN\,2021yja, SN\,2023axu and ASASSN-14kg, selected here only by their slow rise, also show significantly higher expansion velocities compared to normal SNe IIP (Fig. \ref{fig:Velo_comparison}). This points to a strong correlation between high expansion velocities and slow light curve rise, which may indicate a progenitor mass-loss -- explosion-parameter connection.

The observational properties and modeling results for SN\,2020bij and the other slowly rising Type IIP SNe discussed here are summarized qualitatively in Table \ref{tab:sample_summary}. We propose that this suggests a new subclass of Type IIP SNe characterized by a slow rise to the plateau and high expansion velocities. These observational properties might be tied to particular pre-explosion CSM profiles, but a larger sample, with well covered early phases, is needed to determine the nature of this new subclass.

\section{Summary and Conclusions} \label{sec:Summary}

We presented the photometric and spectroscopic evolution of the Type IIP SN\,2020bij. SN\,2020bij is generally similar both photometrically and spectroscopically to normal SNe IIP, except that it shows a slow rise of roughly 14 days to the plateau, high expansion velocities, and low blackbody temperatures during the plateau.

We find that both the \citet{Sapir_and_Waxman} and \citet{Morag_2023} analytic shock cooling models reproduce the early light curve of SN\,2020bij fairly well, indicating that shock cooling alone can explain the early emission, in contrast to other SNe IIP which require also CSM interaction power to explain the rise to the plateau \citep{Morozova_2018}. We further fit numerical stellar evolution and explosion models to the bolometric light curve of SN\,2020bij using SNEmcee. Here too we find a good fit to the data without requiring a CSM interaction component. In fact, adding CSM to the fit does not significantly affect the results, and the fit with CSM is marginally consistent with a CSM thickness of zero.

To analyze SN\,2020bij in context, we searched the literature for other well-observed Type IIP SNe with slow light curve rises, and identify ASASSN-14kg, SN\,2018fif, SN\,2021yja and SN\,2023axu. Interestingly, these events, which were selected only by their slow rise to the plateau, also show significantly higher expansion velocities compared to normal SNe IIP. Moreover, for the events with sufficient data, shock cooling alone is enough to explain their early emission (as for SN\,2020bij). 
Using SNEmcee, we find that models with CSM can also fit the data, but are limited to a corner of CSM-interaction parameter space representing high-density and low radial-extent CSM, with SN\,2020bij the event with the least CSM required.

We therefore propose that the five events studied here constitute a new subclass of Type IIP SNe, characterized by a slow rise to the plateau and high expansion velocities, and associated with a confined CSM configuration.

Early and dense photometric observations are critical for mapping the rise to the plateau, and early spectroscopic data can be used to independently map the presence of confined CSM. Together, such observations for more slow-rising SNe IIP could better constrain the mass-loss history of the progenitors of this new subclass.
 ~\\ 

We thank M. Fraser for insightful and helpful discussions, and S. Srivastav for assistance.

S.M.-T. and I.A. acknowledge support from the European Research Council (ERC) under the European Union’s Horizon 2020 research and innovation program (grant agreement number 852097). S.M.-T. acknowledges further support from the Pazy Foundation (grant number 216312) and the Neubauer Family Foundation.
C.P.G. acknowledges financial support from the Secretary of Universities and Research (Government of Catalonia) and by the Horizon 2020 Research and Innovation Programme of the European Union under the Marie Sk\l{}odowska-Curie and the Beatriu de Pin\'os 2021 BP 00168 programme, from the Spanish Ministerio de Ciencia e Innovaci\'on (MCIN) and the Agencia Estatal de Investigaci\'on (AEI) 10.13039/501100011033 under the PID2023-151307NB-I00 SNNEXT project, from Centro Superior de Investigaciones Cient\'ificas (CSIC) under the PIE project 20215AT016 and the program Unidad de Excelencia Mar\'ia de Maeztu CEX2020-001058-M, and from the Departament de Recerca i Universitats de la Generalitat de Catalunya through the 2021-SGR-01270 grant.
T.-W.C. acknowledges financial support from the Yushan Fellow Program of the Ministry of Education, Taiwan (MOE-111-YSFMS-0008-001-P1), and from the National Science and Technology Council, Taiwan (NSTC 114-2112-M-008-021-MY3).
The Las Cumbres Observatory team is supported by National Science Foundation (NSF) grants AST-2308113 and AST-1911151.

This work makes use of observations from the Las Cumbres Observatory network obtained through the Global Supernova Project, from the European Organisation for Astronomical Research in the Southern Hemisphere (ESO), Chile, as part of ePESSTO+ (the advanced Public ESO Spectroscopic Survey for Transient Objects) on the ESO NTT 3.58\,m Telescope, and from the ATLAS project. ATLAS is primarily funded to search for near-earth objects (NEOs) through NASA grants NN12AR55G, 80NSSC18K0284, and 80NSSC18K1575; by-products of the NEO search include images and catalogs from the survey area.
The ATLAS science products have been made possible through the contributions of the University of Hawaii Institute for Astronomy, the Queen’s University Belfast, the Space Telescope Science Institute, the South African Astronomical Observatory, and The Millennium Institute of Astrophysics (MAS), Chile.

This work also has made use of the NASA/IPAC Extragalactic Database (NED), which is funded by the National Aeronautics and Space Administration and operated by the Jet Propulsion Laboratory, California Institute of Technology, under contract with NASA.

\bibliography{sample631}{}

@ARTICLE{Brown_2013,
       author = {{Brown}, T.~M. and {Baliber}, N. and {Bianco}, F.~B. and {Bowman}, M. and {Burleson}, B. and {Conway}, P. and {Crellin}, M. and {Depagne}, {\'E}. and {De Vera}, J. and {Dilday}, B. and {Dragomir}, D. and {Dubberley}, M. and {Eastman}, J.~D. and {Elphick}, M. and {Falarski}, M. and {Foale}, S. and {Ford}, M. and {Fulton}, B.~J. and {Garza}, J. and {Gomez}, E.~L. and {Graham}, M. and {Greene}, R. and {Haldeman}, B. and {Hawkins}, E. and {Haworth}, B. and {Haynes}, R. and {Hidas}, M. and {Hjelstrom}, A.~E. and {Howell}, D.~A. and {Hygelund}, J. and {Lister}, T.~A. and {Lobdill}, R. and {Martinez}, J. and {Mullins}, D.~S. and {Norbury}, M. and {Parrent}, J. and {Paulson}, R. and {Petry}, D.~L. and {Pickles}, A. and {Posner}, V. and {Rosing}, W.~E. and {Ross}, R. and {Sand}, D.~J. and {Saunders}, E.~S. and {Shobbrook}, J. and {Shporer}, A. and {Street}, R.~A. and {Thomas}, D. and {Tsapras}, Y. and {Tufts}, J.~R. and {Valenti}, S. and {Vander Horst}, K. and {Walker}, Z. and {White}, G. and {Willis}, M.},
        title = "{Las Cumbres Observatory Global Telescope Network}",
      journal = {\pasp},
         year = 2013,
        month = sep,
       volume = {125},
       number = {931},
        pages = {1031},
          doi = {10.1086/673168},
archivePrefix = {arXiv},
       eprint = {1305.2437},
 primaryClass = {astro-ph.IM},
       adsurl = {https://ui.adsabs.harvard.edu/abs/2013PASP..125.1031B}
}

@ARTICLE{Burrows_2024,
       author = {{Burrows}, Adam and {Wang}, Tianshu and {Vartanyan}, David},
        title = "{Physical Correlations and Predictions Emerging from Modern Core-collapse Supernova Theory}",
      journal = {\apjl},
         year = 2024,
        month = mar,
       volume = {964},
       number = {1},
          eid = {L16},
        pages = {L16},
          doi = {10.3847/2041-8213/ad319e},
archivePrefix = {arXiv},
       eprint = {2401.06840},
 primaryClass = {astro-ph.HE},
       adsurl = {https://ui.adsabs.harvard.edu/abs/2024ApJ...964L..16B}
}

@ARTICLE{Poznanski_2012,
       author = {{Poznanski}, Dovi and {Prochaska}, J. Xavier and {Bloom}, Joshua S.},
        title = "{An empirical relation between sodium absorption and dust extinction}",
      journal = {\mnras},
         year = 2012,
        month = oct,
       volume = {426},
       number = {2},
        pages = {1465-1474},
          doi = {10.1111/j.1365-2966.2012.21796.x},
archivePrefix = {arXiv},
       eprint = {1206.6107},
 primaryClass = {astro-ph.IM},
       adsurl = {https://ui.adsabs.harvard.edu/abs/2012MNRAS.426.1465P}
}

@ARTICLE{Itagaki_2020bij_discovery,
       author = {{Itagaki}, K.},
        title = "{Transient Discovery Report for 2020-01-29}",
      journal = {Transient Name Server Discovery Report},
         year = 2020,
        month = jan,
       volume = {2020-310},
        pages = {1},
       adsurl = {https://ui.adsabs.harvard.edu/abs/2020TNSTR.310....1I}
}

@ARTICLE{ATLAS_data1_Tonry2018,
       author = {{Tonry}, J.~L. and {Denneau}, L. and {Heinze}, A.~N. and {Stalder}, B. and {Smith}, K.~W. and {Smartt}, S.~J. and {Stubbs}, C.~W. and {Weiland}, H.~J. and {Rest}, A.},
        title = "{ATLAS: A High-cadence All-sky Survey System}",
      journal = {\pasp},
         year = 2018,
        month = jun,
       volume = {130},
       number = {988},
        pages = {064505},
          doi = {10.1088/1538-3873/aabadf},
archivePrefix = {arXiv},
       eprint = {1802.00879},
 primaryClass = {astro-ph.IM},
       adsurl = {https://ui.adsabs.harvard.edu/abs/2018PASP..130f4505T}
}

@ARTICLE{ATLAS_data2_Smith2020,
       author = {{Smith}, M. and {Sullivan}, M. and {Wiseman}, P. and {Kessler}, R. and {Scolnic}, D. and {Brout}, D. and {D'Andrea}, C.~B. and {Davis}, T.~M. and {Foley}, R.~J. and {Frohmaier}, C. and {Galbany}, L. and {Gupta}, R.~R. and {Guti{\'e}rrez}, C.~P. and {Hinton}, S.~R. and {Kelsey}, L. and {Lidman}, C. and {Macaulay}, E. and {M{\"o}ller}, A. and {Nichol}, R.~C. and {Nugent}, P. and {Palmese}, A. and {Pursiainen}, M. and {Sako}, M. and {Swann}, E. and {Thomas}, R.~C. and {Tucker}, B.~E. and {Vincenzi}, M. and {Carollo}, D. and {Lewis}, G.~F. and {Sommer}, N.~E. and {Abbott}, T.~M.~C. and {Aguena}, M. and {Allam}, S. and {Avila}, S. and {Bertin}, E. and {Bhargava}, S. and {Brooks}, D. and {Buckley-Geer}, E. and {Burke}, D.~L. and {Carnero Rosell}, A. and {Carrasco Kind}, M. and {Costanzi}, M. and {da Costa}, L.~N. and {De Vicente}, J. and {Desai}, S. and {Diehl}, H.~T. and {Doel}, P. and {Eifler}, T.~F. and {Everett}, S. and {Flaugher}, B. and {Fosalba}, P. and {Frieman}, J. and {Garc{\'\i}a-Bellido}, J. and {Gaztanaga}, E. and {Glazebrook}, K. and {Gruen}, D. and {Gruendl}, R.~A. and {Gschwend}, J. and {Gutierrez}, G. and {Hartley}, W.~G. and {Hollowood}, D.~L. and {Honscheid}, K. and {James}, D.~J. and {Krause}, E. and {Kuehn}, K. and {Kuropatkin}, N. and {Lima}, M. and {MacCrann}, N. and {Maia}, M.~A.~G. and {Marshall}, J.~L. and {Martini}, P. and {Melchior}, P. and {Menanteau}, F. and {Miquel}, R. and {Paz-Chinch{\'o}n}, F. and {Plazas}, A.~A. and {Romer}, A.~K. and {Roodman}, A. and {Rykoff}, E.~S. and {Sanchez}, E. and {Scarpine}, V. and {Schubnell}, M. and {Serrano}, S. and {Sevilla-Noarbe}, I. and {Suchyta}, E. and {Swanson}, M.~E.~C. and {Tarle}, G. and {Thomas}, D. and {Tucker}, D.~L. and {Varga}, T.~N. and {Walker}, A.~R. and {DES Collaboration}},
        title = "{First cosmology results using type Ia supernovae from the Dark Energy Survey: the effect of host galaxy properties on supernova luminosity}",
      journal = {\mnras},
         year = 2020,
        month = may,
       volume = {494},
       number = {3},
        pages = {4426-4447},
          doi = {10.1093/mnras/staa946},
archivePrefix = {arXiv},
       eprint = {2001.11294},
 primaryClass = {astro-ph.CO},
       adsurl = {https://ui.adsabs.harvard.edu/abs/2020MNRAS.494.4426S}
}

@article{Gaia_alerts,
	doi = {10.1051/0004-6361/202140735},
  
	url = {https://doi.org/10.1051%2F0004-6361%2F202140735},
  
	year = 2021,
	month = {aug},
  
	publisher = {{EDP} Sciences},
  
	volume = {652},
  
	pages = {A76},
  
	author = {S. T. Hodgkin and D. L. Harrison and E. Breedt and T. Wevers and G. Rixon and A. Delgado and A. Yoldas and Z. Kostrzewa-Rutkowska and {\L}. Wyrzykowski and M. van Leeuwen and N. Blagorodnova and H. Campbell and D. Eappachen and M. Fraser and N. Ihanec and S. E. Koposov and K. Kruszy{\'{n}
}ska and G. Marton and K. A. Rybicki and A. G. A. Brown and P. W. Burgess and G. Busso and S. Cowell and F. De Angeli and C. Diener and D. W. Evans and G. Gilmore and G. Holland and P. G. Jonker and F. van Leeuwen and F. Mignard and P. J. Osborne and J. Portell and T. Prusti and P. J. Richards and M. Riello and G. M. Seabroke and N. A. Walton and P. {\'{A}}brah{\'{a}}m and G. Altavilla and S. G. Baker and U. Bastian and P. O'Brien and J. de Bruijne and T. Butterley and J. M. Carrasco and J. Casta{\~{n}}eda and J. S. Clark and G. Clementini and C. M. Copperwheat and M. Cropper and G. Damljanovic and M. Davidson and C. J. Davis and M. Dennefeld and V. S. Dhillon and C. Dolding and M. Dominik and P. Esquej and L. Eyer and C. Fabricius and M. Fridman and D. Froebrich and N. Garralda and A. Gomboc and J. J. Gonz{\'{a}}lez-Vidal and R. Guerra and N. C. Hambly and L. K. Hardy and B. Holl and A. Hourihane and J. Japelj and D. A. Kann and C. Kiss and C. Knigge and U. Kolb and S. Komossa and {\'{A}}. K{\'{o}}sp{\'{a}}l and G. Kov{\'{a}}cs and M. Kun and G. Leto and F. Lewis and S. P. Littlefair and A. A. Mahabal and C. G. Mundell and Z. Nagy and D. Padeletti and L. Palaversa and A. Pigulski and M. L. Pretorius and W. van Reeven and V. A. R. M. Ribeiro and M. Roelens and N. Rowell and N. Schartel and A. Scholz and A. Schwope and B. M. Sip{\H{o}}cz and S. J. Smartt and M. D. Smith and I. Serraller and D. Steeghs and M. Sullivan and L. Szabados and E. Szegedi-Elek and P. Tisserand and L. Tomasella and S. van Velzen and P. A. Whitelock and R. W. Wilson and D. R. Young},
  
	title = {$\less$i$\greater$Gaia$\less$/i$\greater$ Early Data Release 3},
  
	journal = {Astronomy \& Astrophysics}
}

@article{Zackay_2016,
   title={PROPER IMAGE SUBTRACTION—OPTIMAL TRANSIENT DETECTION, PHOTOMETRY, AND HYPOTHESIS TESTING},
   volume={830},
   ISSN={1538-4357},
   url={http://dx.doi.org/10.3847/0004-637X/830/1/27},
   DOI={10.3847/0004-637x/830/1/27},
   number={1},
   journal={The Astrophysical Journal},
   publisher={American Astronomical Society},
   author={Zackay, Barak and Ofek, Eran O. and Gal-Yam, Avishay},
   year={2016},
   month=oct, pages={27} }

@ARTICLE{WMAP9,
       author = {{Hinshaw}, G. and {Larson}, D. and {Komatsu}, E. and {Spergel}, D.~N. and {Bennett}, C.~L. and {Dunkley}, J. and {Nolta}, M.~R. and {Halpern}, M. and {Hill}, R.~S. and {Odegard}, N. and {Page}, L. and {Smith}, K.~M. and {Weiland}, J.~L. and {Gold}, B. and {Jarosik}, N. and {Kogut}, A. and {Limon}, M. and {Meyer}, S.~S. and {Tucker}, G.~S. and {Wollack}, E. and {Wright}, E.~L.},
        title = "{Nine-year Wilkinson Microwave Anisotropy Probe (WMAP) Observations: Cosmological Parameter Results}",
      journal = {\apjs},
         year = 2013,
        month = oct,
       volume = {208},
       number = {2},
          eid = {19},
        pages = {19},
          doi = {10.1088/0067-0049/208/2/19},
archivePrefix = {arXiv},
       eprint = {1212.5226},
 primaryClass = {astro-ph.CO},
       adsurl = {https://ui.adsabs.harvard.edu/abs/2013ApJS..208...19H}
}

@ARTICLE{Schlafly,
       author = {{Schlafly}, Edward F. and {Finkbeiner}, Douglas P.},
        title = "{Measuring Reddening with Sloan Digital Sky Survey Stellar Spectra and Recalibrating SFD}",
      journal = {\apj},
         year = 2011,
        month = aug,
       volume = {737},
       number = {2},
          eid = {103},
        pages = {103},
          doi = {10.1088/0004-637X/737/2/103},
archivePrefix = {arXiv},
       eprint = {1012.4804},
 primaryClass = {astro-ph.GA},
       adsurl = {https://ui.adsabs.harvard.edu/abs/2011ApJ...737..103S}
}

@ARTICLE{Fitzpatrick_1999,
       author = {{Fitzpatrick}, Edward L.},
        title = "{Correcting for the Effects of Interstellar Extinction}",
      journal = {\pasp},
         year = 1999,
        month = jan,
       volume = {111},
       number = {755},
        pages = {63-75},
          doi = {10.1086/316293},
archivePrefix = {arXiv},
       eprint = {astro-ph/9809387},
 primaryClass = {astro-ph},
       adsurl = {https://ui.adsabs.harvard.edu/abs/1999PASP..111...63F}
}

@ARTICLE{Irani_SN2020bij_classification,
       author = {{Irani}, I. and {Zimmerman}, E. and {Bruch}, R. and {Yaron}, O.},
        title = "{ePESSTO+ Transient Classification Report for 2020-02-02}",
      journal = {Transient Name Server Classification Report},
         year = 2020,
        month = feb,
       volume = {2020-369},
        pages = {1},
       adsurl = {https://ui.adsabs.harvard.edu/abs/2020TNSCR.369....1I}
}

@ARTICLE{Zimmerman_SN2020bij_classification,
       author = {{Zimmerman}, E. and {Irani}, I. and {Bruch}, R. and {Yaron}, O. and {Gal-Yam}, A. and {Williams}, S.~C. and {Taubenberger}, S. and {Ihanec}, N. and {Anderson}, J. and {Bravo}, T.~M. and {Chen}, T.~W. and {Gromadzki}, M. and {Inserra}, C. and {Kankare}, E. and {Nicholl}, M. and {Young}, D.~R. and {Manulis}, I. and {Tonry}, J. and {Denneau}, L. and {Heinze}, A. and {Weiland}, H. and {Flewelling}, H. and {Stalder}, B. and {Rest}, A. and {Smith}, K.~W. and {Smartt}, S.~J. and {McBrien}, O. and {Srivastav}, S.},
        title = "{ePESSTO+ spectroscopic classification of optical transients}",
      journal = {Transient Name Server AstroNote},
         year = 2020,
        month = feb,
       volume = {33},
        pages = {1},
       adsurl = {https://ui.adsabs.harvard.edu/abs/2020TNSAN..33....1Z}
}

@ARTICLE{Schlegel_extinction,
       author = {{Schlegel}, David J. and {Finkbeiner}, Douglas P. and {Davis}, Marc},
        title = "{Maps of Dust Infrared Emission for Use in Estimation of Reddening and Cosmic Microwave Background Radiation Foregrounds}",
      journal = {\apj},
         year = 1998,
        month = jun,
       volume = {500},
       number = {2},
        pages = {525-553},
          doi = {10.1086/305772},
archivePrefix = {arXiv},
       eprint = {astro-ph/9710327},
 primaryClass = {astro-ph},
       adsurl = {https://ui.adsabs.harvard.edu/abs/1998ApJ...500..525S}
}

@article{Valenti_2016,
   title={The diversity of Type II supernova versus the similarity in their progenitors},
   volume={459},
   ISSN={1365-2966},
   url={http://dx.doi.org/10.1093/mnras/stw870},
   DOI={10.1093/mnras/stw870},
   number={4},
   journal={Monthly Notices of the Royal Astronomical Society},
   publisher={Oxford University Press (OUP)},
   author={Valenti, S. and Howell, D. A. and Stritzinger, M. D. and Graham, M. L. and Hosseinzadeh, G. and Arcavi, I. and Bildsten, L. and Jerkstrand, A. and McCully, C. and Pastorello, A. and Piro, A. L. and Sand, D. and Smartt, S. J. and Terreran, G. and Baltay, C. and Benetti, S. and Brown, P. and Filippenko, A. V. and Fraser, M. and Rabinowitz, D. and Sullivan, M. and Yuan, F.},
   year={2016},
   month={Apr},
   pages={3939–3962} }

@misc{Morozova_2015,
  doi = {10.48550/ARXIV.1505.06746},
  
  url = {https://arxiv.org/abs/1505.06746},
  
  author = {Morozova, V. and Piro, A. L. and Renzo, M. and Ott, C. D. and Clausen, D. and Couch, S. M. and Ellis, J. and Roberts, L. F.},
  
  title = {Light Curves of Core-Collapse Supernovae with Substantial Mass Loss using the New Open-Source SuperNova Explosion Code (SNEC)},
  
  publisher = {arXiv},
  
  year = {2015},
  
  copyright = {arXiv.org perpetual, non-exclusive license}
}

@article{Morozova_2016,
   title={NUMERICAL MODELING OF THE EARLY LIGHT CURVES OF TYPE IIP SUPERNOVAE},
   volume={829},
   ISSN={1538-4357},
   url={http://dx.doi.org/10.3847/0004-637X/829/2/109},
   DOI={10.3847/0004-637x/829/2/109},
   number={2},
   journal={The Astrophysical Journal},
   publisher={American Astronomical Society},
   author={Morozova, Viktoriya and Piro, Anthony L. and Renzo, Mathieu and Ott, Christian D.},
   year={2016},
   month=sep, pages={109} }

@article{Morozova_2017,
	doi = {10.3847/1538-4357/aa6251},
  
	url = {https://doi.org/10.3847%2F1538-4357%2Faa6251},
  
	year = 2017,
	month = {mar},
  
	publisher = {American Astronomical Society},
  
	volume = {838},
  
	number = {1},
  
	pages = {28},
  
	author = {Viktoriya Morozova and Anthony L. Piro and Stefano Valenti},
  
	title = {Unifying Type {II} Supernova Light Curves with Dense Circumstellar Material},
  
	journal = {The Astrophysical Journal}
}

@software{david_guevel_pyzogy,
  author       = {David Guevel and
                  Griffin Hosseinzadeh},
  title        = {dguevel/PyZOGY: Initial release},
  month        = nov,
  year         = 2017,
  publisher    = {Zenodo},
  version      = {v0.0.1},
  doi          = {10.5281/zenodo.1043973},
  url          = {https://doi.org/10.5281/zenodo.1043973}
}

@misc{Lukasz_2012,
  doi = {10.48550/ARXIV.1210.5007},
  
  url = {https://arxiv.org/abs/1210.5007},
  
  author = {Wyrzykowski, Lukasz and Hodgkin, Simon and Blogorodnova, Nadejda and Koposov, Sergey and Burgon, Ross},
  
  title = {Photometric Science Alerts from Gaia},
  
  publisher = {arXiv},
  
  year = {2012},
  
  copyright = {arXiv.org perpetual, non-exclusive license}
}

@article{Hodgkin_2021,
	doi = {10.1051/0004-6361/202140735},
  
	url = {https://doi.org/10.1051%2F0004-6361%2F202140735},
  
	year = 2021,
	month = {aug},
  
	publisher = {{EDP} Sciences},
  
	volume = {652},
  
	pages = {A76},
  
	author = {S. T. Hodgkin and D. L. Harrison and E. Breedt and T. Wevers and G. Rixon and A. Delgado and A. Yoldas and Z. Kostrzewa-Rutkowska and {\L}. Wyrzykowski and M. van Leeuwen and N. Blagorodnova and H. Campbell and D. Eappachen and M. Fraser and N. Ihanec and S. E. Koposov and K. Kruszy{\'{n}
}ska and G. Marton and K. A. Rybicki and A. G. A. Brown and P. W. Burgess and G. Busso and S. Cowell and F. De Angeli and C. Diener and D. W. Evans and G. Gilmore and G. Holland and P. G. Jonker and F. van Leeuwen and F. Mignard and P. J. Osborne and J. Portell and T. Prusti and P. J. Richards and M. Riello and G. M. Seabroke and N. A. Walton and P. {\'{A}}brah{\'{a}}m and G. Altavilla and S. G. Baker and U. Bastian and P. O'Brien and J. de Bruijne and T. Butterley and J. M. Carrasco and J. Casta{\~{n}}eda and J. S. Clark and G. Clementini and C. M. Copperwheat and M. Cropper and G. Damljanovic and M. Davidson and C. J. Davis and M. Dennefeld and V. S. Dhillon and C. Dolding and M. Dominik and P. Esquej and L. Eyer and C. Fabricius and M. Fridman and D. Froebrich and N. Garralda and A. Gomboc and J. J. Gonz{\'{a}}lez-Vidal and R. Guerra and N. C. Hambly and L. K. Hardy and B. Holl and A. Hourihane and J. Japelj and D. A. Kann and C. Kiss and C. Knigge and U. Kolb and S. Komossa and {\'{A}}. K{\'{o}}sp{\'{a}}l and G. Kov{\'{a}}cs and M. Kun and G. Leto and F. Lewis and S. P. Littlefair and A. A. Mahabal and C. G. Mundell and Z. Nagy and D. Padeletti and L. Palaversa and A. Pigulski and M. L. Pretorius and W. van Reeven and V. A. R. M. Ribeiro and M. Roelens and N. Rowell and N. Schartel and A. Scholz and A. Schwope and B. M. Sip{\H{o}}cz and S. J. Smartt and M. D. Smith and I. Serraller and D. Steeghs and M. Sullivan and L. Szabados and E. Szegedi-Elek and P. Tisserand and L. Tomasella and S. van Velzen and P. A. Whitelock and R. W. Wilson and D. R. Young},
  
	title = {$\less$i$\greater$Gaia$\less$/i$\greater$ Early Data Release 3},
  
	journal = {Astronomy \& Astrophysics}
}

@ARTICLE{Stetson,
       author = {{Stetson}, Peter B.},
        title = "{Homogeneous Photometry for Star Clusters and Resolved Galaxies. II. Photometric Standard Stars}",
      journal = {\pasp},
         year = 2000,
        month = jul,
       volume = {112},
       number = {773},
        pages = {925-931},
          doi = {10.1086/316595},
archivePrefix = {arXiv},
       eprint = {astro-ph/0004144},
 primaryClass = {astro-ph},
       adsurl = {https://ui.adsabs.harvard.edu/abs/2000PASP..112..925S}
}

@ARTICLE{Landolt,
       author = {{Landolt}, Arlo U.},
        title = "{UBVRI Photometric Standard Stars in the Magnitude Range 11.5 < V < 16.0 Around the Celestial Equator}",
      journal = {\aj},
         year = 1992,
        month = jul,
       volume = {104},
        pages = {340},
          doi = {10.1086/116242},
       adsurl = {https://ui.adsabs.harvard.edu/abs/1992AJ....104..340L}
}

@ARTICLE{SDSS-survey,
       author = {{Albareti}, Franco D. and {Allende Prieto}, Carlos and {Almeida}, Andres. and {et al.}},
        title = "{The 13th Data Release of the Sloan Digital Sky Survey: First Spectroscopic Data from the SDSS-IV Survey Mapping Nearby Galaxies at Apache Point Observatory}",
      journal = {\apjs},
         year = 2017,
        month = dec,
       volume = {233},
       number = {2},
          eid = {25},
        pages = {25},
          doi = {10.3847/1538-4365/aa8992},
archivePrefix = {arXiv},
       eprint = {1608.02013},
 primaryClass = {astro-ph.GA},
       adsurl = {https://ui.adsabs.harvard.edu/abs/2017ApJS..233...25A}
}

@article{Morozova_2018,
	doi = {10.3847/1538-4357/aab9a6},
  
	url = {https://doi.org/10.3847%2F1538-4357%2Faab9a6},
  
	year = 2018,
	month = {apr},
  
	publisher = {American Astronomical Society},
  
	volume = {858},
  
	number = {1},
  
	pages = {15},
  
	author = {Viktoriya Morozova and Anthony L. Piro and Stefano Valenti},
  
	title = {Measuring the Progenitor Masses and Dense Circumstellar Material of Type {II} Supernovae},
  
	journal = {The Astrophysical Journal}
}

@article{Guillochon_2017,
   title={An Open Catalog for Supernova Data},
   volume={835},
   ISSN={1538-4357},
   url={http://dx.doi.org/10.3847/1538-4357/835/1/64},
   DOI={10.3847/1538-4357/835/1/64},
   number={1},
   journal={The Astrophysical Journal},
   publisher={American Astronomical Society},
   author={Guillochon, James and Parrent, Jerod and Kelley, Luke Zoltan and Margutti, Raffaella},
   year={2017},
   month=jan, pages={64} }

@ARTICLE{Galbany_2016,
       author = {{Galbany}, Llu{\'\i}s and {Hamuy}, Mario and {Phillips}, Mark M. and {Suntzeff}, Nicholas B. and {Maza}, Jos{\'e} and {de Jaeger}, Thomas and {Moraga}, Tania and {Gonz{\'a}lez-Gait{\'a}n}, Santiago and {Krisciunas}, Kevin and {Morrell}, Nidia I. and {Thomas-Osip}, Joanna and {Krzeminski}, Wojtek and {Gonz{\'a}lez}, Luis and {Antezana}, Roberto and {Wishnjewski}, Marina and {McCarthy}, Patrick and {Anderson}, Joseph P. and {Guti{\'e}rrez}, Claudia P. and {Stritzinger}, Maximilian and {Folatelli}, Gast{\'o}n and {Anguita}, Claudio and {Galaz}, Gaspar and {Green}, Elisabeth M. and {Impey}, Chris and {Kim}, Yong-Cheol and {Kirhakos}, Sofia and {Malkan}, Mathew A. and {Mulchaey}, John S. and {Phillips}, Andrew C. and {Pizzella}, Alessandro and {Prosser}, Charles F. and {Schmidt}, Brian P. and {Schommer}, Robert A. and {Sherry}, William and {Strolger}, Louis-Gregory and {Wells}, Lisa A. and {Williger}, Gerard M.},
        title = "{UBVRIz Light Curves of 51 Type II Supernovae}",
      journal = {\aj},
         year = 2016,
        month = feb,
       volume = {151},
       number = {2},
          eid = {33},
        pages = {33},
          doi = {10.3847/0004-6256/151/2/33},
archivePrefix = {arXiv},
       eprint = {1511.08402},
 primaryClass = {astro-ph.SR},
       adsurl = {https://ui.adsabs.harvard.edu/abs/2016AJ....151...33G}
}

@article{Zimmerman_2024,
   title={The complex circumstellar environment of supernova 2023ixf},
   volume={627},
   ISSN={1476-4687},
   url={http://dx.doi.org/10.1038/s41586-024-07116-6},
   DOI={10.1038/s41586-024-07116-6},
   number={8005},
   journal={Nature},
   publisher={Springer Science and Business Media LLC},
   author={Zimmerman, E. A. and Irani, I. and Chen, P. and Gal-Yam, A. and Schulze, S. and Perley, D. A. and Sollerman, J. and Filippenko, A. V. and Shenar, T. and Yaron, O. and Shahaf, S. and Bruch, R. J. and Ofek, E. O. and De Cia, A. and Brink, T. G. and Yang, Y. and Vasylyev, S. S. and Ben Ami, S. and Aubert, M. and Badash, A. and Bloom, J. S. and Brown, P. J. and De, K. and Dimitriadis, G. and Fransson, C. and Fremling, C. and Hinds, K. and Horesh, A. and Johansson, J. P. and Kasliwal, M. M. and Kulkarni, S. R. and Kushnir, D. and Martin, C. and Matuzewski, M. and McGurk, R. C. and Miller, A. A. and Morag, J. and Neil, J. D. and Nugent, P. E. and Post, R. S. and Prusinski, N. Z. and Qin, Y. and Raichoor, A. and Riddle, R. and Rowe, M. and Rusholme, B. and Sfaradi, I. and Sjoberg, K. M. and Soumagnac, M. and Stein, R. D. and Strotjohann, N. L. and Terwel, J. H. and Wasserman, T. and Wise, J. and Wold, A. and Yan, L. and Zhang, K.},
   year={2024},
   month=mar, pages={759–762} }

@ARTICLE{Hosseinzadeh_2022,
       author = {{Hosseinzadeh}, Griffin and {Kilpatrick}, Charles D. and {Dong}, Yize and {Sand}, David J. and {Andrews}, Jennifer E. and {Bostroem}, K. Azalee and {Janzen}, Daryl and {Jencson}, Jacob E. and {Lundquist}, Michael and {Meza Retamal}, Nicolas E. and {Pearson}, Jeniveve and {Valenti}, Stefano and {Wyatt}, Samuel and {Burke}, Jamison and {Hiramatsu}, Daichi and {Howell}, D. Andrew and {McCully}, Curtis and {Newsome}, Megan and {Gonzalez}, Estefania Padilla and {Pellegrino}, Craig and {Terreran}, Giacomo and {Auchettl}, Katie and {Davis}, Kyle W. and {Foley}, Ryan J. and {Miao}, Hao-Yu and {Pan}, Yen-Chen and {Rest}, Armin and {Siebert}, Matthew R. and {Taggart}, Kirsty and {Tucker}, Brad E. and {Cyrus Leung}, Feng Lin and {Swift}, Jonathan J. and {Yang}, Grace and {Anderson}, Joseph P. and {Ashall}, Chris and {Benetti}, Stefano and {Brown}, Peter J. and {Cartier}, R{\'e}gis and {Chen}, Ting-Wan and {Valle}, Massimo Della and {Galbany}, Llu{\'\i}s and {Gomez}, Sebastian and {Gromadzki}, Mariusz and {Haislip}, Joshua and {Hsiao}, Eric Y. and {Inserra}, Cosimo and {Jha}, Saurabh W. and {Killestein}, Thomas L. and {Kouprianov}, Vladimir and {Kozyreva}, Alexandra and {M{\"u}ller-Bravo}, Tom{\'a}s E. and {Nicholl}, Matt and {Paraskeva}, Emmy and {Reichart}, Daniel E. and {Ryder}, Stuart and {Shahbandeh}, Melissa and {Shappee}, Ben and {Smith}, Nathan and {Young}, David R.},
        title = "{Weak Mass Loss from the Red Supergiant Progenitor of the Type II SN 2021yja}",
      journal = {\apj},
         year = 2022,
        month = aug,
       volume = {935},
       number = {1},
          eid = {31},
        pages = {31},
          doi = {10.3847/1538-4357/ac75f0},
archivePrefix = {arXiv},
       eprint = {2203.08155},
 primaryClass = {astro-ph.HE},
       adsurl = {https://ui.adsabs.harvard.edu/abs/2022ApJ...935...31H}
}

@ARTICLE{Gutierrez,
       author = {{Guti{\'e}rrez}, Claudia P. and {Anderson}, Joseph P. and {Hamuy}, Mario and {Morrell}, Nidia and {Gonz{\'a}lez-Gaitan}, Santiago and {Stritzinger}, Maximilian D. and {Phillips}, Mark M. and {Galbany}, Lluis and {Folatelli}, Gast{\'o}n and {Dessart}, Luc and {Contreras}, Carlos and {Della Valle}, Massimo and {Freedman}, Wendy L. and {Hsiao}, Eric Y. and {Krisciunas}, Kevin and {Madore}, Barry F. and {Maza}, Jos{\'e} and {Suntzeff}, Nicholas B. and {Prieto}, Jose Luis and {Gonz{\'a}lez}, Luis and {Cappellaro}, Enrico and {Navarrete}, Mauricio and {Pizzella}, Alessandro and {Ruiz}, Maria T. and {Smith}, R. Chris and {Turatto}, Massimo},
        title = "{Type II Supernova Spectral Diversity. I. Observations, Sample Characterization, and Spectral Line Evolution}",
      journal = {\apj},
         year = 2017,
        month = nov,
       volume = {850},
       number = {1},
          eid = {89},
        pages = {89},
          doi = {10.3847/1538-4357/aa8f52},
archivePrefix = {arXiv},
       eprint = {1709.02487},
 primaryClass = {astro-ph.HE},
       adsurl = {https://ui.adsabs.harvard.edu/abs/2017ApJ...850...89G}
}

@ARTICLE{Leonard_2002,
       author = {{Leonard}, Douglas C. and {Filippenko}, Alexei V. and {Gates}, Elinor L. and {Li}, Weidong and {Eastman}, Ronald G. and {Barth}, Aaron J. and {Bus}, Schelte J. and {Chornock}, Ryan and {Coil}, Alison L. and {Frink}, Sabine and {Grady}, Carol A. and {Harris}, Alan W. and {Malkan}, Matthew A. and {Matheson}, Thomas and {Quirrenbach}, Andreas and {Treffers}, Richard R.},
        title = "{The Distance to SN 1999em in NGC 1637 from the Expanding Photosphere Method}",
      journal = {\pasp},
         year = 2002,
        month = jan,
       volume = {114},
       number = {791},
        pages = {35-64},
          doi = {10.1086/324785},
archivePrefix = {arXiv},
       eprint = {astro-ph/0109535},
 primaryClass = {astro-ph},
       adsurl = {https://ui.adsabs.harvard.edu/abs/2002PASP..114...35L}
}

@software{Hosseinzadeh_2024_11405219,
  author       = {Hosseinzadeh, Griffin and
                  Bostroem, K. Azalee and
                  Ben-Ami, Tom and
                  Gomez, Sebastian},
  title        = {Light Curve Fitting v0.10.0},
  month        = may,
  year         = 2024,
  publisher    = {Zenodo},
  version      = {v0.10.0},
  doi          = {10.5281/zenodo.11405219},
  url          = {https://doi.org/10.5281/zenodo.11405219},
}

@ARTICLE{Faran_2018,
       author = {{Faran}, T. and {Nakar}, E. and {Poznanski}, D.},
        title = "{The evolution of temperature and bolometric luminosity in Type II supernovae}",
      journal = {\mnras},
         year = 2018,
        month = jan,
       volume = {473},
       number = {1},
        pages = {513-537},
          doi = {10.1093/mnras/stx2288},
archivePrefix = {arXiv},
       eprint = {1707.07695},
 primaryClass = {astro-ph.HE},
       adsurl = {https://ui.adsabs.harvard.edu/abs/2018MNRAS.473..513F}
}

@ARTICLE{Nakar2010,
       author = {{Nakar}, Ehud and {Sari}, Re'em},
        title = "{Early Supernovae Light Curves Following the Shock Breakout}",
      journal = {\apj},
         year = 2010,
        month = dec,
       volume = {725},
       number = {1},
        pages = {904-921},
          doi = {10.1088/0004-637X/725/1/904},
archivePrefix = {arXiv},
       eprint = {1004.2496},
 primaryClass = {astro-ph.HE},
       adsurl = {https://ui.adsabs.harvard.edu/abs/2010ApJ...725..904N}
}

@ARTICLE{Maayane,
       author = {{Soumagnac}, Maayane T. and {Ganot}, Noam and {Irani}, Ido and {Gal-yam}, Avishay and {Ofek}, Eran O. and {Waxman}, Eli and {Morag}, Jonathan and {Yaron}, Ofer and {Schulze}, Steve and {Yang}, Yi and {Rubin}, Adam and {Cenko}, S. Bradley and {Sollerman}, Jesper and {Perley}, Daniel A. and {Fremling}, Christoffer and {Nugent}, Peter and {Neill}, James D. and {Karamehmetoglu}, Emir and {Bellm}, Eric C. and {Bruch}, Rachel J. and {Burruss}, Rick and {Cunningham}, Virginia and {Dekany}, Richard and {Golkhou}, V. Zach and {Graham}, Matthew J. and {Kasliwal}, Mansi M. and {Konidaris}, Nicholas P. and {Kulkarni}, Shrinivas R. and {Kupfer}, Thomas and {Laher}, Russ R. and {Masci}, Frank J. and {Riddle}, Reed and {Rigault}, Mickael and {Rusholme}, Ben and {van Roestel}, Jan and {Zackay}, Barak},
        title = "{SN 2018fif: The Explosion of a Large Red Supergiant Discovered in Its Infancy by the Zwicky Transient Facility}",
      journal = {\apj},
         year = 2020,
        month = oct,
       volume = {902},
       number = {1},
          eid = {6},
        pages = {6},
          doi = {10.3847/1538-4357/abb247},
archivePrefix = {arXiv},
       eprint = {1907.11252},
 primaryClass = {astro-ph.HE},
       adsurl = {https://ui.adsabs.harvard.edu/abs/2020ApJ...902....6S}
}

@ARTICLE{EFOSC2,
       author = {{Buzzoni}, B. and {Delabre}, B. and {Dekker}, H. and {Dodorico}, S. and {Enard}, D. and {Focardi}, P. and {Gustafsson}, B. and {Nees}, W. and {Paureau}, J. and {Reiss}, R.},
        title = "{The ESO Faint Object Spectrograph and Camera / EFOSC}",
      journal = {The Messenger},
         year = 1984,
        month = dec,
       volume = {38},
        pages = {9},
       adsurl = {https://ui.adsabs.harvard.edu/abs/1984Msngr..38....9B}
}

@ARTICLE{FLOYDS_REDUCTION,
       author = {{Valenti}, S. and {Sand}, D. and {Pastorello}, A. and {Graham}, M.~L. and {Howell}, D.~A. and {Parrent}, J.~T. and {Tomasella}, L. and {Ochner}, P. and {Fraser}, M. and {Benetti}, S. and {Yuan}, F. and {Smartt}, S.~J. and {Maund}, J.~R. and {Arcavi}, I. and {Gal-Yam}, A. and {Inserra}, C. and {Young}, D.},
        title = "{The first month of evolution of the slow-rising Type IIP SN 2013ej in M74$^{★}$}",
      journal = {\mnras},
         year = 2014,
        month = feb,
       volume = {438},
       number = {1},
        pages = {L101-L105},
          doi = {10.1093/mnrasl/slt171},
archivePrefix = {arXiv},
       eprint = {1309.4269},
 primaryClass = {astro-ph.CO},
       adsurl = {https://ui.adsabs.harvard.edu/abs/2014MNRAS.438L.101V}
}

@ARTICLE{PESSTO_Smartt_2015,
       author = {{Smartt}, S.~J. and {Valenti}, S. and {Fraser}, M. and {Inserra}, C. and {Young}, D.~R. and {Sullivan}, M. and {Pastorello}, A. and {Benetti}, S. and {Gal-Yam}, A. and {Knapic}, C. and {Molinaro}, M. and {Smareglia}, R. and {Smith}, K.~W. and {Taubenberger}, S. and {Yaron}, O. and {Anderson}, J.~P. and {Ashall}, C. and {Balland}, C. and {Baltay}, C. and {Barbarino}, C. and {Bauer}, F.~E. and {Baumont}, S. and {Bersier}, D. and {Blagorodnova}, N. and {Bongard}, S. and {Botticella}, M.~T. and {Bufano}, F. and {Bulla}, M. and {Cappellaro}, E. and {Campbell}, H. and {Cellier-Holzem}, F. and {Chen}, T.-W. and {Childress}, M.~J. and {Clocchiatti}, A. and {Contreras}, C. and {Dall'Ora}, M. and {Danziger}, J. and {de Jaeger}, T. and {De Cia}, A. and {Della Valle}, M. and {Dennefeld}, M. and {Elias-Rosa}, N. and {Elman}, N. and {Feindt}, U. and {Fleury}, M. and {Gall}, E. and {Gonzalez-Gaitan}, S. and {Galbany}, L. and {Morales Garoffolo}, A. and {Greggio}, L. and {Guillou}, L.~L. and {Hachinger}, S. and {Hadjiyska}, E. and {Hage}, P.~E. and {Hillebrandt}, W. and {Hodgkin}, S. and {Hsiao}, E.~Y. and {James}, P.~A. and {Jerkstrand}, A. and {Kangas}, T. and {Kankare}, E. and {Kotak}, R. and {Kromer}, M. and {Kuncarayakti}, H. and {Leloudas}, G. and {Lundqvist}, P. and {Lyman}, J.~D. and {Hook}, I.~M. and {Maguire}, K. and {Manulis}, I. and {Margheim}, S.~J. and {Mattila}, S. and {Maund}, J.~R. and {Mazzali}, P.~A. and {McCrum}, M. and {McKinnon}, R. and {Moreno-Raya}, M.~E. and {Nicholl}, M. and {Nugent}, P. and {Pain}, R. and {Pignata}, G. and {Phillips}, M.~M. and {Polshaw}, J. and {Pumo}, M.~L. and {Rabinowitz}, D. and {Reilly}, E. and {Romero-Ca{\~n}izales}, C. and {Scalzo}, R. and {Schmidt}, B. and {Schulze}, S. and {Sim}, S. and {Sollerman}, J. and {Taddia}, F. and {Tartaglia}, L. and {Terreran}, G. and {Tomasella}, L. and {Turatto}, M. and {Walker}, E. and {Walton}, N.~A. and {Wyrzykowski}, L. and {Yuan}, F. and {Zampieri}, L.},
        title = "{PESSTO: survey description and products from the first data release by the Public ESO Spectroscopic Survey of Transient Objects}",
      journal = {\aap},
         year = 2015,
        month = jul,
       volume = {579},
          eid = {A40},
        pages = {A40},
          doi = {10.1051/0004-6361/201425237},
archivePrefix = {arXiv},
       eprint = {1411.0299},
 primaryClass = {astro-ph.SR},
       adsurl = {https://ui.adsabs.harvard.edu/abs/2015A&A...579A..40S}
}

@ARTICLE{Faran_2014,
       author = {{Faran}, T. and {Poznanski}, D. and {Filippenko}, A.~V. and {Chornock}, R. and {Foley}, R.~J. and {Ganeshalingam}, M. and {Leonard}, D.~C. and {Li}, W. and {Modjaz}, M. and {Nakar}, E. and {Serduke}, F.~J.~D. and {Silverman}, J.~M.},
        title = "{Photometric and spectroscopic properties of Type II-P supernovae}",
      journal = {\mnras},
         year = 2014,
        month = jul,
       volume = {442},
       number = {1},
        pages = {844-861},
          doi = {10.1093/mnras/stu955},
archivePrefix = {arXiv},
       eprint = {1404.0378},
 primaryClass = {astro-ph.HE},
       adsurl = {https://ui.adsabs.harvard.edu/abs/2014MNRAS.442..844F}
}

@article{Hamuy_2001,
	doi = {10.1086/322450},
  
	url = {https://doi.org/10.1086%2F322450},
  
	year = 2001,
	month = {sep},
  
	publisher = {American Astronomical Society},
  
	volume = {558},
  
	number = {2},
  
	pages = {615--642},
  
	author = {Mario Hamuy and Philip A. Pinto and Jose Maza and Nicholas B. Suntzeff and M. M. Phillips and Ronald G. Eastman and R. C. Smith and C. J. Corbally and D. Burstein and Yong Li and Valentin Ivanov and Amaya Moro-Martin and L. G. Strolger and R. E. de Souza and S. dos Anjos and Elizabeth M. Green and T. E. Pickering and Luis Gonzalez and Roberto Antezana and Marina Wischnjewsky and G. Galaz and M. Roth and S. E. Persson and R. A. Schommer},
  
	title = {The Distance to {SN} 1999em from the Expanding Photosphere Method},
  
	journal = {The Astrophysical Journal}
}

@article{Woosley_2002,
  title = {The evolution and explosion of massive stars},
  author = {Woosley, S. E. and Heger, A. and Weaver, T. A.},
  journal = {Rev. Mod. Phys.},
  volume = {74},
  issue = {4},
  pages = {1015--1071},
  numpages = {0},
  year = {2002},
  month = {Nov},
  publisher = {American Physical Society},
  doi = {10.1103/RevModPhys.74.1015},
  url = {https://link.aps.org/doi/10.1103/RevModPhys.74.1015}
}

@incollection{Arcavi_2017,
	doi = {10.1007/978-3-319-21846-5_39},
  
	url = {https://doi.org/10.1007%2F978-3-319-21846-5_39},
  
	year = 2017,
	publisher = {Springer International Publishing},
  
	pages = {239--276},
  
	author = {Iair Arcavi},
  
	title = {Hydrogen-Rich Core-Collapse Supernovae},
  
	booktitle = {Handbook of Supernovae}
}

@ARTICLE{Arcavi2017gkg,
       author = {{Arcavi}, Iair and {Hosseinzadeh}, Griffin and {Brown}, Peter J. and {Smartt}, Stephen J. and {Valenti}, Stefano and {Tartaglia}, Leonardo and {Piro}, Anthony L. and {Sanchez}, Jos{\'e} L. and {Nicholls}, Brent and {Monard}, Berto L.~A.~G. and {Howell}, D. Andrew and {McCully}, Curtis and {Sand}, David J. and {Tonry}, John and {Denneau}, Larry and {Stalder}, Brian and {Heinze}, Ari and {Rest}, Armin and {Smith}, Ken W. and {Bishop}, David},
        title = "{Constraints on the Progenitor of SN 2016gkg from Its Shock-cooling Light Curve}",
      journal = {\apjl},
         year = 2017,
        month = mar,
       volume = {837},
       number = {1},
          eid = {L2},
        pages = {L2},
          doi = {10.3847/2041-8213/aa5be1},
archivePrefix = {arXiv},
       eprint = {1611.06451},
 primaryClass = {astro-ph.HE},
       adsurl = {https://ui.adsabs.harvard.edu/abs/2017ApJ...837L...2A}
}

@ARTICLE{Smith_2011,
       author = {{Smith}, Nathan and {Li}, Weidong and {Filippenko}, Alexei V. and {Chornock}, Ryan},
        title = "{Observed fractions of core-collapse supernova types and initial masses of their single and binary progenitor stars}",
      journal = {\mnras},
         year = 2011,
        month = apr,
       volume = {412},
       number = {3},
        pages = {1522-1538},
          doi = {10.1111/j.1365-2966.2011.17229.x},
archivePrefix = {arXiv},
       eprint = {1006.3899},
 primaryClass = {astro-ph.HE},
       adsurl = {https://ui.adsabs.harvard.edu/abs/2011MNRAS.412.1522S}
}

@ARTICLE{Li_2011,
       author = {{Li}, Weidong and {Leaman}, Jesse and {Chornock}, Ryan and {Filippenko}, Alexei V. and {Poznanski}, Dovi and {Ganeshalingam}, Mohan and {Wang}, Xiaofeng and {Modjaz}, Maryam and {Jha}, Saurabh and {Foley}, Ryan J. and {Smith}, Nathan},
        title = "{Nearby supernova rates from the Lick Observatory Supernova Search - II. The observed luminosity functions and fractions of supernovae in a complete sample}",
      journal = {\mnras},
         year = 2011,
        month = apr,
       volume = {412},
       number = {3},
        pages = {1441-1472},
          doi = {10.1111/j.1365-2966.2011.18160.x},
archivePrefix = {arXiv},
       eprint = {1006.4612},
 primaryClass = {astro-ph.SR},
       adsurl = {https://ui.adsabs.harvard.edu/abs/2011MNRAS.412.1441L}
}

@ARTICLE{Smartt_2009,
       author = {{Smartt}, Stephen J.},
        title = "{Progenitors of Core-Collapse Supernovae}",
      journal = {\araa},
         year = 2009,
        month = sep,
       volume = {47},
       number = {1},
        pages = {63-106},
          doi = {10.1146/annurev-astro-082708-101737},
archivePrefix = {arXiv},
       eprint = {0908.0700},
 primaryClass = {astro-ph.SR},
       adsurl = {https://ui.adsabs.harvard.edu/abs/2009ARA&A..47...63S}
}

@ARTICLE{Smartt_2015,
       author = {{Smartt}, S.~J.},
        title = "{Observational Constraints on the Progenitors of Core-Collapse Supernovae: The Case for Missing High-Mass Stars}",
      journal = {\pasa},
         year = 2015,
        month = apr,
       volume = {32},
          eid = {e016},
        pages = {e016},
          doi = {10.1017/pasa.2015.17},
archivePrefix = {arXiv},
       eprint = {1504.02635},
 primaryClass = {astro-ph.SR},
       adsurl = {https://ui.adsabs.harvard.edu/abs/2015PASA...32...16S}
}

@article{Smith_2014,
	doi = {10.1146/annurev-astro-081913-040025},
  
	url = {https://doi.org/10.1146%2Fannurev-astro-081913-040025},
  
	year = 2014,
	month = {aug},
  
	publisher = {Annual Reviews},
  
	volume = {52},
  
	number = {1},
  
	pages = {487--528},
  
	author = {Nathan Smith},
  
	title = {Mass Loss: Its Effect on the Evolution and Fate of High-Mass Stars},
  
	journal = {Annual Review of Astronomy and Astrophysics}
}

@ARTICLE{Schlegel_1990,
       author = {{Schlegel}, Eric M.},
        title = "{A new subclass of type II supernovae ?}",
      journal = {\mnras},
         year = 1990,
        month = may,
       volume = {244},
        pages = {269-271},
       adsurl = {https://ui.adsabs.harvard.edu/abs/1990MNRAS.244..269S}
}

@ARTICLE{Chugai_1991,
       author = {{Chugai}, N.~N.},
        title = "{Evidence for energizing of H alpha emission in type II supernovae by ejecta-wind interaction.}",
      journal = {\mnras},
         year = 1991,
        month = jun,
       volume = {250},
        pages = {513},
          doi = {10.1093/mnras/250.3.513},
       adsurl = {https://ui.adsabs.harvard.edu/abs/1991MNRAS.250..513C}
}

@article{Shivvers_2015,
	doi = {10.1088/0004-637x/806/2/213},
  
	url = {https://doi.org/10.1088%2F0004-637x%2F806%2F2%2F213},
  
	year = 2015,
	month = {jun},
  
	publisher = {American Astronomical Society},
  
	volume = {806},
  
	number = {2},
  
	pages = {213},
  
	author = {Isaac Shivvers and Jose H. Groh and Jon C. Mauerhan and Ori D. Fox and Douglas C. Leonard and Alexei V. Filippenko},
  
	title = {{EARLY} {EMISSION} {FROM} {THE} {TYPE} {IIn} {SUPERNOVA} 1998S {AT} {HIGH} {RESOLUTION}
},
  
	journal = {The Astrophysical Journal}
}

@ARTICLE{Sukhbold_2016,
       author = {{Sukhbold}, Tuguldur and {Ertl}, T. and {Woosley}, S.~E. and {Brown}, Justin M. and {Janka}, H. -T.},
        title = "{Core-collapse Supernovae from 9 to 120 Solar Masses Based on Neutrino-powered Explosions}",
      journal = {\apj},
         year = 2016,
        month = apr,
       volume = {821},
       number = {1},
          eid = {38},
        pages = {38},
          doi = {10.3847/0004-637X/821/1/38},
archivePrefix = {arXiv},
       eprint = {1510.04643},
 primaryClass = {astro-ph.HE},
       adsurl = {https://ui.adsabs.harvard.edu/abs/2016ApJ...821...38S}
}

@article{Bose_2015,
   title={SN 2013ab: a normal Type IIP supernova in NGC 5669},
   volume={450},
   ISSN={1365-2966},
   url={http://dx.doi.org/10.1093/mnras/stv759},
   DOI={10.1093/mnras/stv759},
   number={3},
   journal={Monthly Notices of the Royal Astronomical Society},
   publisher={Oxford University Press (OUP)},
   author={Bose, Subhash and Valenti, Stefano and Misra, Kuntal and Pumo, Maria Letizia and Zampieri, Luca and Sand, David and Kumar, Brijesh and Pastorello, Andrea and Sutaria, Firoza and Maccarone, Thomas J. and Kumar, Brajesh and Graham, M. L. and Howell, D. Andrew and Ochner, Paolo and Chandola, H. C. and Pandey, Shashi B.},
   year={2015},
   month=may, pages={2373–2392} }

@ARTICLE{Sukhbold_2014,
       author = {{Sukhbold}, Tuguldur and {Woosley}, S.~E.},
        title = "{The Compactness of Presupernova Stellar Cores}",
      journal = {\apj},
         year = 2014,
        month = mar,
       volume = {783},
       number = {1},
          eid = {10},
        pages = {10},
          doi = {10.1088/0004-637X/783/1/10},
archivePrefix = {arXiv},
       eprint = {1311.6546},
 primaryClass = {astro-ph.SR},
       adsurl = {https://ui.adsabs.harvard.edu/abs/2014ApJ...783...10S}
}

@ARTICLE{Woosley_Heger_2015,
       author = {{Woosley}, S.~E. and {Heger}, Alexander},
        title = "{The Remarkable Deaths of 9-11 Solar Mass Stars}",
      journal = {\apj},
         year = 2015,
        month = sep,
       volume = {810},
       number = {1},
          eid = {34},
        pages = {34},
          doi = {10.1088/0004-637X/810/1/34},
archivePrefix = {arXiv},
       eprint = {1505.06712},
 primaryClass = {astro-ph.SR},
       adsurl = {https://ui.adsabs.harvard.edu/abs/2015ApJ...810...34W}
}

@ARTICLE{Woosley_Heger_2007,
       author = {{Woosley}, S.~E. and {Heger}, A.},
        title = "{Nucleosynthesis and remnants in massive stars of solar metallicity}",
      journal = {\physrep},
         year = 2007,
        month = apr,
       volume = {442},
       number = {1-6},
        pages = {269-283},
          doi = {10.1016/j.physrep.2007.02.009},
archivePrefix = {arXiv},
       eprint = {astro-ph/0702176},
 primaryClass = {astro-ph},
       adsurl = {https://ui.adsabs.harvard.edu/abs/2007PhR...442..269W}
}

@ARTICLE{Weaver_1978,
       author = {{Weaver}, T.~A. and {Zimmerman}, G.~B. and {Woosley}, S.~E.},
        title = "{Presupernova evolution of massive stars.}",
      journal = {\apj},
         year = 1978,
        month = nov,
       volume = {225},
        pages = {1021-1029},
          doi = {10.1086/156569},
       adsurl = {https://ui.adsabs.harvard.edu/abs/1978ApJ...225.1021W}
}

@ARTICLE{Bellm_2019,
       author = {{Bellm}, Eric C. and {Kulkarni}, Shrinivas R. and {Graham}, Matthew J. and {Dekany}, Richard and {Smith}, Roger M. and {Riddle}, Reed and {Masci}, Frank J. and {Helou}, George and {Prince}, Thomas A. and {Adams}, Scott M. and {Barbarino}, C. and {Barlow}, Tom and {Bauer}, James and {Beck}, Ron and {Belicki}, Justin and {Biswas}, Rahul and {Blagorodnova}, Nadejda and {Bodewits}, Dennis and {Bolin}, Bryce and {Brinnel}, Valery and {Brooke}, Tim and {Bue}, Brian and {Bulla}, Mattia and {Burruss}, Rick and {Cenko}, S. Bradley and {Chang}, Chan-Kao and {Connolly}, Andrew and {Coughlin}, Michael and {Cromer}, John and {Cunningham}, Virginia and {De}, Kishalay and {Delacroix}, Alex and {Desai}, Vandana and {Duev}, Dmitry A. and {Eadie}, Gwendolyn and {Farnham}, Tony L. and {Feeney}, Michael and {Feindt}, Ulrich and {Flynn}, David and {Franckowiak}, Anna and {Frederick}, S. and {Fremling}, C. and {Gal-Yam}, Avishay and {Gezari}, Suvi and {Giomi}, Matteo and {Goldstein}, Daniel A. and {Golkhou}, V. Zach and {Goobar}, Ariel and {Groom}, Steven and {Hacopians}, Eugean and {Hale}, David and {Henning}, John and {Ho}, Anna Y.~Q. and {Hover}, David and {Howell}, Justin and {Hung}, Tiara and {Huppenkothen}, Daniela and {Imel}, David and {Ip}, Wing-Huen and {Ivezi{\'c}}, {\v{Z}}eljko and {Jackson}, Edward and {Jones}, Lynne and {Juric}, Mario and {Kasliwal}, Mansi M. and {Kaspi}, S. and {Kaye}, Stephen and {Kelley}, Michael S.~P. and {Kowalski}, Marek and {Kramer}, Emily and {Kupfer}, Thomas and {Landry}, Walter and {Laher}, Russ R. and {Lee}, Chien-De and {Lin}, Hsing Wen and {Lin}, Zhong-Yi and {Lunnan}, Ragnhild and {Giomi}, Matteo and {Mahabal}, Ashish and {Mao}, Peter and {Miller}, Adam A. and {Monkewitz}, Serge and {Murphy}, Patrick and {Ngeow}, Chow-Choong and {Nordin}, Jakob and {Nugent}, Peter and {Ofek}, Eran and {Patterson}, Maria T. and {Penprase}, Bryan and {Porter}, Michael and {Rauch}, Ludwig and {Rebbapragada}, Umaa and {Reiley}, Dan and {Rigault}, Mickael and {Rodriguez}, Hector and {van Roestel}, Jan and {Rusholme}, Ben and {van Santen}, Jakob and {Schulze}, S. and {Shupe}, David L. and {Singer}, Leo P. and {Soumagnac}, Maayane T. and {Stein}, Robert and {Surace}, Jason and {Sollerman}, Jesper and {Szkody}, Paula and {Taddia}, F. and {Terek}, Scott and {Van Sistine}, Angela and {van Velzen}, Sjoert and {Vestrand}, W. Thomas and {Walters}, Richard and {Ward}, Charlotte and {Ye}, Quan-Zhi and {Yu}, Po-Chieh and {Yan}, Lin and {Zolkower}, Jeffry},
        title = "{The Zwicky Transient Facility: System Overview, Performance, and First Results}",
      journal = {\pasp},
         year = 2019,
        month = jan,
       volume = {131},
       number = {995},
        pages = {018002},
          doi = {10.1088/1538-3873/aaecbe},
archivePrefix = {arXiv},
       eprint = {1902.01932},
 primaryClass = {astro-ph.IM},
       adsurl = {https://ui.adsabs.harvard.edu/abs/2019PASP..131a8002B}
}

@ARTICLE{Smith_2015,
       author = {{Smith}, Nathan and {Mauerhan}, Jon C. and {Cenko}, S. Bradley and {Kasliwal}, Mansi M. and {Silverman}, Jeffrey M. and {Filippenko}, Alexei V. and {Gal-Yam}, Avishay and {Clubb}, Kelsey I. and {Graham}, Melissa L. and {Leonard}, Douglas C. and {Horst}, J. Chuck and {Williams}, G. Grant and {Andrews}, Jennifer E. and {Kulkarni}, Shrinivas R. and {Nugent}, Peter and {Sullivan}, Mark and {Maguire}, Kate and {Xu}, Dong and {Ben-Ami}, Sagi},
        title = "{PTF11iqb: cool supergiant mass-loss that bridges the gap between Type IIn and normal supernovae}",
      journal = {\mnras},
         year = 2015,
        month = may,
       volume = {449},
       number = {2},
        pages = {1876-1896},
          doi = {10.1093/mnras/stv35410.48550/arXiv.1501.02820},
archivePrefix = {arXiv},
       eprint = {1501.02820},
 primaryClass = {astro-ph.HE},
       adsurl = {https://ui.adsabs.harvard.edu/abs/2015MNRAS.449.1876S}
}

@ARTICLE{Gal-Yam_2014,
       author = {{Gal-Yam}, Avishay and {Arcavi}, I. and {Ofek}, E.~O. and {Ben-Ami}, S. and {Cenko}, S.~B. and {Kasliwal}, M.~M. and {Cao}, Y. and {Yaron}, O. and {Tal}, D. and {Silverman}, J.~M. and {Horesh}, A. and {De Cia}, A. and {Taddia}, F. and {Sollerman}, J. and {Perley}, D. and {Vreeswijk}, P.~M. and {Kulkarni}, S.~R. and {Nugent}, P.~E. and {Filippenko}, A.~V. and {Wheeler}, J.~C.},
        title = "{A Wolf-Rayet-like progenitor of SN 2013cu from spectral observations of a stellar wind}",
      journal = {\nat},
         year = 2014,
        month = may,
       volume = {509},
       number = {7501},
        pages = {471-474},
          doi = {10.1038/nature13304},
archivePrefix = {arXiv},
       eprint = {1406.7640},
 primaryClass = {astro-ph.HE},
       adsurl = {https://ui.adsabs.harvard.edu/abs/2014Natur.509..471G}
}

@article{Yaron_2017,
	doi = {10.1038/nphys4025},
  
	url = {https://doi.org/10.1038%2Fnphys4025},
  
	year = 2017,
	month = {feb},
  
	publisher = {Springer Science and Business Media {LLC}
},
  
	volume = {13},
  
	number = {5},
  
	pages = {510--517},
  
	author = {O. Yaron and D. A. Perley and A. Gal-Yam and J. H. Groh and A. Horesh and E. O. Ofek and S. R. Kulkarni and J. Sollerman and C. Fransson and A. Rubin and P. Szabo and N. Sapir and F. Taddia and S. B. Cenko and S. Valenti and I. Arcavi and D. A. Howell and M. M. Kasliwal and P. M. Vreeswijk and D. Khazov and O. D. Fox and Y. Cao and O. Gnat and P. L. Kelly and P. E. Nugent and A. V. Filippenko and R. R. Laher and P. R. Wozniak and W. H. Lee and U. D. Rebbapragada and K. Maguire and M. Sullivan and M. T. Soumagnac},
  
	title = {Confined dense circumstellar material surrounding a regular type {II} supernova},
  
	journal = {Nature Physics}
}

@article{Pastorello_2009,
   title={SN 2005cs in M51 - II. Complete evolution in the optical and the near-infrared},
   volume={394},
   ISSN={1365-2966},
   url={http://dx.doi.org/10.1111/j.1365-2966.2009.14505.x},
   DOI={10.1111/j.1365-2966.2009.14505.x},
   number={4},
   journal={Monthly Notices of the Royal Astronomical Society},
   publisher={Oxford University Press (OUP)},
   author={Pastorello, A. and Valenti, S. and Zampieri, L. and Navasardyan, H. and Taubenberger, S. and Smartt, S. J. and Arkharov, A. A. and Bärnbantner, O. and Barwig, H. and Benetti, S. and Birtwhistle, P. and Botticella, M. T. and Cappellaro, E. and Del Principe, M. and Di Mille, F. and Di Rico, G. and Dolci, M. and Elias-Rosa, N. and Efimova, N. V. and Fiedler, M. and Harutyunyan, A. and Höflich, P. A. and Kloehr, W. and Larionov, V. M. and Lorenzi, V. and Maund, J. R. and Napoleone, N. and Ragni, M. and Richmond, M. and Ries, C. and Spiro, S. and Temporin, S. and Turatto, M. and Wheeler, J. C.},
   year={2009},
   month=apr, pages={2266–2282} }

@ARTICLE{yaron_2012,
       author = {{Yaron}, Ofer and {Gal-Yam}, Avishay},
        title = "{WISeREP{\textemdash}An Interactive Supernova Data Repository}",
      journal = {\pasp},
         year = 2012,
        month = jul,
       volume = {124},
       number = {917},
        pages = {668},
          doi = {10.1086/666656},
archivePrefix = {arXiv},
       eprint = {1204.1891},
 primaryClass = {astro-ph.IM},
       adsurl = {https://ui.adsabs.harvard.edu/abs/2012PASP..124..668Y}
}

@ARTICLE{Law_2009,
       author = {{Law}, Nicholas M. and {Kulkarni}, Shrinivas R. and {Dekany}, Richard G. and {Ofek}, Eran O. and {Quimby}, Robert M. and {Nugent}, Peter E. and {Surace}, Jason and {Grillmair}, Carl C. and {Bloom}, Joshua S. and {Kasliwal}, Mansi M. and {Bildsten}, Lars and {Brown}, Tim and {Cenko}, S. Bradley and {Ciardi}, David and {Croner}, Ernest and {Djorgovski}, S. George and {van Eyken}, Julian and {Filippenko}, Alexei V. and {Fox}, Derek B. and {Gal-Yam}, Avishay and {Hale}, David and {Hamam}, Nouhad and {Helou}, George and {Henning}, John and {Howell}, D. Andrew and {Jacobsen}, Janet and {Laher}, Russ and {Mattingly}, Sean and {McKenna}, Dan and {Pickles}, Andrew and {Poznanski}, Dovi and {Rahmer}, Gustavo and {Rau}, Arne and {Rosing}, Wayne and {Shara}, Michael and {Smith}, Roger and {Starr}, Dan and {Sullivan}, Mark and {Velur}, Viswa and {Walters}, Richard and {Zolkower}, Jeff},
        title = "{The Palomar Transient Factory: System Overview, Performance, and First Results}",
      journal = {\pasp},
         year = 2009,
        month = dec,
       volume = {121},
       number = {886},
        pages = {1395},
          doi = {10.1086/648598},
archivePrefix = {arXiv},
       eprint = {0906.5350},
 primaryClass = {astro-ph.IM},
       adsurl = {https://ui.adsabs.harvard.edu/abs/2009PASP..121.1395L}
}

@ARTICLE{Rau_2009,
       author = {{Rau}, Arne and {Kulkarni}, Shrinivas R. and {Law}, Nicholas M. and {Bloom}, Joshua S. and {Ciardi}, David and {Djorgovski}, George S. and {Fox}, Derek B. and {Gal-Yam}, Avishay and {Grillmair}, Carl C. and {Kasliwal}, Mansi M. and {Nugent}, Peter E. and {Ofek}, Eran O. and {Quimby}, Robert M. and {Reach}, William T. and {Shara}, Michael and {Bildsten}, Lars and {Cenko}, S. Bradley and {Drake}, Andrew J. and {Filippenko}, Alexei V. and {Helfand}, David J. and {Helou}, George and {Howell}, D. Andrew and {Poznanski}, Dovi and {Sullivan}, Mark},
        title = "{Exploring the Optical Transient Sky with the Palomar Transient Factory}",
      journal = {\pasp},
         year = 2009,
        month = dec,
       volume = {121},
       number = {886},
        pages = {1334},
          doi = {10.1086/605911},
archivePrefix = {arXiv},
       eprint = {0906.5355},
 primaryClass = {astro-ph.CO},
       adsurl = {https://ui.adsabs.harvard.edu/abs/2009PASP..121.1334R}
}

@article{Tartaglia_2017,
   title={The Progenitor and Early Evolution of the Type IIb SN 2016gkg},
   volume={836},
   ISSN={2041-8213},
   url={http://dx.doi.org/10.3847/2041-8213/aa5c7f},
   DOI={10.3847/2041-8213/aa5c7f},
   number={1},
   journal={The Astrophysical Journal},
   publisher={American Astronomical Society},
   author={Tartaglia, L. and Fraser, M. and Sand, D. J. and Valenti, S. and Smartt, S. J. and McCully, C. and Anderson, J. P. and Arcavi, I. and Elias-Rosa, N. and Galbany, L. and Gal-Yam, A. and Haislip, J. B. and Hosseinzadeh, G. and Howell, D. A. and Inserra, C. and Jha, S. W. and Kankare, E. and Lundqvist, P. and Maguire, K. and Mattila, S. and Reichart, D. and Smith, K. W. and Smith, M. and Stritzinger, M. and Sullivan, M. and Taddia, F. and Tomasella, L.},
   year={2017},
   month=feb, pages={L12} }

@ARTICLE{Willick1997,
       author = {{Willick}, Jeffrey A. and {Courteau}, St{\'e}phane and {Faber}, S.~M. and {Burstein}, David and {Dekel}, Avishai and {Strauss}, Michael A.},
        title = "{Homogeneous Velocity-Distance Data for Peculiar Velocity Analysis. III. The Mark III Catalog of Galaxy Peculiar Velocities}",
      journal = {\apjs},
         year = 1997,
        month = apr,
       volume = {109},
       number = {2},
        pages = {333-366},
          doi = {10.1086/312983},
archivePrefix = {arXiv},
       eprint = {astro-ph/9610202},
 primaryClass = {astro-ph},
       adsurl = {https://ui.adsabs.harvard.edu/abs/1997ApJS..109..333W}
}

@ARTICLE{Khazov_2016,
       author = {{Khazov}, D. and {Yaron}, O. and {Gal-Yam}, A. and {Manulis}, I. and {Rubin}, A. and {Kulkarni}, S.~R. and {Arcavi}, I. and {Kasliwal}, M.~M. and {Ofek}, E.~O. and {Cao}, Y. and {Perley}, D. and {Sollerman}, J. and {Horesh}, A. and {Sullivan}, M. and {Filippenko}, A.~V. and {Nugent}, P.~E. and {Howell}, D.~A. and {Cenko}, S.~B. and {Silverman}, J.~M. and {Ebeling}, H. and {Taddia}, F. and {Johansson}, J. and {Laher}, R.~R. and {Surace}, J. and {Rebbapragada}, U.~D. and {Wozniak}, P.~R. and {Matheson}, T.},
        title = "{Flash Spectroscopy: Emission Lines from the Ionized Circumstellar Material around <10-day-old Type II Supernovae}",
      journal = {\apj},
         year = 2016,
        month = feb,
       volume = {818},
       number = {1},
          eid = {3},
        pages = {3},
          doi = {10.3847/0004-637X/818/1/3},
archivePrefix = {arXiv},
       eprint = {1512.00846},
 primaryClass = {astro-ph.HE},
       adsurl = {https://ui.adsabs.harvard.edu/abs/2016ApJ...818....3K}
}

@ARTICLE{Hosseinzadeh_2018,
       author = {{Hosseinzadeh}, Griffin and {Valenti}, Stefano and {McCully}, Curtis and {Howell}, D. Andrew and {Arcavi}, Iair and {Jerkstrand}, Anders and {Guevel}, David and {Tartaglia}, Leonardo and {Rui}, Liming and {Mo}, Jun and {Wang}, Xiaofeng and {Huang}, Fang and {Song}, Hao and {Zhang}, Tianmeng and {Itagaki}, Koichi},
        title = "{Short-lived Circumstellar Interaction in the Low-luminosity Type IIP SN 2016bkv}",
      journal = {\apj},
         year = 2018,
        month = jul,
       volume = {861},
       number = {1},
          eid = {63},
        pages = {63},
          doi = {10.3847/1538-4357/aac5f6},
archivePrefix = {arXiv},
       eprint = {1801.00015},
 primaryClass = {astro-ph.HE},
       adsurl = {https://ui.adsabs.harvard.edu/abs/2018ApJ...861...63H}
}

@ARTICLE{Sapir_and_Waxman,
       author = {{Sapir}, Nir and {Waxman}, Eli},
        title = "{UV/Optical Emission from the Expanding Envelopes of Type II Supernovae}",
      journal = {\apj},
         year = 2017,
        month = apr,
       volume = {838},
       number = {2},
          eid = {130},
        pages = {130},
          doi = {10.3847/1538-4357/aa64df},
archivePrefix = {arXiv},
       eprint = {1607.03700},
 primaryClass = {astro-ph.HE},
       adsurl = {https://ui.adsabs.harvard.edu/abs/2017ApJ...838..130S}
}

@ARTICLE{kasen,
       author = {{Kasen}, Daniel and {Woosley}, S.~E.},
        title = "{Type II Supernovae: Model Light Curves and Standard Candle Relationships}",
      journal = {\apj},
         year = 2009,
        month = oct,
       volume = {703},
       number = {2},
        pages = {2205-2216},
          doi = {10.1088/0004-637X/703/2/2205},
archivePrefix = {arXiv},
       eprint = {0910.1590},
 primaryClass = {astro-ph.CO},
       adsurl = {https://ui.adsabs.harvard.edu/abs/2009ApJ...703.2205K}
}

@BOOK{Levesque_2017,
       author = {{Levesque}, Emily M.},
        title = "{Astrophysics of Red Supergiants}",
         year = 2017,
          doi = {10.1088/978-0-7503-1329-2},
       adsurl = {https://ui.adsabs.harvard.edu/abs/2017ars..book.....L}
}

@INCOLLECTION{Gal-Yam2017,
       author = {{Gal-Yam}, Avishay},
        title = "{Observational and Physical Classification of Supernovae}",
    booktitle = {Handbook of Supernovae},
         year = 2017,
       editor = {{Alsabti}, Athem W. and {Murdin}, Paul},
        pages = {195},
          doi = {10.1007/978-3-319-21846-5_35},
       adsurl = {https://ui.adsabs.harvard.edu/abs/2017hsn..book..195G}
}

@ARTICLE{Leonard_2011,
       author = {{Leonard}, Douglas C.},
        title = "{On the progenitors of core-collapse supernovae}",
      journal = {\apss},
         year = 2011,
        month = nov,
       volume = {336},
       number = {1},
        pages = {117-122},
          doi = {10.1007/s10509-010-0530-8},
archivePrefix = {arXiv},
       eprint = {1011.0203},
 primaryClass = {astro-ph.SR},
       adsurl = {https://ui.adsabs.harvard.edu/abs/2011Ap&SS.336..117L}
}

@article{Gal_Yam_2010id,
	doi = {10.1088/0004-637x/736/2/159},
  
	url = {https://doi.org/10.1088%2F0004-637x%2F736%2F2%2F159},
  
	year = 2011,
	month = {jul},
  
	publisher = {American Astronomical Society},
  
	volume = {736},
  
	number = {2},
  
	pages = {159},
  
	author = {Avishay Gal-Yam and Mansi M. Kasliwal and Iair Arcavi and Yoav Green and Ofer Yaron and Sagi Ben-Ami and Dong Xu and Assaf Sternberg and Robert M. Quimby and Shrinivas R. Kulkarni and Eran O. Ofek and Richard Walters and Peter E. Nugent and Dovi Poznanski and Joshua S. Bloom and S. Bradley Cenko and Alexei V. Filippenko and Weidong Li and Jeffrey M. Silverman and Emma S. Walker and Mark Sullivan and K. Maguire and D. Andrew Howell and Paolo A. Mazzali and Dale A. Frail and David Bersier and Phil A. James and C. W. Akerlof and Fang Yuan and Nicholas Law and Derek B. Fox and Neil Gehrels},
  
	title = {{REAL}-{TIME} {DETECTION} {AND} {RAPID} {MULTIWAVELENGTH} {FOLLOW}-{UP} {OBSERVATIONS} {OF} A {HIGHLY} {SUBLUMINOUS} {TYPE} {II}-P {SUPERNOVA} {FROM} {THE} {PALOMAR} {TRANSIENT} {FACTORY} {SURVEY}
},
  
	journal = {The Astrophysical Journal}
}

@misc{Jacobson_2023,
      title={SN 2023ixf in Messier 101: Photo-ionization of Dense, Close-in Circumstellar Material in a Nearby Type II Supernova}, 
      author={W. V. Jacobson-Galan and L. Dessart and R. Margutti and R. Chornock and R. J. Foley and C. D. Kilpatrick and D. O. Jones and K. Taggart and C. R. Angus and S. Bhattacharjee and L. A. Braff and D. Brethauer and A. J. Burgasser and F. Cao and C. M. Carlile and K. C. Chambers and D. A. Coulter and E. Dominguez-Ruiz and C. B. Dickinson and T. de Boer and A. Gagliano and C. Gall and H. Gao and E. L. Gates and S. Gomez and M. Guolo and M. R. J. Halford and J. Hjorth and M. E. Huber and M. N. Johnson and P. R. Karpoor and T. Laskar and N LeBaron and Z. Li and Y. Lin and S. D. Loch and P. D. Lynam and E. A. Magnier and P. Maloney and D. J. Matthews and M. McDonald and H. -Y. Miao and D. Milisavljevic and Y. -C. Pan and S. Pradyumna and C. L. Ransome and J. M. Rees and A. Rest and C. Rojas-Bravo and N. R. Sandford and L. Sandoval Ascencio and S. Sanjaripour and A. Savino and H. Sears and N. Sharei and S. J. Smartt and E. R. Softich and C. A. Theissen and S. Tinyanont and H. Tohfa and V. A. Villar and Q. Wang and R. J. Wainscoat and A. L. Westerling and E. Wiston and M. A. Wozniak and S. K. Yadavalli and Y. Zenati},
      year={2023},
      eprint={2306.04721},
      archivePrefix={arXiv},
      primaryClass={astro-ph.HE}
}

@ARTICLE{Zacharias2013,
       author = {{Zacharias}, N. and {Finch}, C.~T. and {Girard}, T.~M. and {Henden}, A. and {Bartlett}, J.~L. and {Monet}, D.~G. and {Zacharias}, M.~I.},
        title = "{The Fourth US Naval Observatory CCD Astrograph Catalog (UCAC4)}",
      journal = {\aj},
         year = 2013,
        month = feb,
       volume = {145},
       number = {2},
          eid = {44},
        pages = {44},
          doi = {10.1088/0004-6256/145/2/44},
archivePrefix = {arXiv},
       eprint = {1212.6182},
 primaryClass = {astro-ph.IM},
       adsurl = {https://ui.adsabs.harvard.edu/abs/2013AJ....145...44Z}
}

@INPROCEEDINGS{McCully_2018,
       author = {{McCully}, Curtis and {Volgenau}, Nikolaus H. and {Harbeck}, Daniel-Rolf and {Lister}, Tim A. and {Saunders}, Eric S. and {Turner}, Monica L. and {Siiverd}, Robert J. and {Bowman}, Mark},
        title = "{Real-time processing of the imaging data from the network of Las Cumbres Observatory Telescopes using BANZAI}",
    booktitle = {Software and Cyberinfrastructure for Astronomy V},
         year = 2018,
       editor = {{Guzman}, Juan C. and {Ibsen}, Jorge},
       series = {Society of Photo-Optical Instrumentation Engineers (SPIE) Conference Series},
       volume = {10707},
        month = jul,
          eid = {107070K},
        pages = {107070K},
          doi = {10.1117/12.2314340},
archivePrefix = {arXiv},
       eprint = {1811.04163},
 primaryClass = {astro-ph.IM},
       adsurl = {https://ui.adsabs.harvard.edu/abs/2018SPIE10707E..0KM}
}

@ARTICLE{redshift_NED,
       author = {{Kaldare}, Raven and {Colless}, Matthew and {Raychaudhury}, Somak and {Peterson}, B.~A.},
        title = "{FLASH redshift survey - I. Observations and catalogue}",
      journal = {\mnras},
         year = 2003,
        month = mar,
       volume = {339},
       number = {3},
        pages = {652-662},
          doi = {10.1046/j.1365-8711.2003.05695.x},
archivePrefix = {arXiv},
       eprint = {astro-ph/0109415},
 primaryClass = {astro-ph},
       adsurl = {https://ui.adsabs.harvard.edu/abs/2003MNRAS.339..652K}
}

@article{Tully_2016,
   title={COSMICFLOWS-3},
   volume={152},
   ISSN={1538-3881},
   url={http://dx.doi.org/10.3847/0004-6256/152/2/50},
   DOI={10.3847/0004-6256/152/2/50},
   number={2},
   journal={The Astronomical Journal},
   publisher={American Astronomical Society},
   author={Tully, R. Brent and Courtois, Hélène M. and Sorce, Jenny G.},
   year={2016},
   month=aug, pages={50} }

@ARTICLE{data_release14,
       author = {{Abolfathi}, Bela and {Aguado}, D.~S. and {Aguilar}, Gabriela and {Allende Prieto}, Carlos and {Almeida}, Andres and {Ananna}, Tonima Tasnim and {Anders}, Friedrich and {Anderson}, Scott F. and {Andrews}, Brett H. and {Anguiano}, Borja and {Arag{\'o}n-Salamanca}, Alfonso and {Argudo-Fern{\'a}ndez}, Maria and {Armengaud}, Eric and {Ata}, Metin and {Aubourg}, Eric and {Avila-Reese}, Vladimir and {Badenes}, Carles and {Bailey}, Stephen and {Balland}, Christophe and {Barger}, Kathleen A. and {Barrera-Ballesteros}, Jorge and {Bartosz}, Curtis and {Bastien}, Fabienne and {Bates}, Dominic and {Baumgarten}, Falk and {Bautista}, Julian and {Beaton}, Rachael and {Beers}, Timothy C. and {Belfiore}, Francesco and {Bender}, Chad F. and {Bernardi}, Mariangela and {Bershady}, Matthew A. and {Beutler}, Florian and {Bird}, Jonathan C. and {Bizyaev}, Dmitry and {Blanc}, Guillermo A. and {Blanton}, Michael R. and {Blomqvist}, Michael and {Bolton}, Adam S. and {Boquien}, M{\'e}d{\'e}ric and {Borissova}, Jura and {Bovy}, Jo and {Bradna Diaz}, Christian Andres and {Brandt}, William Nielsen and {Brinkmann}, Jonathan and {Brownstein}, Joel R. and {Bundy}, Kevin and {Burgasser}, Adam J. and {Burtin}, Etienne and {Busca}, Nicol{\'a}s G. and {Ca{\~n}as}, Caleb I. and {Cano-D{\'\i}az}, Mariana and {Cappellari}, Michele and {Carrera}, Ricardo and {Casey}, Andrew R. and {Cervantes Sodi}, Bernardo and {Chen}, Yanping and {Cherinka}, Brian and {Chiappini}, Cristina and {Choi}, Peter Doohyun and {Chojnowski}, Drew and {Chuang}, Chia-Hsun and {Chung}, Haeun and {Clerc}, Nicolas and {Cohen}, Roger E. and {Comerford}, Julia M. and {Comparat}, Johan and {Correa do Nascimento}, Janaina and {da Costa}, Luiz and {Cousinou}, Marie-Claude and {Covey}, Kevin and {Crane}, Jeffrey D. and {Cruz-Gonzalez}, Irene and {Cunha}, Katia and {da Silva Ilha}, Gabriele and {Damke}, Guillermo J. and {Darling}, Jeremy and {Davidson}, James W., Jr. and {Dawson}, Kyle and {de Icaza Lizaola}, Miguel Angel C. and {de la Macorra}, Axel and {de la Torre}, Sylvain and {De Lee}, Nathan and {de Sainte Agathe}, Victoria and {Deconto Machado}, Alice and {Dell'Agli}, Flavia and {Delubac}, Timoth{\'e}e and {Diamond-Stanic}, Aleksandar M. and {Donor}, John and {Downes}, Juan Jos{\'e} and {Drory}, Niv and {du Mas des Bourboux}, H{\'e}lion and {Duckworth}, Christopher J. and {Dwelly}, Tom and {Dyer}, Jamie and {Ebelke}, Garrett and {Davis Eigenbrot}, Arthur and {Eisenstein}, Daniel J. and {Elsworth}, Yvonne P. and {Emsellem}, Eric and {Eracleous}, Michael and {Erfanianfar}, Ghazaleh and {Escoffier}, Stephanie and {Fan}, Xiaohui and {Fern{\'a}ndez Alvar}, Emma and {Fernandez-Trincado}, J.~G. and {Fernando Cirolini}, Rafael and {Feuillet}, Diane and {Finoguenov}, Alexis and {Fleming}, Scott W. and {Font-Ribera}, Andreu and {Freischlad}, Gordon and {Frinchaboy}, Peter and {Fu}, Hai and {G{\'o}mez Maqueo Chew}, Yilen and {Galbany}, Llu{\'\i}s and {Garc{\'\i}a P{\'e}rez}, Ana E. and {Garcia-Dias}, R. and {Garc{\'\i}a-Hern{\'a}ndez}, D.~A. and {Garma Oehmichen}, Luis Alberto and {Gaulme}, Patrick and {Gelfand}, Joseph and {Gil-Mar{\'\i}n}, H{\'e}ctor and {Gillespie}, Bruce A. and {Goddard}, Daniel and {Gonz{\'a}lez Hern{\'a}ndez}, Jonay I. and {Gonzalez-Perez}, Violeta and {Grabowski}, Kathleen and {Green}, Paul J. and {Grier}, Catherine J. and {Gueguen}, Alain and {Guo}, Hong and {Guy}, Julien and {Hagen}, Alex and {Hall}, Patrick and {Harding}, Paul and {Hasselquist}, Sten and {Hawley}, Suzanne and {Hayes}, Christian R. and {Hearty}, Fred and {Hekker}, Saskia and {Hernandez}, Jesus and {Hernandez Toledo}, Hector and {Hogg}, David W. and {Holley-Bockelmann}, Kelly and {Holtzman}, Jon A. and {Hou}, Jiamin and {Hsieh}, Bau-Ching and {Hunt}, Jason A.~S. and {Hutchinson}, Timothy A. and {Hwang}, Ho Seong and {Jimenez Angel}, Camilo Eduardo and {Johnson}, Jennifer A. and {Jones}, Amy and {J{\"o}nsson}, Henrik and {Jullo}, Eric and {Khan}, Fahim Sakil and {Kinemuchi}, Karen and {Kirkby}, David and {Kirkpatrick}, Charles C., IV and {Kitaura}, Francisco-Shu and {Knapp}, Gillian R. and {Kneib}, Jean-Paul and {Kollmeier}, Juna A. and {Lacerna}, Ivan and {Lane}, Richard R. and {Lang}, Dustin and {Law}, David R. and {Le Goff}, Jean-Marc and {Lee}, Young-Bae and {Li}, Hongyu and {Li}, Cheng and {Lian}, Jianhui and {Liang}, Yu and {Lima}, Marcos and {Lin}, Lihwai and {Long}, Dan and {Lucatello}, Sara and {Lundgren}, Britt and {Mackereth}, J. Ted and {MacLeod}, Chelsea L. and {Mahadevan}, Suvrath and {Maia}, Marcio Antonio Geimba and {Majewski}, Steven and {Manchado}, Arturo and {Maraston}, Claudia and {Mariappan}, Vivek and {Marques-Chaves}, Rui and {Masseron}, Thomas and {Masters}, Karen L. and {McDermid}, Richard M. and {McGreer}, Ian D. and {Melendez}, Matthew and {Meneses-Goytia}, Sofia and {Merloni}, Andrea and {Merrifield}, Michael R. and {Meszaros}, Szabolcs and {Meza}, Andres and {Minchev}, Ivan and {Minniti}, Dante and {Mueller}, Eva-Maria and {Muller-Sanchez}, Francisco and {Muna}, Demitri and {Mu{\~n}oz}, Ricardo R. and {Myers}, Adam D. and {Nair}, Preethi and {Nandra}, Kirpal and {Ness}, Melissa and {Newman}, Jeffrey A. and {Nichol}, Robert C. and {Nidever}, David L. and {Nitschelm}, Christian and {Noterdaeme}, Pasquier and {O'Connell}, Julia and {Oelkers}, Ryan James and {Oravetz}, Audrey and {Oravetz}, Daniel and {Ort{\'\i}z}, Erik Aquino and {Osorio}, Yeisson and {Pace}, Zach and {Padilla}, Nelson and {Palanque-Delabrouille}, Nathalie and {Palicio}, Pedro Alonso and {Pan}, Hsi-An and {Pan}, Kaike and {Parikh}, Taniya and {P{\^a}ris}, Isabelle and {Park}, Changbom and {Peirani}, Sebastien and {Pellejero-Ibanez}, Marcos and {Penny}, Samantha and {Percival}, Will J. and {Perez-Fournon}, Ismael and {Petitjean}, Patrick and {Pieri}, Matthew M. and {Pinsonneault}, Marc and {Pisani}, Alice and {Prada}, Francisco and {Prakash}, Abhishek and {Queiroz}, Anna B{\'a}rbara de Andrade and {Raddick}, M. Jordan and {Raichoor}, Anand and {Barboza Rembold}, Sandro and {Richstein}, Hannah and {Riffel}, Rogemar A. and {Riffel}, Rog{\'e}rio and {Rix}, Hans-Walter and {Robin}, Annie C. and {Rodr{\'\i}guez Torres}, Sergio and {Rom{\'a}n-Z{\'u}{\~n}iga}, Carlos and {Ross}, Ashley J. and {Rossi}, Graziano and {Ruan}, John and {Ruggeri}, Rossana and {Ruiz}, Jose and {Salvato}, Mara and {S{\'a}nchez}, Ariel G. and {S{\'a}nchez}, Sebasti{\'a}n F. and {Sanchez Almeida}, Jorge and {S{\'a}nchez-Gallego}, Jos{\'e} R. and {Santana Rojas}, Felipe Antonio and {Santiago}, Bas{\'\i}lio Xavier and {Schiavon}, Ricardo P. and {Schimoia}, Jaderson S. and {Schlafly}, Edward and {Schlegel}, David and {Schneider}, Donald P. and {Schuster}, William J. and {Schwope}, Axel and {Seo}, Hee-Jong and {Serenelli}, Aldo and {Shen}, Shiyin and {Shen}, Yue and {Shetrone}, Matthew and {Shull}, Michael and {Silva Aguirre}, V{\'\i}ctor and {Simon}, Joshua D. and {Skrutskie}, Mike and {Slosar}, An{\v{z}}e and {Smethurst}, Rebecca and {Smith}, Verne and {Sobeck}, Jennifer and {Somers}, Garrett and {Souter}, Barbara J. and {Souto}, Diogo and {Spindler}, Ashley and {Stark}, David V. and {Stassun}, Keivan and {Steinmetz}, Matthias and {Stello}, Dennis and {Storchi-Bergmann}, Thaisa and {Streblyanska}, Alina and {Stringfellow}, Guy S. and {Su{\'a}rez}, Genaro and {Sun}, Jing and {Szigeti}, Laszlo and {Taghizadeh-Popp}, Manuchehr and {Talbot}, Michael S. and {Tang}, Baitian and {Tao}, Charling and {Tayar}, Jamie and {Tembe}, Mita and {Teske}, Johanna and {Thakar}, Aniruddha R. and {Thomas}, Daniel and {Tissera}, Patricia and {Tojeiro}, Rita and {Tremonti}, Christy and {Troup}, Nicholas W. and {Urry}, Meg and {Valenzuela}, O. and {van den Bosch}, Remco and {Vargas-Gonz{\'a}lez}, Jaime and {Vargas-Maga{\~n}a}, Mariana and {Vazquez}, Jose Alberto and {Villanova}, Sandro and {Vogt}, Nicole and {Wake}, David and {Wang}, Yuting and {Weaver}, Benjamin Alan and {Weijmans}, Anne-Marie and {Weinberg}, David H. and {Westfall}, Kyle B. and {Whelan}, David G. and {Wilcots}, Eric and {Wild}, Vivienne and {Williams}, Rob A. and {Wilson}, John and {Wood-Vasey}, W.~M. and {Wylezalek}, Dominika and {Xiao}, Ting and {Yan}, Renbin and {Yang}, Meng and {Ybarra}, Jason E. and {Y{\`e}che}, Christophe and {Zakamska}, Nadia and {Zamora}, Olga and {Zarrouk}, Pauline and {Zasowski}, Gail and {Zhang}, Kai and {Zhao}, Cheng and {Zhao}, Gong-Bo and {Zheng}, Zheng and {Zheng}, Zheng and {Zhou}, Zhi-Min and {Zhu}, Guangtun and {Zinn}, Joel C. and {Zou}, Hu},
        title = "{The Fourteenth Data Release of the Sloan Digital Sky Survey: First Spectroscopic Data from the Extended Baryon Oscillation Spectroscopic Survey and from the Second Phase of the Apache Point Observatory Galactic Evolution Experiment}",
      journal = {\apjs},
         year = 2018,
        month = apr,
       volume = {235},
       number = {2},
          eid = {42},
        pages = {42},
          doi = {10.3847/1538-4365/aa9e8a},
archivePrefix = {arXiv},
       eprint = {1707.09322},
 primaryClass = {astro-ph.GA},
       adsurl = {https://ui.adsabs.harvard.edu/abs/2018ApJS..235...42A}
}

@ARTICLE{Kiewe2012,
       author = {{Kiewe}, Michael and {Gal-Yam}, Avishay and {Arcavi}, Iair and {Leonard}, Douglas C. and {Emilio Enriquez}, J. and {Cenko}, S. Bradley and {Fox}, Derek B. and {Moon}, Dae-Sik and {Sand}, David J. and {Soderberg}, Alicia M. and {CCCP}, The},
        title = "{Caltech Core-Collapse Project (CCCP) Observations of Type IIn Supernovae: Typical Properties and Implications for Their Progenitor Stars}",
      journal = {\apj},
         year = 2012,
        month = jan,
       volume = {744},
       number = {1},
          eid = {10},
        pages = {10},
          doi = {10.1088/0004-637X/744/1/10},
archivePrefix = {arXiv},
       eprint = {1010.2689},
 primaryClass = {astro-ph.CO},
       adsurl = {https://ui.adsabs.harvard.edu/abs/2012ApJ...744...10K}
}

@article{Teja_2023,
   title={Far-ultraviolet to Near-infrared Observations of SN 2023ixf: A High-energy Explosion Engulfed in Complex Circumstellar Material},
   volume={954},
   ISSN={2041-8213},
   url={http://dx.doi.org/10.3847/2041-8213/acef20},
   DOI={10.3847/2041-8213/acef20},
   number={1},
   journal={The Astrophysical Journal Letters},
   publisher={American Astronomical Society},
   author={Teja, Rishabh Singh and Singh, Avinash and Basu, Judhajeet and Anupama, G. C. and Sahu, D. K. and Dutta, Anirban and Swain, Vishwajeet and Nakaoka, Tatsuya and Pathak, Utkarsh and Bhalerao, Varun and Barway, Sudhanshu and Kumar, Harsh and A. J., Nayana and Imazawa, Ryo and Kumar, Brajesh and Kawabata, Koji S.},
   year={2023},
   month=aug, pages={L12} }

@ARTICLE{2023ixf_discovery,
       author = {{Itagaki}, K.},
        title = "{Transient Discovery Report for 2023-05-19}",
      journal = {Transient Name Server Discovery Report},
         year = 2023,
        month = may,
       volume = {2023-1158},
        pages = {1},
       adsurl = {https://ui.adsabs.harvard.edu/abs/2023TNSTR1158....1I}
}

@ARTICLE{2023ixf_classification,
       author = {{Perley}, D. and {Gal-Yam}, A.},
        title = "{Transient Classification Report for 2023-05-19}",
      journal = {Transient Name Server Classification Report},
         year = 2023,
        month = may,
       volume = {2023-1164},
        pages = {1},
       adsurl = {https://ui.adsabs.harvard.edu/abs/2023TNSCR1164....1P}
}

@article{Dyk_2017,
author = {Van Dyk, Schuyler D. },
title = {The direct identification of core-collapse supernova progenitors},
journal = {Philosophical Transactions of the Royal Society A: Mathematical, Physical and Engineering Sciences},
volume = {375},
number = {2105},
pages = {20160277},
year = {2017},
doi = {10.1098/rsta.2016.0277},

URL = {https://royalsocietypublishing.org/doi/abs/10.1098/rsta.2016.0277},
eprint = {https://royalsocietypublishing.org/doi/pdf/10.1098/rsta.2016.0277}
}

@ARTICLE{Beasor_2020,
       author = {{Beasor}, Emma R. and {Davies}, Ben and {Smith}, Nathan and {van Loon}, Jacco Th and {Gehrz}, Robert D. and {Figer}, Donald F.},
        title = "{A new mass-loss rate prescription for red supergiants}",
      journal = {\mnras},
         year = 2020,
        month = mar,
       volume = {492},
       number = {4},
        pages = {5994-6006},
          doi = {10.1093/mnras/staa255},
archivePrefix = {arXiv},
       eprint = {2001.07222},
 primaryClass = {astro-ph.SR},
       adsurl = {https://ui.adsabs.harvard.edu/abs/2020MNRAS.492.5994B}
}

@article{Massey_2023,
   title={The Time-averaged Mass-loss Rates of Red Supergiants as Revealed by Their Luminosity Functions in M31 and M33},
   volume={942},
   ISSN={1538-4357},
   url={http://dx.doi.org/10.3847/1538-4357/aca665},
   DOI={10.3847/1538-4357/aca665},
   number={2},
   journal={The Astrophysical Journal},
   publisher={American Astronomical Society},
   author={Massey, Philip and Neugent, Kathryn F. and Ekström, Sylvia and Georgy, Cyril and Meynet, Georges},
   year={2023},
   month=jan, pages={69} }

@ARTICLE{RabinakWaxman_2011,
       author = {{Rabinak}, Itay and {Waxman}, Eli},
        title = "{The Early UV/Optical Emission from Core-collapse Supernovae}",
      journal = {\apj},
         year = 2011,
        month = feb,
       volume = {728},
       number = {1},
          eid = {63},
        pages = {63},
          doi = {10.1088/0004-637X/728/1/63},
archivePrefix = {arXiv},
       eprint = {1002.3414},
 primaryClass = {astro-ph.HE},
       adsurl = {https://ui.adsabs.harvard.edu/abs/2011ApJ...728...63R}
}

@ARTICLE{Morag_2023,
       author = {{Morag}, Jonathan and {Sapir}, Nir and {Waxman}, Eli},
        title = "{Shock cooling emission from explosions of red supergiants - I. A numerically calibrated analytic model}",
      journal = {\mnras},
         year = 2023,
        month = jun,
       volume = {522},
       number = {2},
        pages = {2764-2776},
          doi = {10.1093/mnras/stad899},
archivePrefix = {arXiv},
       eprint = {2207.06179},
 primaryClass = {astro-ph.HE},
       adsurl = {https://ui.adsabs.harvard.edu/abs/2023MNRAS.522.2764M}
}

@article{Dong_2020,
   title={Supernova 2018cuf: A Type IIP Supernova with a Slow Fall from Plateau},
   volume={906},
   ISSN={1538-4357},
   url={http://dx.doi.org/10.3847/1538-4357/abc417},
   DOI={10.3847/1538-4357/abc417},
   number={1},
   journal={The Astrophysical Journal},
   publisher={American Astronomical Society},
   author={Dong , Yize  and Valenti, S. and Bostroem, K. A. and Sand, D. J. and Andrews, Jennifer E. and Galbany, L. and Jha, Saurabh W. and Eweis, Youssef and Kwok, Lindsey and Hsiao, E. Y. and Davis, Scott and Brown, Peter J. and Kuncarayakti, H. and Maeda, Keiichi and Rho, Jeonghee and Amaro, R. C. and Anderson, J. P. and Arcavi, Iair and Burke, Jamison and Dastidar, Raya and Folatelli, Gastón and Haislip, Joshua and Hiramatsu, Daichi and Hosseinzadeh, Griffin and Howell, D. Andrew and Jencson, J. and Kouprianov, Vladimir and Lundquist, M. and Lyman, J. D. and McCully, Curtis and Misra, Kuntal and Reichart, Daniel E. and Sánchez, S. F. and Smith, Nathan and Wang, Xiaofeng and Wang, Lingzhi and Wyatt, S.},
   year={2020},
   month=jan, pages={56} }

@article{Hosseinzadeh_2023,
   title={Shock Cooling and Possible Precursor Emission in the Early Light Curve of the Type II SN 2023ixf},
   volume={953},
   ISSN={2041-8213},
   url={http://dx.doi.org/10.3847/2041-8213/ace4c4},
   DOI={10.3847/2041-8213/ace4c4},
   number={1},
   journal={The Astrophysical Journal Letters},
   publisher={American Astronomical Society},
   author={Hosseinzadeh, Griffin and Farah, Joseph and Shrestha, Manisha and Sand, David J. and Dong 董, Yize 一泽 and Brown, Peter J. and Bostroem, K. Azalee and Valenti, Stefano and Jha, Saurabh W. and Andrews, Jennifer E. and Arcavi, Iair and Haislip, Joshua and Hiramatsu, Daichi and Hoang, Emily and Howell, D. Andrew and Janzen, Daryl and Jencson, Jacob E. and Kouprianov, Vladimir and Lundquist, Michael and McCully, Curtis and Meza Retamal, Nicolas E. and Modjaz, Maryam and Newsome, Megan and Gonzalez, Estefania Padilla and Pearson, Jeniveve and Pellegrino, Craig and Ravi, Aravind P. and Reichart, Daniel E. and Smith, Nathan and Terreran, Giacomo and Vinkó, József},
   year={2023},
   month=aug, pages={L16} }

@ARTICLE{Heger_2003,
       author = {{Heger}, A. and {Fryer}, C.~L. and {Woosley}, S.~E. and {Langer}, N. and {Hartmann}, D.~H.},
        title = "{How Massive Single Stars End Their Life}",
      journal = {\apj},
         year = 2003,
        month = jul,
       volume = {591},
       number = {1},
        pages = {288-300},
          doi = {10.1086/375341},
archivePrefix = {arXiv},
       eprint = {astro-ph/0212469},
 primaryClass = {astro-ph},
       adsurl = {https://ui.adsabs.harvard.edu/abs/2003ApJ...591..288H}
}

@ARTICLE{Filippenko_1997,
       author = {{Filippenko}, Alexei V.},
        title = "{Optical Spectra of Supernovae}",
      journal = {\araa},
         year = 1997,
        month = jan,
       volume = {35},
        pages = {309-355},
          doi = {10.1146/annurev.astro.35.1.309},
       adsurl = {https://ui.adsabs.harvard.edu/abs/1997ARA&A..35..309F}
}

@ARTICLE{Moriya_2011,
       author = {{Moriya}, Takashi and {Tominaga}, Nozomu and {Blinnikov}, Sergei I. and {Baklanov}, Petr V. and {Sorokina}, Elena I.},
        title = "{Supernovae from red supergiants with extensive mass loss}",
      journal = {\mnras},
         year = 2011,
        month = jul,
       volume = {415},
       number = {1},
        pages = {199-213},
          doi = {10.1111/j.1365-2966.2011.18689.x},
archivePrefix = {arXiv},
       eprint = {1009.5799},
 primaryClass = {astro-ph.SR},
       adsurl = {https://ui.adsabs.harvard.edu/abs/2011MNRAS.415..199M}
}
\bibliographystyle{aasjournal}
\appendix \label{chap:app}
\section{SNEmcee Fits} \label{sec:snemcee_posteriors}

Figure \ref{fig:snemcee_sample} presents the SNEmcee fits (purple lines) drawn from the MCMC posterior to the observed bolometric luminosity of the slowly rising SNe ASASSN-14kg, SN\,2018fif, SN\,2021yja, and SN\,2023axu with (left) and without (right) CSM interaction. Figures \ref{fig:cornerplot_ASASSN14kg}--\ref{fig:cornerplot_SN2023axu} show the posterior probability distributions of the fit parameters with (red) and without (blue) CSM.

\begin{figure*}
\centering
\includegraphics[width=0.8\textwidth]{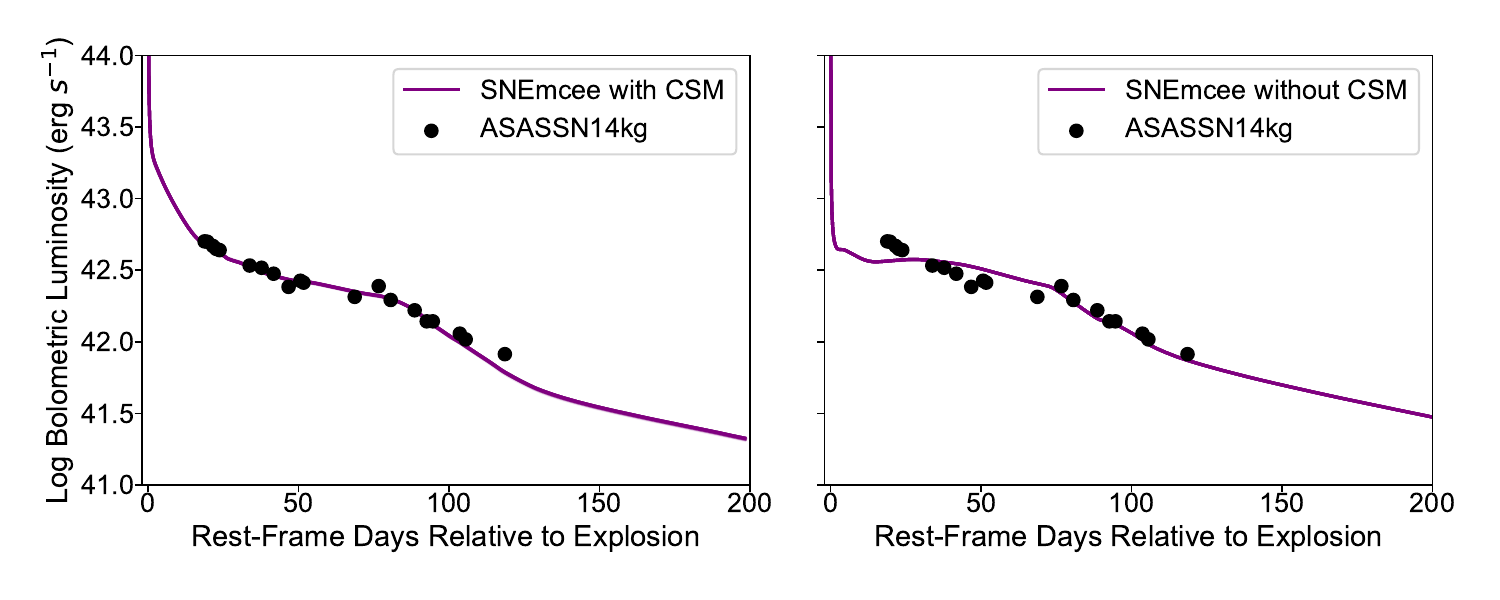} \\
\vspace{-0.5cm}
\includegraphics[width=0.8\textwidth]{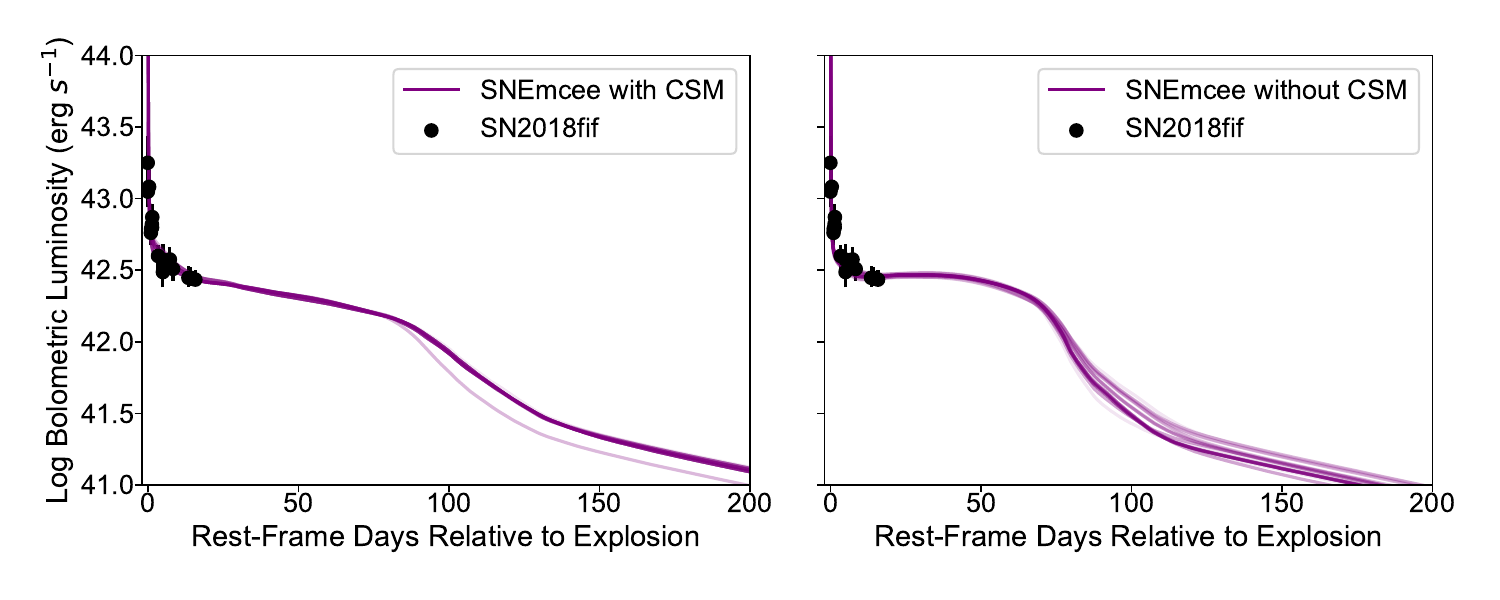} \\
\vspace{-0.5cm}
\includegraphics[width=0.8\textwidth]{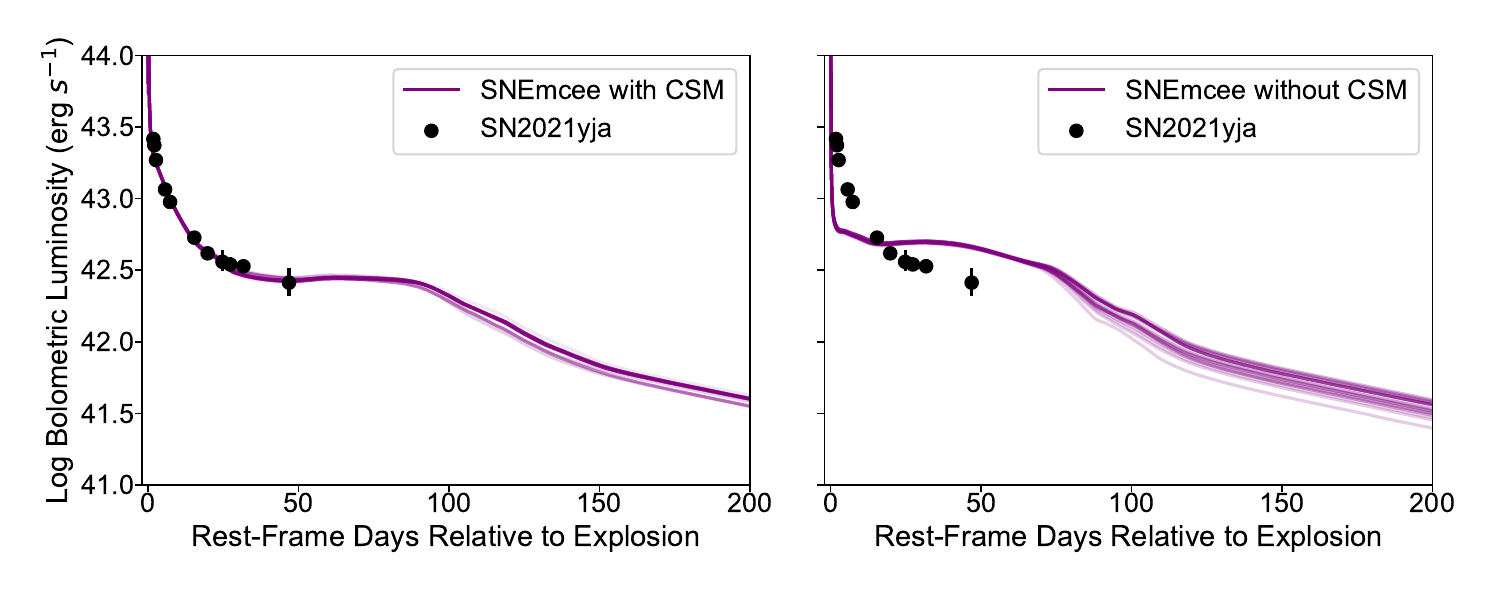} \\
\vspace{-0.5cm}
\includegraphics[width=0.8\textwidth]{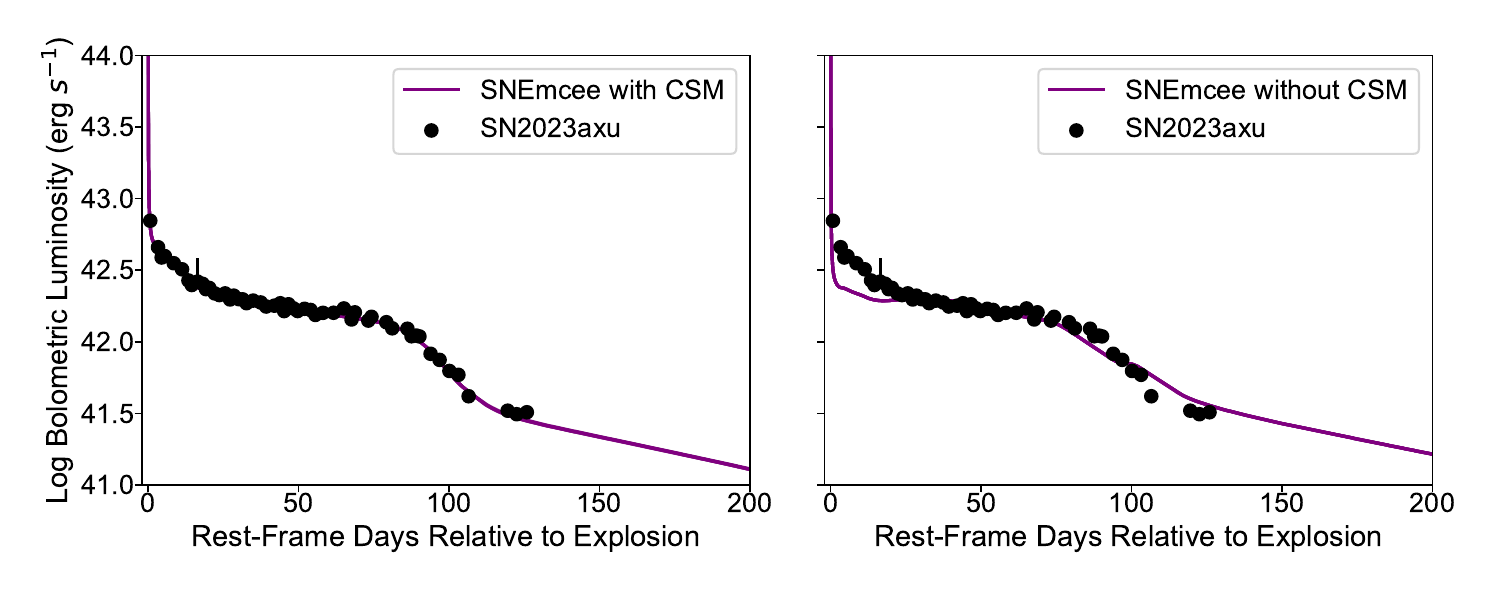}
\caption{Same as Figure \ref{fig:comparison_snemcee} for the additional four slowly rising SNe studied here.}
\label{fig:snemcee_sample}
\end{figure*}

\begin{figure*}
\centering
\includegraphics[width=1\textwidth]{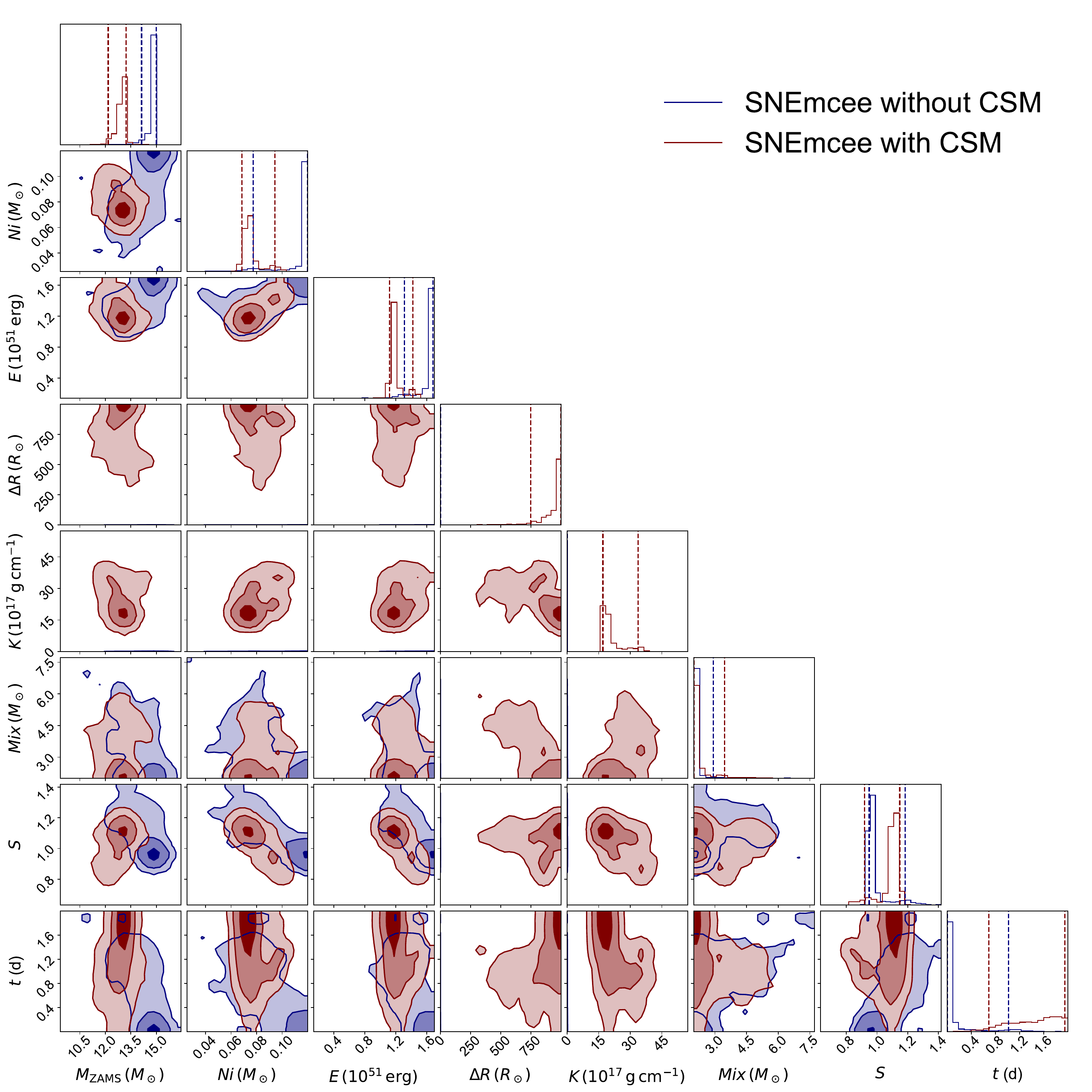}
\caption{Same as Figure \ref{fig:Snemcee_with_and_without_CSM} for ASASSN-14kg.}
\label{fig:cornerplot_ASASSN14kg}
\end{figure*}

\begin{figure*}
\centering
\includegraphics[width=1\textwidth]{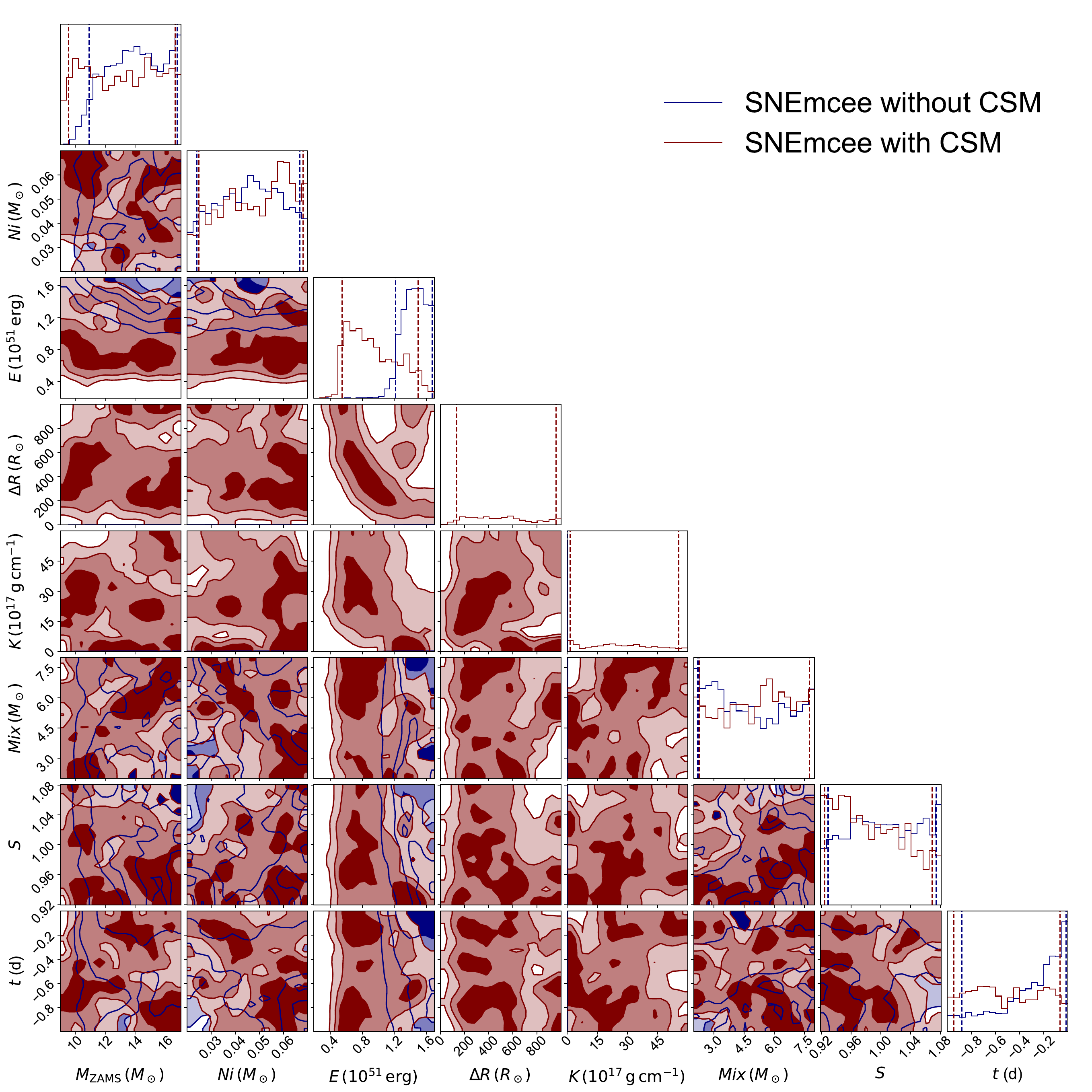}
\caption{Same as Figure \ref{fig:Snemcee_with_and_without_CSM} for SN\,2018fif.}
\label{fig:cornerplot_SN2018fif}
\end{figure*}

\begin{figure*}
\centering
\includegraphics[width=1\textwidth]{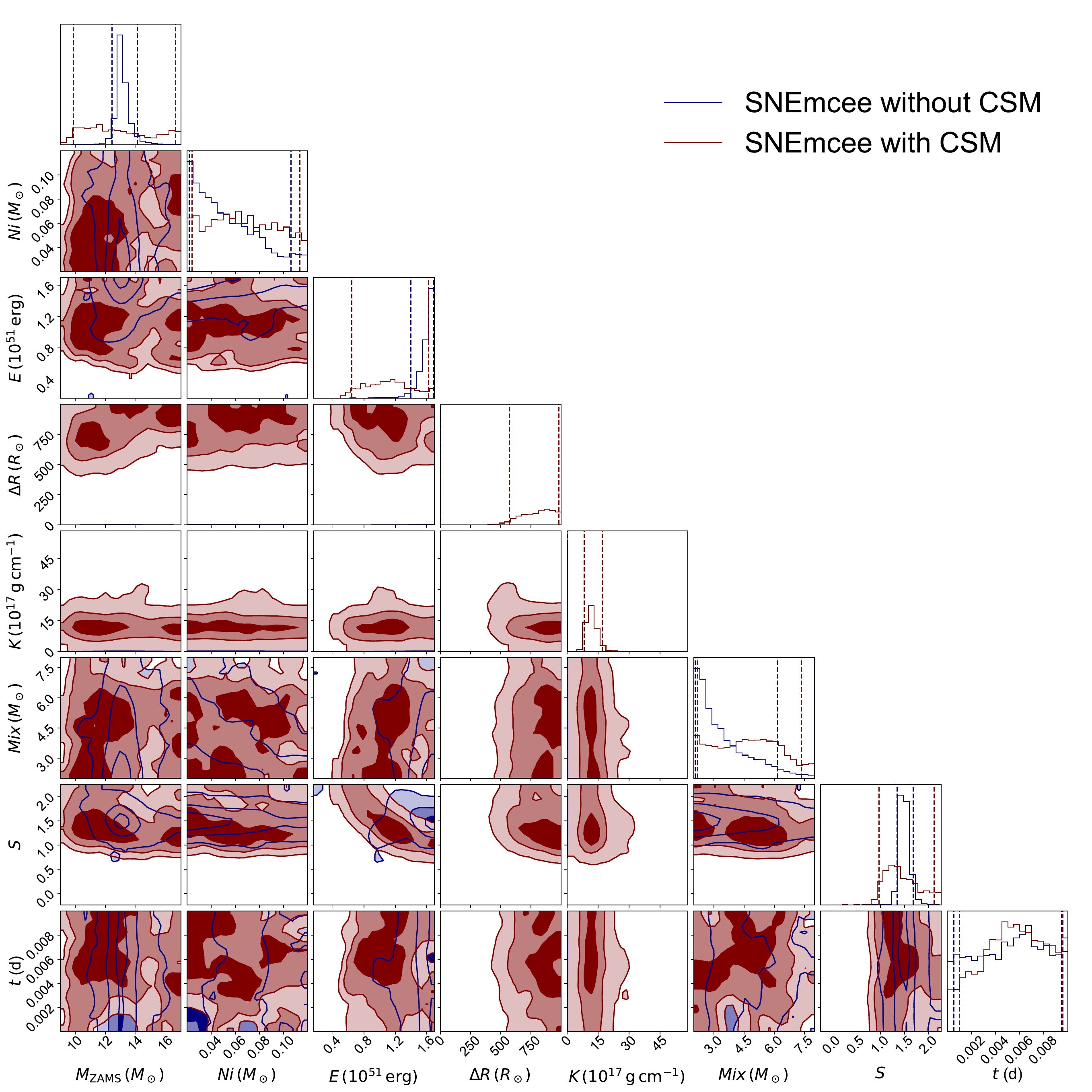}
\caption{Same as Figure \ref{fig:Snemcee_with_and_without_CSM} for SN\,2021yja.}
\label{fig:cornerplot_SN2021yja}
\end{figure*}

\begin{figure*}
\centering
\includegraphics[width=1\textwidth]{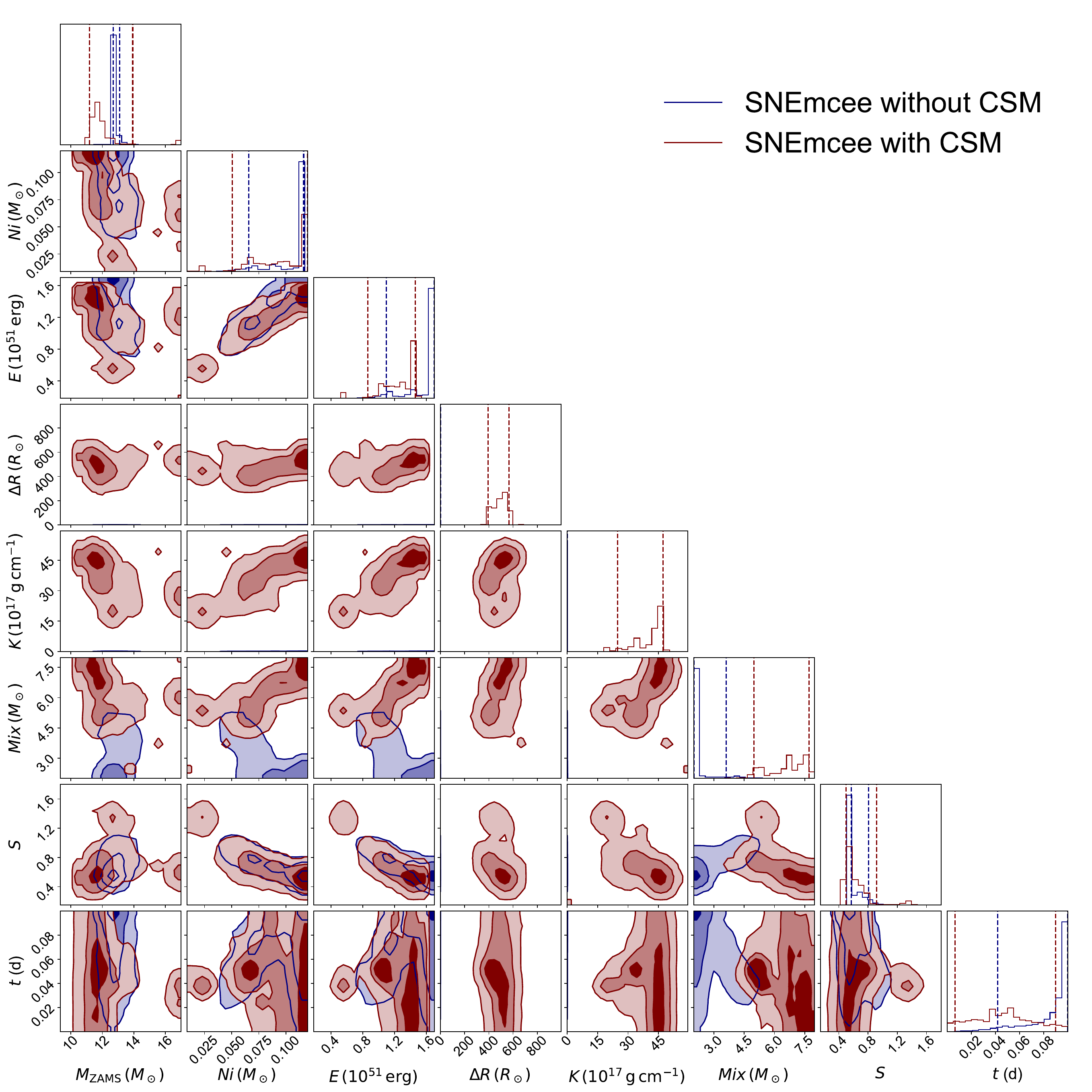}
\caption{Same as Figure \ref{fig:Snemcee_with_and_without_CSM} for SN\,2023axu.}
\label{fig:cornerplot_SN2023axu}
\end{figure*}

\section{Observations of ASASSN-14kg} \label{sec:ASASSN-14kg_spectra}

We obtained spectra of ASASSN-14kg through the GSP with the Las Cumbres Observatory FLOYDS spectrographs mounted on the 2-meter FTN at Haleakalā observatory. They were reduced in the same way as our SN\,2020bij spectra (see Section \ref{sec:Observations}). 
A log of our spectroscopic observations of ASASSN-14kg is available in Table \ref{tab:Spectra_ASASSN-14kg} and the spectra are displayed in Figure \ref{fig:Spectra_ASASSN_14kg}. All spectra will be made available through WISeREP.

\begin{deluxetable}{ccccc}[h]
\caption{Log of the spectroscopic observations of ASASSN-14kg.}
\label{tab:Spectra_ASASSN-14kg}
\centering
\tablehead{
 \colhead{MJD} & \colhead{Phase} & \colhead{Telescope} & \colhead{Slit Width} & \colhead{Exposure Time}
 \\ & {(days)} & & (\arcsec) & (s)
}
\startdata 
56980.45 & 10.79 & OGG 2m & 1.6 & 1800 \\
56983.44 & 13.74 & OGG 2m & 1.6 & 2700 \\
57007.42 & 37.37 & OGG 2m & 1.6 & 2700 \\
57009.40 & 39.33 & OGG 2m & 1.6 & 3600 \\
57030.32 & 59.95 & OGG 2m & 1.6 & 3600 \\
57048.29 & 77.66 & OGG 2m & 1.6 & 3600 \\
57070.233 & 99.29 & OGG 2m & 1.6 & 3600 \\
\enddata
\tablecomments{The phase of each spectrum is listed in rest-frame days relative to explosion. OGG denotes the Haleakalā site.}
\end{deluxetable}

\begin{figure}
\includegraphics[width=0.7\textwidth]{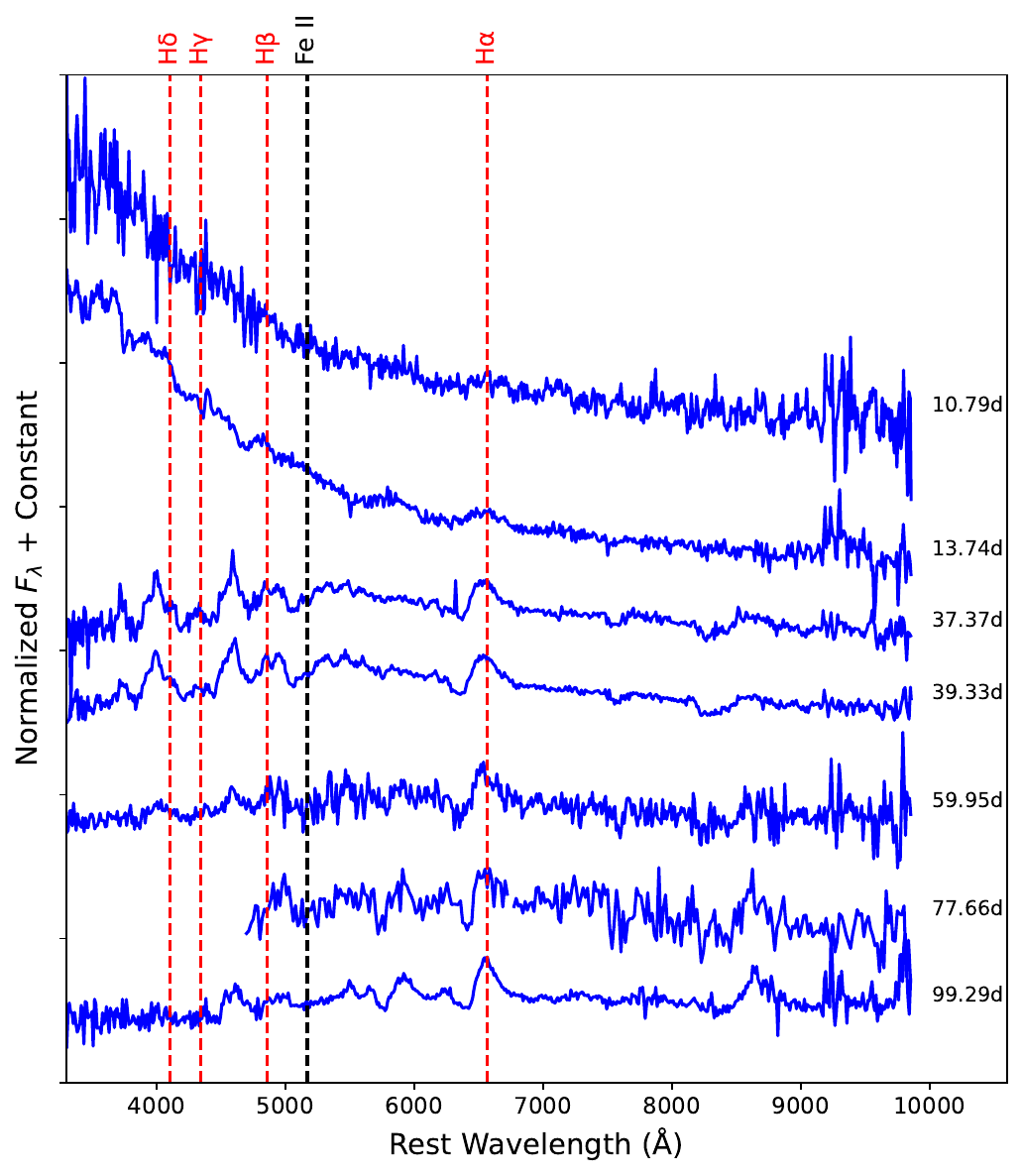}
\centering
\caption{The spectroscopic evolution of ASASSN-14kg. The phase of each spectrum in rest-frame days relative to explosion is shown on the right. All spectra are shifted in flux for clarity. The vertical colored lines at the rest wavelengths of hydrogen (\hd\ 4101, \hg\ 4340, \hb\ 4861, and \ha\ 6563\,\AA) and iron (\Feii\ 5169\,\AA) denote spectral features common in Type II SNe.}
\label{fig:Spectra_ASASSN_14kg}
\end{figure}

\end{document}